\ifdefined\VersionLong%
	\newcommand{\LongVersion}[1]{\#1}
	\newcommand{\ShortVersion}[1]{}
\else
	\newcommand{\LongVersion}[1]{}
	\newcommand{\ShortVersion}[1]{#1}
\fi

\ifdefined\VersionFinal%
  \newcommand{\FinalVersion}[1]{#1}
  \newcommand{\ArxivVersion}[1]{}
  \newcommand{\FinalOrArxivVersion}[2]{#1}
\else
  \newcommand{\FinalVersion}[1]{}
  \newcommand{\ArxivVersion}[1]{#1}
  \newcommand{\FinalOrArxivVersion}[2]{#2}
\fi

\RequirePackage{amsmath} 
\documentclass[runningheads]{llncs-mod}

\usepackage[utf8]{inputenc}
\usepackage{lmodern}

\usepackage{wrapfig}

\usepackage[pdftex]{graphicx}
\usepackage[table]{xcolor}
\usepackage{tikz}
\usetikzlibrary{arrows.meta, automata, calc, graphs, positioning}
\tikzstyle{every node}=[initial text=]
\tikzstyle{location}=[rectangle, rounded corners, minimum size=12pt, draw=black, fill=blue!10, inner sep=2pt]

\usepackage{mathtools}
\usepackage{amssymb}

\usepackage{orcidlink}
\usepackage{cite}
\usepackage{url}

\usepackage[inline]{enumitem}
\setlist{nolistsep,leftmargin=1.5em}

\usepackage[amsmath]{ntheorem}
\usepackage{cleveref}

\theoremnumbering{arabic}
\theoremseparator{.}
\theorembodyfont{\itshape}
\theorempreskip{0.8\topsep}
\theorempostskip{0.8\topsep}
\newtheorem{mytheorem}{Theorem}[section]

\theorembodyfont{\rmfamily}

\newtheorem{myremark}[mytheorem]{Remark}
\newtheorem{myexample}[mytheorem]{Example}
\newtheorem{mydefinition}[mytheorem]{Definition}

\newtheorem{myproblem}[mytheorem]{Problem}

\creflabelformat{enumi}{#2(#1)#3}
\crefname{mytheorem}{Thm.}{Thms}
\crefname{mydefinition}{Def.}{Defs}
\crefname{myproposition}{Prop.}{Props}
\crefname{myremark}{Rem.}{Remarks}
\crefname{mylemma}{Lem.}{Lemmas}
\crefname{myproof}{Proof.}{Proofs}
\crefname{myproblem}{Prob.}{Probs}
\crefname{myexample}{Ex.}{Exs}
\crefname{appendix}{Appendix}{Appendixes}
\crefname{algorithm}{Alg.}{Algs}
\crefformat{section}{{\S}#2#1#3}
\crefname{figure}{Fig.}{Figs}
\crefname{equation}{Eq.}{Eqs}
\crefname{table}{Table}{Tables}

\usepackage{algorithm}
\usepackage[commentColor=red,noEnd]{algpseudocodex}

\algrenewcommand\alglinenumber[1]{\scriptsize #1:}

\algnewcommand\Continue{\textbf{continue}}

\usepackage{comment}

\ifdefined\VersionFinal%
  \excludecomment{ArxivBlock}
\else
  \includecomment{ArxivBlock}
\fi

\usepackage{booktabs}
\usepackage{array}
\newcolumntype{L}[1]{>{\raggedright\arraybackslash}p{#1}}
\newcolumntype{R}[1]{>{\raggedleft\arraybackslash}p{#1}}

\usepackage{adjustbox}
\usepackage{blindtext}

\ifdefined\VersionWithComments%
  \usepackage[colorinlistoftodos,textsize=footnotesize]{todonotes}
\else
  \usepackage[disable]{todonotes}
\fi

\newcommand{\gennote}[3]{\todo[linecolor=#2,backgroundcolor=#2!25,bordercolor=#2]{#3: #1}}

\newcommand{\ih}[1]{\gennote{#1}{blue}{IH}}

\specialcomment{auxproof}
{\mbox{}\newline\textbf{BEGIN: AUX-PROOF}\dotfill\newline}
{\mbox{}\newline\textbf{END: AUX-PROOF}\dotfill\newline}
\excludecomment{auxproof}

\newcommand{\word}{w}
\newcommand{\alphabet}{\Sigma}
\newcommand{\Lg}{\mathcal{L}}
\newcommand{\targetLg}{\mathcal{L}_{\mathrm{tgt}}}
\newcommand{\hypothesisA}{\mathcal{A}_{\mathrm{hyp}}}

\newcommand{\cex}{\mathit{cex}}
\newcommand{\setdiff}{\triangle}

\newcommand{\seqIndex}[2]{{#1}_{[#2]}}
\newcommand{\seqSlice}[3]{{#1}_{[#2,#3)}}
\newcommand{\restrictTo}[2]{{#1}|_{#2}}

\newcommand{\incomingEdges}[1]{E^\mathsf{in}_{#1}}
\newcommand{\outgoingEdges}[1]{E^\mathsf{out}_{#1}}
\newcommand{\inputNodes}{V^\mathsf{in}}
\newcommand{\outputNodes}{V^\mathsf{out}}
\newcommand{\componentNodes}{V^\mathsf{c}}
\newcommand{\inputEdges}{E^\mathsf{in}}
\newcommand{\outputEdges}{E^\mathsf{out}}

\newcommand{\semMoore}[1]{[\![#1]\!]}

\newcommand{\componentInputAlphabet}[1]{\Sigma^\mathsf{in}_{#1}}
\newcommand{\componentOutputAlphabet}[1]{\Sigma^\mathsf{out}_{#1}}
\newcommand{\componentOutputFunction}[1]{\lambda_{#1}}
\newcommand{\inputAlphabet}{\Sigma^\mathsf{in}}
\newcommand{\outputAlphabet}{\Sigma^\mathsf{out}}
\newcommand{\totalOutputAlphabet}{\overline{\Sigma}}
\newcommand{\totalOutputFunction}{\overline{\lambda}}
\newcommand{\outputFunction}{\lambda}

\newcommand{\mmnMoore}[1]{{[#1]}}

\newcommand{\OneExtER}{\textsc{1Ext}_{\mathcal{E},\mathcal{R}}}

\newcommand{\lighting}{$\mathtt{MQTT\_Lighting}$}
\newcommand{\MMNlighting}{\mathcal{M}_{\textsf{light}}}
\newcommand{\brokerComponent}{b}
\newcommand{\brightnessComponent}{s_1}
\newcommand{\motionComponent}{s_2}
\newcommand{\lightComponent}{\ell}
\newcommand{\brokerMoore}{M_{\brokerComponent}}
\newcommand{\brightnessMoore}{M_{\brightnessComponent}}
\newcommand{\motionMoore}{M_{\motionComponent}}
\newcommand{\lightMoore}{M_{\lightComponent}}

\usepackage{subcaption}
\usepackage{xspace}

\newenvironment{renumeration}
	{\ifdefined\VersionLong\begin{enumerate}\else\begin{enumerate*}[label=\roman*)]\fi}
	{\ifdefined\VersionLong\end{enumerate}\else\end{enumerate*}\fi}

\ifdefined \VersionWithComments
 	\definecolor{colorok}{RGB}{80,80,150}
\else
	\definecolor{colorok}{RGB}{0,0,0}
\fi

\newcommand{\ie}{\textcolor{colorok}{i.e.,}\xspace}

\newcommand{\myparagraph}[1]{\smallskip\noindent \textbf{#1}\;}

\makeatletter
\renewcommand\section{\@startsection{section}{1}{\z@}%
                       {-16\p@ \@plus -4\p@ \@minus -4\p@}%
                       {10\p@ \@plus 4\p@ \@minus 4\p@}%
                       {\normalfont\large\bfseries\boldmath
                        \rightskip=\z@ \@plus 8em\pretolerance=10000 }}
\makeatother

\begin{document}

\title{
Componentwise
 Automata Learning\\ for System Integration\FinalOrArxivVersion{}{ (Extended Version)}}
\author{
\ifdefined\VersionAnonymous
\else
    Hiroya Fujinami\inst{1,5}\orcidlink{0009-0007-0794-5743}
  \and
    Masaki Waga\inst{2,1}\orcidlink{0000-0001-9360-7490}
  \and
    Jie An \inst{3}\thanks{J.A.'s technical contribution was made when he was at National Institute of Informatics, Tokyo, Japan.}\orcidlink{0000-0001-9260-9697}
  \and
    Kohei Suenaga\inst{2,1}\orcidlink{0000-0002-7466-8789}
  \and
    Nayuta Yanagisawa\inst{4}\thanks{This paper presents a theoretical investigation that is independent of any testing procedures conducted at the institutions or companies with which the industry-affiliated co-authors are associated.}
  \and
    Hiroki Iseri\inst{4}${}^{\star\star}$
  \and
    Ichiro Hasuo\inst{1,5,6}\orcidlink{0000-0002-8300-4650}
  \fi
}
\authorrunning{
\ifdefined\VersionAnonymous
\else
\fi
}
\institute{
\ifdefined\VersionAnonymous
\else
    National Institute of Informatics, Tokyo, Japan\\
       \email{\{makenowjust,hasuo\}@nii.ac.jp}
   \and
     Kyoto University, Kyoto, Japan\\
       \email{\{mwaga,ksuenaga\}@fos.kuis.kyoto-u.ac.jp}
   \and
     Institute of Software, Chinese Academy of Sciences, Beijing, China\\
       \email{anjie@iscas.ac.cn}
   \and
     Toyota Motor Corporation, Tokyo, Japan\\
       \email{\{nayuta\_yanagisawa,hiroki\_iseri\}@mail.toyota.co.jp}
   \and
     SOKENDAI (The Graduate University for Advanced Studies), Kanagawa, Japan
   \and
     Imiron Co., Ltd., Tokyo, Japan
  \fi
}

\maketitle

\begin{abstract}
\emph{Compositional automata learning} is attracting attention as an analysis technique for  complex black-box systems. It exploits a target system's internal compositional structure to reduce complexity. In this paper, we identify \emph{system integration}---the process of building a new system as a composite of potentially third-party and black-box components---as a new application domain of compositional automata learning. Accordingly, we propose a new problem setting, where the learner has direct access to black-box components. This is in contrast with the usual problem settings of compositional learning, where the target is a legacy black-box system and queries can only be made to the whole system (but not to components). We call our problem \emph{componentwise automata learning} for distinction. We identify a challenge there called \emph{component redundancies}: some parts of components may not contribute to system-level behaviors, and learning them incurs unnecessary effort. We introduce a \emph{contextual componentwise learning} algorithm that systematically removes such redundancies. We experimentally evaluate our proposal and show its practical relevance.

\keywords{automata learning \and compositional automata learning \and systems engineering \and Moore machine}
\end{abstract}


\section{Introduction}\label{sec:intro}
\myparagraph{Automata Learning}
\emph{(Active) automata learning} is a problem to infer an automaton recognizing the target language $\targetLg \subseteq \alphabet^*$ via a finite number of queries to an oracle.
The L* algorithm~\cite{DBLP:journals/iandc/Angluin87}, the best known active automata learning algorithm by Angluin,
infers the minimum DFA recognizing the target regular language $\targetLg$ via two kinds of queries: \emph{membership} and \emph{equivalence} queries.
\begin{itemize}
 \item In a membership query, the learner asks if a word $\word \in \alphabet^*$ is in $\targetLg$. 
(For machines with output (e.g.\ Mealy and Moore), membership queries are called \emph{output queries}, a term we will be using in this paper.) The answers to those membership/output queries are recorded in an \emph{(observation) table}; once the table is \emph{closed} it induces a \emph{hypothesis DFA}.
 \item In an equivalence query, the learner asks if a hypothesis DFA $\hypothesisA$ recognizes $\targetLg$. If not,
the oracle returns a \emph{counterexample} $\cex \in \targetLg \setdiff \Lg(\hypothesisA)$ that witnesses the deviation of 
 $\hypothesisA$'s language
from $\targetLg$. 
\end{itemize}
A target  of automata learning is commonly called a \emph{system under learning (SUL)}.

After the seminal work~\cite{DBLP:journals/iandc/Angluin87}, various algorithms have been proposed,
for example, to improve the efficiency~\cite{DBLP:journals/iandc/RivestS93,DBLP:conf/rv/IsbernerHS14,DBLP:conf/tacas/VaandragerGRW22} and
to learn other classes of automata (e.g.\ Mealy machines~\cite{DBLP:phd/de/Niese2003}, weighted automata~\cite{DBLP:conf/cai/2015,DBLP:conf/fossacs/HeerdtKR020}, symbolic automata~\cite{DBLP:conf/tacas/DrewsD17,DBLP:conf/cav/ArgyrosD18,DBLP:journals/lmcs/FismanFZ23}, and visibly pushdown automata~\cite{DBLP:conf/stoc/AlurM04}). The LearnLib library offers an open source framework for automata learning~\cite{DBLP:conf/cav/IsbernerHS15}. 
Many real-world applications of automata learning have been reported, too. See e.g.~\cite{DBLP:conf/birthday/BainczykSSH17,DBLP:conf/icse/DuhaibyG20}.

In the context of verification and testing,
active automata learning is used to approximate \emph{black-box} systems and obtain a surrogate model amenable to white-box analysis.
For example, automata learning of Moore or Mealy machines has been applied for
 model checking~\cite{DBLP:conf/forte/PeledVY99,MP19,DBLP:journals/tecs/ShijuboWS23} and controller synthesis~\cite{DBLP:journals/tase/ZhangFL20}.

\myparagraph{Compositional Automata Learning}
Recently, algorithms for \emph{compositional automata learning} are attracting attention~\cite{DBLP:conf/fossacs/LabbafGHM23,DBLP:conf/fase/NeeleS23,DBLP:journals/sttt/FrohmeS21,DBLP:journals/corr/abs-2405-08647,DBLP:conf/icse/DuhaibyG20}. Assuming that the SUL $M$ is a composition of some subsystems $M_{1}, \dotsc, M_{n}$ (called \emph{components}), those algorithms try to 
 learn individual components $M_{i}$ and construct a model of $M$ as their composition, 
 rather than \emph{monolithically} learning the SUL $M$ itself.

A major benefit of such compositional approaches is \emph{complexity}: if each $M_{i}$ has $k_{i}$ states, the SUL has $k_{1}\times \cdots\times k_{n}$ states and the monolithic learning has to learn these, while the compositional learning has to learn only  $k_{1}+ \cdots+ k_{n}$ states in total. Since many real-world systems are constructed using components, compositional automata learning is  a promising approach to scalable learning.

It is important to note that different compositional automata learning algorithms assume very different problem settings. 
The differences lie in the type of automata to learn, how they are composed, the learning interface,
etc. We will make a detailed comparison later; its summary is  in \cref{table:compositionalAutomOverview}. 

In most  existing works including~\cite{DBLP:conf/fossacs/LabbafGHM23,DBLP:conf/fase/NeeleS23,DBLP:journals/sttt/FrohmeS21,DBLP:journals/corr/abs-2405-08647,DBLP:conf/icse/DuhaibyG20}, the learner has no access to individual components to make queries.
It thus tries to learn components indirectly via system-level queries. A typical application scenario is  where the SUL $M$ is a \emph{legacy} black-box system: $M$'s  compositional structure may be known, e.g.\ via  old documentation; yet $M$'s components are buried in the black-box system  $M$ and their interface is not exposed. In this case, the technical challenge is \emph{how to throw component-level queries indirectly}, that is, to translate component-level queries (that the learner wants to ask) to system-level queries (that the learner can ask in reality). 
The works~\cite{DBLP:conf/fossacs/LabbafGHM23,DBLP:conf/fase/NeeleS23,DBLP:journals/sttt/FrohmeS21,DBLP:journals/corr/abs-2405-08647,DBLP:conf/icse/DuhaibyG20} propose different solutions to this challenge, specializing in each problem setting.

Our problem setting---we call it \emph{componentwise learning} for distinction---is very different from the above; so is the main technical challenge there. 
 We first motivate our problem setting with \emph{system integration} as application.

\myparagraph{Motivation: System Integration with Black-Box Components}
\emph{System integration (SI)} in ICT industry refers to ``the process of creating a complex information system that may include designing or building a customized architecture or application, integrating it with new or existing hardware, packaged and custom software, and communications.''\footnote{Gartner Information Technology Glossary, \url{https://www.gartner.com/en/information-technology/glossary/system-integration}} SI is nowadays a norm in various layers of ICT system development:
\begin{itemize}
 \item  Large-scale ICT systems for banks, e-commerce, and other business processes are products of SI where different software components, typically developed by different parties, get integrated.
 \item  Smaller  software pieces  also rely on existing software components offered as libraries (e.g.\ $\mathtt{pip}$ for Python). They can be thus seen as products of SI.  
\end{itemize}
SI is not unique to ICT. In fact, our original motivation comes from the automotive industry, where various systems  (a car, an engine, control software, etc.) get built by assembling parts that are often manufactured by other parties.

In this paper,  a body \todo{KS:entity?} that conducts SI  is called  a \emph{system integrator (SIer)}.
 SIers have to make sure that the composite system behaves as expected. This is not easy, however, since components that constitute the composite system are usually black-box systems.
This situation  thus makes SI a natural target of automata learning.
Moreover, \emph{compositional} automata learning can be used, since an SIer knows the compositional structure that combines black-box components.




\myparagraph{Contribution: Contextual Componentwise Automata Learning}
In  SI, the learner (an SIer) is building a \emph{new} composite system. The learner has (raw) component in its hands, and thus has direct interface for  component-level queries. This is in stark contrast with other works~\cite{DBLP:conf/fossacs/LabbafGHM23,DBLP:conf/fase/NeeleS23,DBLP:journals/sttt/FrohmeS21,DBLP:journals/corr/abs-2405-08647,DBLP:conf/icse/DuhaibyG20} where the target is an 
\emph{old} (legacy) system and the main challenge is indirect component-level queries.
To highlight this difference, we use the term \emph{componentwise automata learning} for compositional learning where direct  component-level queries are available. 

A new challenge that we face in  componentwise automata learning is  \emph{component redundancies}.
In an SI scenario (no matter if it is ICT or automotive), components are rarely \emph{lean}, meaning that most of the time they come with more functionalities than an SIer needs for the composite system. Learning those \emph{rich} components holistically, including redundancies, is costly and wasteful.

This problem of redundancy is practically relevant. It is widely recognized in software engineering, resulting in  active research on \emph{dead code identification} and \emph{elimination} (see e.g.~\cite{MalavoltaNSRLSL23}). As a specific example, in our \lighting{} benchmark (\cref{sec:experiments}), there is a component that can handle many different modes of a communication protocol, but the composite system uses only one mode. 


%
%
%

Towards the goal of eliminating component redundancies, we devise a \emph{contextual} componentwise automata learning algorithm, where observation tables for automata learning are pruned to system-level relevant behaviors.


 To describe how our algorithm works, we first introduce our system model.

\myparagraph{Formalization by Moore Machine Networks}
In this paper, we model each component as a Moore machine (MM), and their composition as what we call a  \emph{Moore machine network (MMN)}. The latter arranges Moore machine components as nodes of a graph,  and  edges of the graph designate either system-level  or inter-component input/output. 
An example is in \cref{ex:counterInit}, where two component MMs operate, driven by system-level input words. The component $M_{c_1}$ passes its output to $M_{c_2}$, and $M_{c_2}$ produces  system-level output.

 Components of an MMN operate in a fully synchronized manner. They share the same clock, and at each tick of it, each component $M_{c}$ produces an output character $a_{e}$ at each outgoing edge $e$ from $M_{c}$. In case the edge $e$ points to another component $M_{c'}$, the character $a_{e}$ becomes (the $e$-component of) the input character to $M_{c'}$ at that tick. System-level input/output characters are consumed/produced synchronously, too. (The choice of Moore machines over Mealy machines is crucial for this operational semantics; see \FinalOrArxivVersion{\cite[Appendix~D.1]{FujinamiH25ATVA_arxiv_extended_ver}}{\cref{appendix:whyMoore}}.)

Therefore our formalism models  \emph{structured}, \emph{synchronized} and \emph{dense} composition of components. This is suited for system integration, where 1) the learner (the SIer)  arranges components with explicit interconnections, and 2) many components continuously receive signals from, and send signals to,  other components (as is the case with many automotive,  cyber-physical, web, and other systems). 

This formalization of ours is in contrast with \emph{flat}, (mostly) \emph{interleaving} and \emph{sparse} composition of components in other compositional works~\cite{DBLP:conf/fossacs/LabbafGHM23,DBLP:conf/fase/NeeleS23,DBLP:journals/corr/abs-2405-08647,DBLP:conf/icse/DuhaibyG20}. This difference mirrors different target applications.  See \cref{table:compositionalAutomOverview}.


\myparagraph{Context Analysis by Reachability Analysis in a Product}
Based on  the MMN formalization, we shall sketch our technique for eliminating component
\begin{wrapfigure}[16]{r}[0pt]{.47\textwidth}
\vspace{-2em}
\centering
 \includegraphics[width=.45\textwidth,trim={3.0cm 0.75cm 2.3cm 1.0cm},clip]{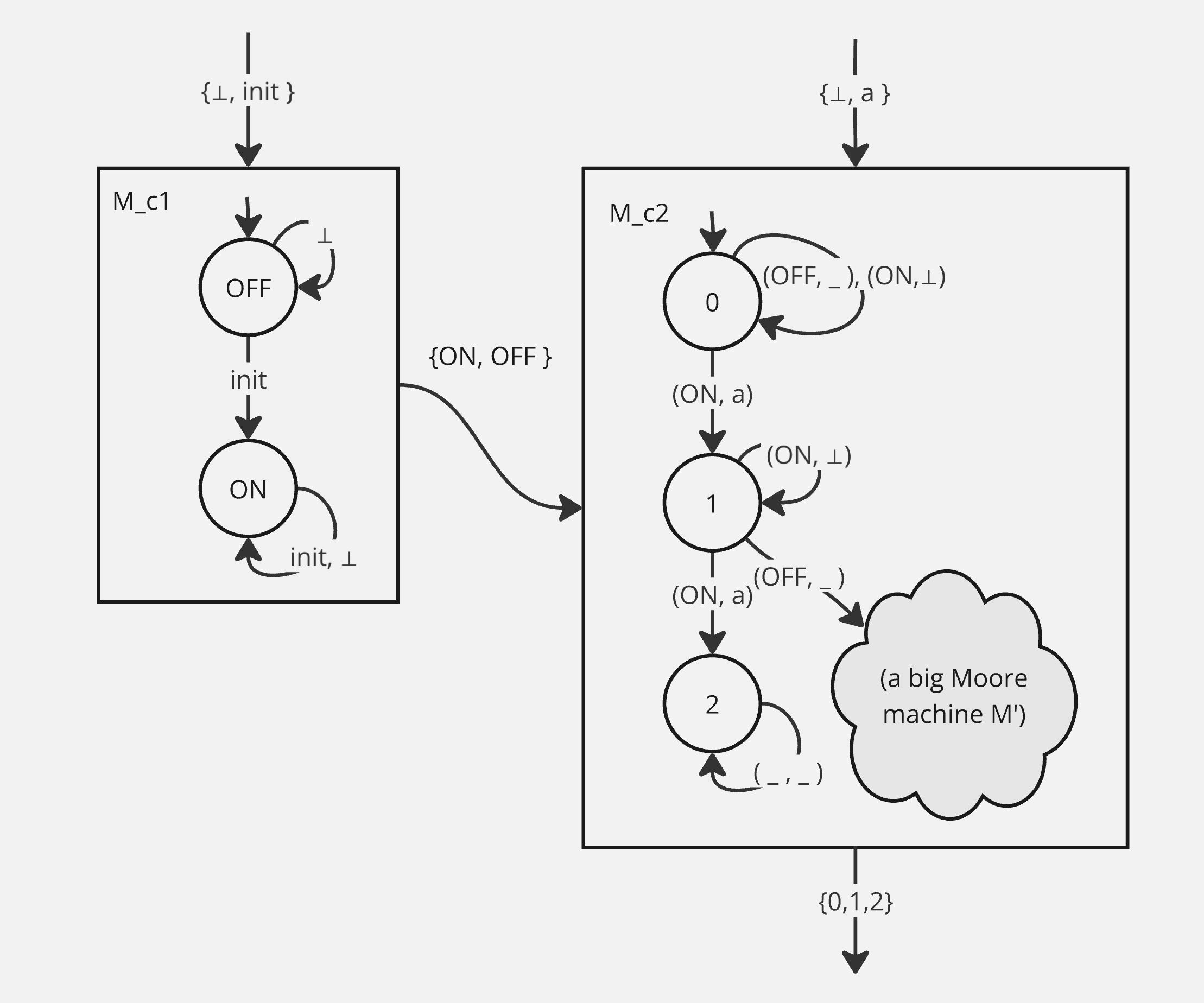}
\caption{an MMN example, a counter with initialization}
\label{fig:contextualityIntro}
\end{wrapfigure}
redundancies. Component redundancy gets formalized as the fact that \emph{system-level input words do not necessarily induce all component-level input words}. To see which input character a component $M_{c}$ can receive at its state $q_{c}$, it suffices to know at which state $q_{c'}$ every other component $M_{c'}$ can be at the same time. 

We conduct this \emph{context analysis (CA)} using the hypothesis automata learned so far for the components. Identifying all state tuples $(q_{c}, (q_{c'})_{c'})$ which can be simultaneously active is done by the reachability analysis in the product automaton of the hypotheses. This can be costly; we therefore introduce two \emph{context analysis parameters (CA-parameters)}, namely $\mathcal{E}$ (for abstracting contexts by quotienting hypothesis automata) and $\mathcal{R}$ (for limiting reachability analysis). 

These parameters give us flexibility in the cost-benefit trade-off of context analysis. This kind of flexibility is important in automata learning since different application scenarios have different  cost models. Specifically, running the SUL can be costly---processing one input character can take some milliseconds (in an embedded system) or even seconds (in a hardware-in-the-loop simulation (HILS) setting). In this case, extensive CA on the learner side (that can use a fast laptop) will pay off. In other cases where the SUL is fast, the cost of CA can be a bottleneck, and we might choose a cheaper and coarser CA-parameters.

We evaluated our contextual componentwise  learning algorithm (called CCwL*) with experiments. Comparison is against two baselines:  1) \emph{monolithic L* (MnL*)} that learns the whole SUL, 2) (naive) \emph{componentwise L* (CwL*)} that learns each component individually (without CA and thus component redundancies).
We used a few realistic benchmarks, including one inspired by robotics application, together with some  toy benchmarks. We also evaluated the effect of different CA-parameters
 ($\mathcal{E}, \mathcal{R}$).
 Overall, the experiment results indicate the value of our algorithm in application scenarios of system integration.

\begin{myexample}[counter with initialization]\label{ex:counterInit}
The  MMN  in \cref{fig:contextualityIntro} consists of two component MMs ($M_{c_1}$ and $M_{c_2}$). Each has a system-level input edge with the designated alphabet, and $M_{c_2}$ has a system-level output edge. The output of $M_{c_1}$ is plugged in as input of $M_{c_2}$, too.

This example exhibits component redundancies: the $M'$ part of $M_{c_2}$ is irrelevant to system-level behaviors. Indeed, $M'$ is never activated---once $M_{c_1}$ moves  to the ON state, it never goes back to OFF. Our contextual componentwise learning algorithm detects and exploits this fact; it learns $M_{c_1}$ and $M_{c_2}$ separately, but in the latter it prunes the unreachable part $M'$. 
\end{myexample}


\myparagraph{Contributions} Our contributions are summarized as follows.
\begin{itemize}
 \item We identify  \emph{system integration} as a new application domain of compositional automata learning.  There, component-level queries are fully available; we use the term \emph{componentwise automata learning} for distinction. 
 \item We formalize the problem using \emph{Moore machine networks} and identify the main challenge to be eliminating component redundancies.
 \item We introduce a \emph{contextual} componentwise learning algorithm. It eliminates component redundancies by pruning observation tables using reachability analysis in the product of (hypothesis automata for) the components. 
 \item We show its practical values through  experiments.
\end{itemize}

\begin{table}[tbp]
\caption{comparison of compositional automata learning frameworks, settings and challenges. Shading is made to signify the span of combined cells}\label{table:compositionalAutomOverview}
\centering\footnotesize
\scalebox{.8}{\begin{tabular}{@{}p{0.15\textwidth}L{\dimexpr0.3\textwidth-2\tabcolsep\relax}*{4}{L{\dimexpr0.12\textwidth-2\tabcolsep\relax}}L{\dimexpr0.2\textwidth-2\tabcolsep\relax}@{}}
\toprule
&
current work
&
\cite{DBLP:conf/fossacs/LabbafGHM23}
&
\cite{DBLP:conf/fase/NeeleS23}
&
\cite{DBLP:conf/icse/DuhaibyG20}
&
\cite{DBLP:journals/corr/abs-2405-08647}
&
\cite{DBLP:journals/sttt/FrohmeS21}
\\
\midrule
typical \qquad
application
&
system integration
&
\multicolumn{4}{c}{\cellcolor{gray!25}analysis of legacy systems}
&
learning  beyond regular
\\\midrule
querying \quad interface
&
system- \& component-level
&
\multicolumn{5}{c}{\cellcolor{gray!25} system-level only}
\\\midrule
target\qquad
 systems
&
Moore machine networks
&
Mealy machines
&
LTSs
&
LTSs
&
Mealy machines
&
SPAs 
\\\midrule
component interaction
&
structured, synchronized, dense
&
\multicolumn{4}{c}{\cellcolor{gray!25} flat, (mostly) interleaving, sparse}
&
procedure calls
\\\midrule
challenge
&
eliminating component redundancies
&
\multicolumn{5}{c}{\cellcolor{gray!25} querying components via system-level queries}
\\\bottomrule
\end{tabular}}
\end{table}

\myparagraph{Related Work}
Many works on automata learning in general have been already discussed; here we focus on compositional approaches.
A comparison of works on compositional automata learning is summarized in \cref{table:compositionalAutomOverview}.  All works but ours allow only system-level queries, and 
many are aimed at learning a \emph{legacy} black-box system. In contrast, in our system integration applications, we are usually building a \emph{new} system.


The compositional  algorithm in~\cite{DBLP:conf/fossacs/LabbafGHM23} assumes that the SUL is a parallel composition of Mealy machines $M_{1},\dotsc, M_{n}$ whose input alphabets $\Sigma_{1},\dotsc, \Sigma_{n}$ are disjoint. The components operate in the interleaved manner, where each component $M_{i}$ takes care of those input characters in $\Sigma_{i}$. The partition   $\Sigma=\Sigma_{1}\sqcup\cdots\sqcup \Sigma_{n}$ as well as the number $n$ of components is not known to the learner, and the challenge is to find them. Their algorithm first assumes the finest partition ($n=|\Sigma|$ and each $\Sigma_{i}$ is a singleton), and merges them in a counterexample-guided manner, when it is found that some input characters must be correlated.

The algorithm in~\cite{DBLP:conf/fase/NeeleS23} relies on more specific assumptions, namely that 1) the SUL is a parallel composition of LTSs which can synchronize by shared input characters, and 2) the output/observation to input words is whether the system gets stuck (i.e.\ there is no outgoing transition). The  challenge here is that, when the SUL gets stuck for an input word $w$ and $w$ contains characters shared by different components, the learner may not know which component to blame. Their solution consists of 1)
constructing an ``access word'' $w'$ that extends $w$ and localizes the blame,
 and 2) allowing ``unknown'' in observation tables in case there is no such $w'$. Unlike in~\cite{DBLP:conf/fossacs/LabbafGHM23}, the number of components and their input alphabets must  be known. The work~\cite{DBLP:conf/fase/NeeleS23}  shows that some Petri nets yield such combinations of LTSs; it is not clear how other types of systems (such as Mealy machines) can be learned by this algorithm. 

The works~\cite{DBLP:journals/corr/abs-2405-08647,DBLP:conf/icse/DuhaibyG20} can be thought of as variations of~\cite{DBLP:conf/fossacs/LabbafGHM23} with similar problem settings. In~\cite{DBLP:conf/icse/DuhaibyG20}, the disjointness assumption in~\cite{DBLP:conf/fossacs/LabbafGHM23} is relaxed, and they give some graph-theoretic conditions that enable compositional learning. These conditions, however, are detailed and the learner
has to somehow know that they hold in the SUL. In~\cite{DBLP:conf/icse/DuhaibyG20}, dually to~\cite{DBLP:conf/fossacs/LabbafGHM23}, they separate components according to output. It is yet to be identified what SULs are suited for this algorithm.

The work~\cite{DBLP:journals/sttt/FrohmeS21} has a different flavor from others. It is in the line of work on learning more expressive formalisms than (usual) automata and regular languages, such as (visibly) pushdown automata. Indeed, their target systems (system of procedural automata, SPA) are described much like context-free grammars. They assume that the invocation of nonterminals (\emph{procedure calls} in the SPA terminology) is observable; this enables application of automata learning. Otherwise the setting is similar to those in~\cite{DBLP:conf/fossacs/LabbafGHM23,DBLP:conf/fase/NeeleS23,DBLP:journals/corr/abs-2405-08647,DBLP:conf/icse/DuhaibyG20}; in particular, the challenge is the same, namely to query components via system-level queries.


Besides the works compared in \cref{table:compositionalAutomOverview}, 
\emph{distributed reactive synthesis}~\cite{DBLP:conf/focs/PnueliR90} and \emph{synthesis from component libraries}~\cite{DBLP:journals/sttt/LustigV13} are related to our work in their emphasis on compositionality. These works target at \emph{synthesis} of automata from given logical specifications, a goal different from ours or the works in \cref{table:compositionalAutomOverview} (namely active automata learning). \ih{revision of what hiroya wrote}

\begin{auxproof}

\myparagraph{Related Work}
As we already announced, while the interest in \emph{compositional automata learning} is growing, precise problem settings as well as intended applications differ a lot from one work to another. A comparison is given in \cref{table:compositionalAutomOverview}.

\begin{table}[tbp]
\caption{Problems of,  and algorithms for,  composition automata learning}\label{table:compositionalAutomOverviewOld}
\centering
\includegraphics[width=\textwidth]{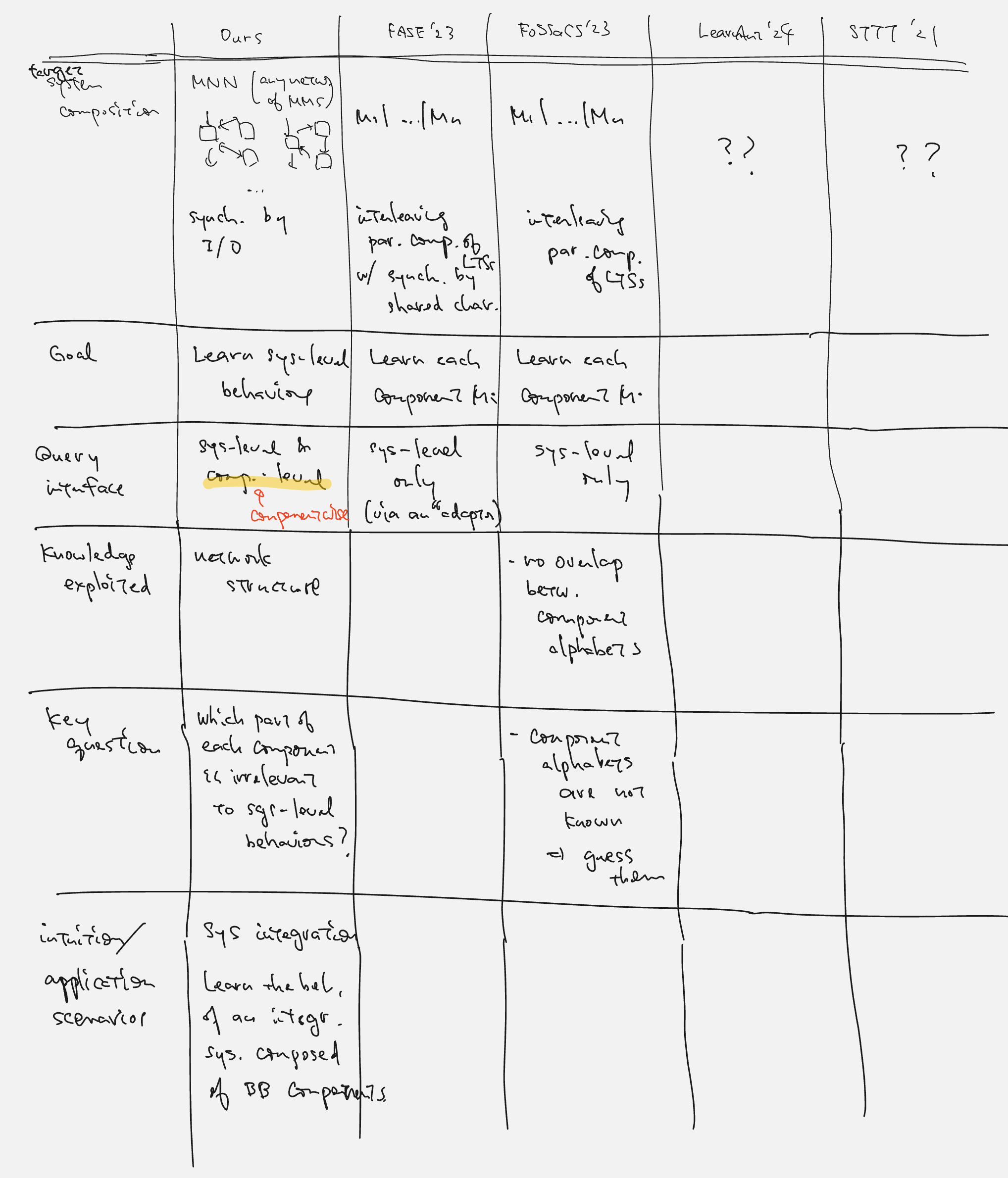}
\end{table}

(Other related works...)

Ichiro's survey. Overall, our problem setting is very different from existing works on compositional automata learning, in that 1) our components operate in a strictly synchronized manner, exhibiting dense interaction between them, and 2) the learner, as a system integrator, has access to each individual component, while in the other works they only can access the whole system. Consequently, the theoretical challenges we encounter are different from those in other works: for us, the biggest challenge is to avoid learning those parts of components that have no system-level consequences; for other works, a major challenge is often how to access a component through the system-level interface.
\begin{itemize}
 \item \cite{DBLP:conf/fossacs/LabbafGHM23}
       \begin{itemize}
	\item They learn the system structure as well as the behavior of each component, for legacy and black-box systems~\cite{DBLP:conf/issse/Koschke08} (we do know the system structure well; the applications are different)
	\item A big assumption is that they assume that 1) the SUL is an interleaving parallel composition of Mealy machines, and 2)  components LTS have pairwise distinct alphabets. Their algorithm first learns a component for each character $a$ (i.e.\ for the singleton alphabet $\{a\}$). The interleaving parallel composition of these usually fails to recreate the behavior of the SUL. In this case the alphabets are merged---in a counter example-driven manner---to account for possible interactions between different characters.
	\item Theoretically, a central notion here is the \emph{independence} of a partition $I_{1},\dotsc, I_{n}$ of the (system-level) input alphabet $I$. Given the (whole) system $M$, one can easily define its \emph{projection} to each restricted alphabet $I_{j}$, and independence asks if the composition of those projections adequately recreates the behavior of $M$.
	\item They prove the termination of their learning algorithm.

       \end{itemize}

 \item \cite{DBLP:conf/fase/NeeleS23}
       \begin{itemize}
	\item (They rely on a specialized setting) They target at learning LTSs, parallelly composed, where components synchronize are forced to synchronize via shared input characters. Another important point of their problem setting is that the observation is the enabledness of each character: given an input character $a$, every component which has $a$ in its input alphabet tries to operate, and if any of these components gets stuck (since $a$ is not enabled in its current state), the whole system gets stuck. This ``stuck'' or ``not stuck'' information is the observation from which they learn. 
	\item They know the system composition (how many components, what are their input alphabets)
	\item They only allow system-level queries. Inter-component communication is invisible
	\item Therefore what they do is to translate component-level queries to system-level queries. Note that the whole learning framework involves $n$ learners, where $n$ is the number of components, where the $i$-th learner  tries to learn the corresponding component $M_{i}$.
	\item But to do so precisely requires knowing how components interact with each other, information that is only discovered as the learning progresses. Concretely, if an input alphabet $a$ from the $i$-th learner is responded with a (necessarily system-level) ``stuck'' observation, it is not necessarily because the component $M_{i}$ got stuck---there may be another component $M_{j}$ that got stuck. 
	\item Their technical solution consists of extending L* with unknowns, and interleaving an input query $w$ to $M_{i}$ with a word $w'$ that 1) synchronizes with $w$ (meaning that their projection to the shared input characters are the same) and 2) is already known to be accepted by all the other components. If this interleaved work leads the whole system to ``stuck,'' any component other than $M_{i}$ cannot be blamed, and thus it is known that $w$ leads $M_{i}$ to ``stuck.'' In case there is no such word $w'$, the query $w$ to $M_{i}$ is responded with ``unknown.''
	\item Their ``realistic example'' comes from Petri nets. It is not clear, due to their specific theoretical framework (of observing deadlocks in LTSs), how they can accommodate other types of machines such as Mealy and Moore machines.
       \end{itemize}

 \item \cite{DBLP:journals/sttt/FrohmeS21}
       \begin{itemize}
	\item target system: a \emph{system of procedural automata (SPA)}, which are essentially DFAs that call each other. Sequential. Aimed at modeling procedure calls. Different from our highly synchronous notion of composition. Those machines serve as word classifiers (i.e.\ they accept or reject input words); accommodating Mealy or Moore machines are described as their future work.
	\item Therefore their ``compositionality'' refers to calling each other. It is sequential and there is no synchronization. 
	\item SPAs are powerful and they can recognize context-free languages. Therefore, in a sense, this work is geared towards ``learning a formalism that denotes more than a regular language''
	\item In their framework, it is notable that calls and returns of procedures are observable---this is much like in visibly pushdown automata~\cite{DBLP:conf/stoc/AlurM04}. 
	\item The gist of their algorithm is remember how to access a component---that is, a DFA that is called in a certain context---by recording so-called \emph{access}, \emph{terminating}, and \emph{return} sequences.
       \end{itemize}
 \item \cite{DBLP:journals/corr/abs-2405-08647}
       \begin{itemize}
	\item They propose \emph{output-decomposed learning} of Mealy machines
	\item Dually to~\cite{DBLP:conf/fossacs/LabbafGHM23} (where it is assumed that the SUL is the interleaving parallel composition of Mealy machines which process pairwise distinct alphabets), in this work they separate the output alphabet. They do not assume any specific structure of the SUL
	\item An upside of the algorithm is that it can be applied to any Mealy machine: unlike the ``input-decomposed'' learning algorithm in~\cite{DBLP:conf/fossacs/LabbafGHM23}, the algorithm returns a correct answer for any Mealy machine as an SUL. A downside is that it may not be faster than monolithic L*: in~\cite{DBLP:journals/corr/abs-2405-08647},  its performance advantages are  clearly observed only in artificial benchmarks that are compositions of output-separate components. 
       \end{itemize}

 \item \cite{DBLP:conf/icse/DuhaibyG20}
       \begin{itemize}
	\item The work's spirit is close to that of \cite{DBLP:conf/fossacs/LabbafGHM23}. The difference is that 1) they, unlike \cite{DBLP:conf/fossacs/LabbafGHM23}, allow synchronization of two components via intersecting alphabets, and 2) as a price, they assume some detailed structural conditions on the SUL, and verifying such conditions require deep knowledge of the SUL's internal working. 
	\item They have a real-world case study.
       \end{itemize}

 \item General remarks
       \begin{itemize}
	\item Resets are time and resource consuming~\cite{DBLP:journals/iandc/RivestS93}. In our experience, there are settings in which a reset takes more than one minute.
	\item Processing one input character in the SUL can take a few seconds, too, especially in a \emph{HILS (Hardware in the Loop Simulation)} setting where a character is communicated physically (e.g.\ by a robot arm).
	\item \cite{DBLP:conf/fase/NeeleS23} has  a nice list of related work on automata learning in general
	\item \cite{DBLP:conf/fase/NeeleS23} has  a nice brief recap of L*, too (it's in general better written than \cite{DBLP:conf/fossacs/LabbafGHM23})
       \end{itemize}

 \item Points to make
       \begin{itemize}
	\item Different applications: in our work, the user is a system integrator, and the emphasis is to learn each component as much as it matters to the whole system
       \end{itemize}
\end{itemize}




\end{auxproof}

\myparagraph{Notations} 
$X \sqcup Y$ denotes the disjoint union of sets $X$ and $Y$.
The powerset of $X$
is denoted by $2^X$.
 The set of all partial functions from $X$ to $Y$ is denoted by $X \rightharpoonup Y$.
For $f\colon X \rightharpoonup Y$ and  $x \in X$, we write $f(x)\!\!\downarrow$ if $f(x)$ is defined, and $f(x)\!\!\uparrow$ otherwise.

Given an equivalence relation ${\sim}\subseteq X\times X$, equivalence classes are denoted by $[x]_{\sim}$ using $x\in X$, and the quotient set is denoted by $X/{\sim}$, as usual.



\section{Problem Formalization by Moore Machine Networks}\label{sec:definitions}


We start by some basic definitions.
\todo{We can send this to the notation section.}
Let $X$ be a nonempty finite set. 
$X^\ast$ denotes the set of finite strings (also called words) over $X$;
 $\varepsilon$ denotes the empty string; 
$|s|$ denotes the length of $s\in X^{\ast}$; and
$s_1 \cdot s_2$ (or simply $s_1 s_{2}$) denotes the concatenation of strings $s_1, s_2\in X^{\ast}$.

In this paper, we use the $0$-based indexing for strings.
For a string $s \in X^\ast$ and an integer $i\in [0,|s|)$, $\seqIndex{s}{i}$ denotes the $(i + 1)$-th character of $s$ (thus $s=\seqIndex{s}{0}\seqIndex{s}{1}\dotsc \seqIndex{s}{|s|-1}$);
for $i,j\in [0,|s|]$ such that $i\le j$, 
$\seqSlice{s}{i}{j}$ denotes the substring $\seqIndex{s}{i}\seqIndex{s}{i+1}\dotsc\seqIndex{s}{j-1}$. Note that $\seqSlice{s}{i}{i} = \varepsilon$ for each $i$.


Let $(X_k)_{k \in K}$ be a ($K$-indexed) family of sets.
 Its product is denoted by $\prod_{k \in K} X_k$, with its element  denoted by a tuple
 $(x_k)_{k \in K}$ (here $x_k \in X_k$).

The restriction of a tuple $t = (x_k)_{k \in K} \in \prod_{k \in K} X_k$ to a subset $K' \subseteq K$, denoted by $\restrictTo{t}{K'}$, is $(x_k)_{k \in K'} \in \prod_{k \in K'} X_k$.
This restriction of tuples $t$ along $K' \subseteq K$ is extended, in a natural pointwise manner, to subsets $S\subseteq \prod_{k \in K} X_k$ and sequences $s\in (\prod_{k \in K} X_k)^{*}$, resulting in the notations $\restrictTo{S}{K'}$ and $\restrictTo{s}{K'}$.

\subsection{Moore Machines}\label{ssec:moore-machines}
Our algorithm learns the following   (deterministic)  Moore machines (MMs).

\begin{mydefinition}[Moore machine]\label{def:moore-machine}
 A  \emph{Moore machine (MM)} is a tuple $M = (Q, q_0, I, O, \Delta, \lambda)$, where
  \begin{itemize}
    \item $Q$ is a finite set of states, $q_0 \in Q$ is an initial state,
    \item $I$ is an input alphabet, $O$ is an output alphabet,
    \item $\Delta\colon Q \times I \rightharpoonup Q$ is a transition (partial) function, and
    \item $\lambda\colon Q \to O$ is an output function that assigns an output symbol to each state.
  \end{itemize}

A  Moore machine is \emph{complete} if $\delta(q, i)\!\!\downarrow$ for all $q \in Q$ and $i \in I$; otherwise, it is called \emph{partial}.
\end{mydefinition}


We will also use \emph{nondeterministic} MMs, later in \cref{sec:ccwlstar}, but only for the purpose of approximate context analysis (CA). It is emphasized that we do \emph{not} learn nondeterministic MMs. The theory of nondeterministic MMs (their definition, semantics, etc.) is obtained in a straightforward manner; it is in \FinalOrArxivVersion{\cite[Appendix~A.1]{FujinamiH25ATVA_arxiv_extended_ver}}{\cref{appendix:mmND}}.

As usual, the transition function $\Delta$ can be extended to
 an input string $w\in I^{\ast}$. Precisely,
$\Delta(q, \varepsilon) = q$ and 
$\Delta(q, wi) =
\Delta\bigl(\Delta(q,w),i\bigr)
$.
Similarly, the output function is extended by 
$\lambda(q, w) = 
\lambda(\Delta(q,w))$.
When starting from the initial state  $q_0$,
often we simply write $\Delta(w) = \Delta(q_0, w)$ and $\lambda(w) = \lambda(q_0, w)$.

Given a Moore machine $M = (Q, q_0, I, O, \Delta, \lambda)$ and 
a state $q\in Q$,
 the \emph{semantics} of $M$,
denoted by  $\semMoore{M}_q\colon I^\ast \to O^\ast$ and defined below, represents  the behavior of the machine when starting from $q$. For each $w \in I^\ast$,
\begin{equation}\label{eq:semMoore}
\begin{aligned}
 &  \semMoore{M}_q(w) \;=\; \lambda(q, \seqSlice{w}{0}{0})\;\lambda(q, \seqSlice{w}{0}{1}) \cdots \lambda(q, \seqSlice{w}{0}{k}),
\end{aligned}
\end{equation}
where $k$ is the smallest number such that  $\lambda(q, \seqSlice{w}{0}{k+1})$ is undefined, or $|w|$ if  no such $k$ exists. 
Note that $\semMoore{M}_q\colon I^\ast \to O^\ast$ is a \emph{total} function: even if $\Delta$ gets stuck (making $\lambda(q, \seqSlice{w}{0}{k+1})$ undefined), it does not make $\semMoore{M}_q(w)$ undefined, while it does make $\semMoore{M}_q(w)$ shorter.
When starting from the initial state $q_0$, we write $\semMoore{M}(w)$ for $\semMoore{M}_{q_0}(w)$ (much like for $\Delta$ and $\lambda$). 

Using this semantics, we define the equivalence of Moore machines.

\begin{mydefinition}[equivalence of Moore machines]\label{def:moore-machine-equivalence}
 Two Moore machines $M_1$ and $M_2$ are said to be \emph{equivalent} if and only if $\semMoore{M_1} = \semMoore{M_2}$.
\end{mydefinition}

\subsection{Moore Machine Networks}\label{ssec:mmn}

A \emph{directed graph} is a tuple $G = (V, E)$ of a finite set $V$ of nodes (or vertices) and a  set  $E \subseteq V \times V$  of (directed) edges.

Let $G = (V, E)$ be a directed graph.
Let $\incomingEdges{v}$ denote the set of \emph{incoming edges} for a node $v \in V$, i.e.\ $\incomingEdges{v} = \{ (u, v) \in E \mid u \in V \}$.
Similarly,  $\outgoingEdges{v}$ denotes the set of \emph{outgoing edges}  ($\outgoingEdges{v} = \{ (v, u) \in E \mid u \in V \}$). 
For a set $U \subseteq V$ of nodes, we define $\incomingEdges{U} = \bigcup_{u \in U} \incomingEdges{u}$ and $\outgoingEdges{U} = \bigcup_{u \in U} \outgoingEdges{u}$.

Towards our definition of MMNs (\cref{def:mmn}; see also \cref{fig:contextualityIntro}), we introduce the following classification of nodes: a node $v\in V$ is
\begin{enumerate*}[label=\arabic*)]
  \item a \emph{system-level input node} if $\incomingEdges{v} = \emptyset$,
  \item a \emph{system-level output node} if $\outgoingEdges{v} = \emptyset$, and
  \item a \emph{component node} otherwise.
\end{enumerate*}
We denote the sets of input, output, and component nodes
by $\inputNodes, \outputNodes$, and $\componentNodes$, respectively. We impose the condition on $G=(V,E)$ that
 $V = \inputNodes \sqcup \outputNodes \sqcup \componentNodes$ is a disjoint union of three nonempty sets. (See \cref{fig:contextualityIntro}, where system-level input and output nodes are implicit.) 

An edge $e = (v, v') \in E$ is called a \emph{system-level input edge} if $v \in \inputNodes$, and a \emph{system-level output edge} if $v' \in \outputNodes$.
The set of system-level input and output edges of $G = (V, E)$ are denoted by $\inputEdges$ and $\outputEdges$, respectively, and are given by $\inputEdges = \outgoingEdges{\inputNodes}$ and $\outputEdges = \incomingEdges{\outputNodes}$.

A \emph{Moore machine network (MMN)} is a directed graph, with a Moore machine associated with each component node $c\in \componentNodes$. Here we present a deterministic definition. Later in \cref{sec:ccwlstar} we also use a nondeterministic version (not to learn, but for CA); this is an easy adaptation. See \FinalOrArxivVersion{\cite[Appendix~A.2]{FujinamiH25ATVA_arxiv_extended_ver}}{\cref{appendix:mmnND}} for explicit definitions.


\begin{mydefinition}[Moore machine network]\label{def:mmn}
  A (deterministic) \emph{Moore machine network (MMN)} is a tuple $\mathcal{M} = (G, {(\Sigma_e)}_{e \in E}, {(M_c)}_{c \in \componentNodes})$, where
  \begin{itemize}
    \item $G = (V, E)$ is a directed graph representing the network structure,
    \item $\Sigma_e$ is an alphabet associated with each edge $e \in E$, and
    \item $M_c$ is a (deterministic) Moore machine associated with a component $c \in \componentNodes$.
  \end{itemize}
 On each component Moore machine $M_c = (Q_c, q_{0,c}, \componentInputAlphabet{c}, \componentOutputAlphabet{c}, \Delta_c, \componentOutputFunction{c})$, we require that its input and output alphabets are in  accordance with the edge alphabets $\Sigma_{e}$. Specifically, we require $\componentInputAlphabet{c} = \prod_{e \in \incomingEdges{c}} \Sigma_e$ (the product of the alphabets of all incoming edges) and, similarly,  $\componentOutputAlphabet{c} = \prod_{e \in \outgoingEdges{c}} \Sigma_e$.
\end{mydefinition}


For an MMN $\mathcal{M}$, we define three 
\emph{system-wide alphabets}:
\begin{enumerate*}[label=\arabic*)]
  \item  $\inputAlphabet = \prod_{e \in \inputEdges} \Sigma_e$ is the  \emph{system-level input alphabet},
\item $\outputAlphabet = \prod_{e \in \outputEdges} \Sigma_e$ is  the \emph{system-level output alphabet}, and 
  \item $\totalOutputAlphabet = \prod_{e \in E \setminus \inputEdges} \Sigma_e = \prod_{c \in \componentNodes} \componentOutputAlphabet{c}$ is the \emph{total output alphabet}. Note that $\totalOutputAlphabet$ collects also those characters sent from a component to another.

\end{enumerate*}

\begin{myexample}
\label{exa:mmn-network-alphabet}
 An example of MMN is in  \cref{fig:mmn-example}. Its detailed formalization is in \FinalOrArxivVersion{\cite[Appendix~D.2]{FujinamiH25ATVA_arxiv_extended_ver}}{\cref{appendix:mmn-network-alphabet-detailed}}.


  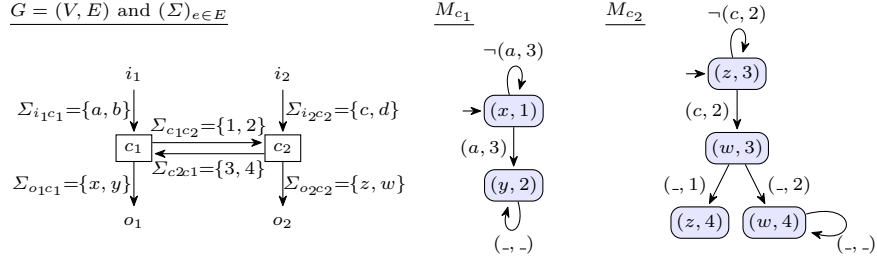
\begin{figure}[tbp]
    \centering
    \begin{tikzpicture}[>={Stealth[round,sep]}, scale=0.99, every node/.style={transform shape}]
      \begin{scope}[xshift=-0.2cm]
        \node (desc) at (-1.2,1.8) {\scriptsize\underline{$G = (V, E)$ and $(\Sigma)_{e \in E}$}};

        \node (i1) at (-1,1) {\scriptsize $i_1$};
        \node (i2) at (1,1)  {\scriptsize $i_2$};
    
        \node (c1) at (-1,0) [draw] {\scriptsize $c_1$};
        \node (c2) at (1,0) [draw] {\scriptsize $c_2$};
    
        \node (o1) at (-1,-1) {\scriptsize $o_1$};
        \node (o2) at (1,-1)  {\scriptsize $o_2$};
    
        \node (i1c1) at (-1.8,0.5) {\scriptsize $\Sigma_{i_1\!c_1}\!\! = \!\!\{ a, b \}$};
        \node (i2c2) at (1.8,0.5) {\scriptsize $\Sigma_{i_2\!c_2}\!\! = \!\!\{ c, d \}$};
    
        \coordinate (c1_1) at ($(c1.north east)!0.3!(c1.south east)$);
        \coordinate (c2_1) at ($(c2.north west)!0.3!(c2.south west)$);
        \coordinate (c1_2) at ($(c1.north east)!0.7!(c1.south east)$);
        \coordinate (c2_2) at ($(c2.north west)!0.7!(c2.south west)$);
    
        \node (c1c2) at (0,0.28) {\scriptsize $\Sigma_{c_1\!c_2}\!\! = \!\!\{ 1, 2 \}$};
        \node (c2c1) at (0,-0.28) {\scriptsize $\Sigma_{c2\!c1}\!\! = \!\!\{ 3, 4 \}$};
    
        \node (o1c1) at (-1.83,-0.5) {\scriptsize $\Sigma_{o_1\!c_1}\!\! = \!\!\{ x, y \}$};
        \node (o2c2) at (1.83,-0.5) {\scriptsize $\Sigma_{o_2\!c_2}\!\! = \!\!\{ z, w \}$};
      \end{scope}
  
      \graph {
        (i1) -> (c1);
        (i2) -> (c2);
        (c1_1) -> (c2_1);
        (c2_2) -> (c1_2);
        (c1) -> (o1);
        (c2) -> (o2);
      };

      \begin{scope}[xshift=-1.1cm]
        \node (mc1) at (4.2,1.8) {\scriptsize\underline{$M_{c_1}$}}; 

        \node (mc1q0) at (4.2,0.5) {};
        \node (mc1q1) at (5,0.5) [location] {\scriptsize $(x,1)$};
        \node (mc1q2) at (5,-0.5) [location] {\scriptsize $(y,2)$};
  
        \node (mc1q1q1) at (5,1.3) {\scriptsize $\lnot (a, 3)$};
        \node (mc1q1q2) at (4.6,0) {\scriptsize $(a,3)$};
        \node (mc1q2q2) at (5,-1.3) {\scriptsize $(\_,\_)$};
  
        \path
          (mc1q0) edge [->] (mc1q1)
          (mc1q1) edge [loop above] (mc1q1)
          (mc1q1) edge [->] (mc1q2)
          (mc1q2) edge [loop below] (mc1q2);

        \node (mc2) at (6.5,1.8) {\scriptsize\underline{$M_{c_2}$}};

        \node (mc2q0) at (7.2,1) {}; 
        \node (mc2q1) at (8,1) [location] {\scriptsize $(z,3)$};
        \node (mc2q2) at (8,0) [location] {\scriptsize $(w,3)$};
        \node (mc2q3) at (7.5,-1) [location] {\scriptsize $(z,4)$};
        \node (mc2q4) at (8.5,-1) [location] {\scriptsize $(w,4)$};
  
        \node (mc2q1q1) at (8,1.8) {\scriptsize $\lnot (c, 2)$};
        \node (mc2q1q2) at (7.6,0.5) {\scriptsize $(c,2)$};
        \node (mc2q2q3) at (7.3,-0.5) {\scriptsize $(\_,1)$};
        \node (mc2q2q4) at (8.7,-0.5) {\scriptsize $(\_,2)$};
        \node (mc2q4q4) at (9.6,-1.3) {\scriptsize $(\_,\_)$};
  
        \path
          (mc2q0) edge [->] (mc2q1)
          (mc2q1) edge [loop above] (mc2q1)
          (mc2q1) edge [->] (mc2q2)
          (mc2q2) edge [->] (mc2q3)
          (mc2q2) edge [->] (mc2q4)
          (mc2q4) edge [loop right] (mc2q4);
      \end{scope}
    \end{tikzpicture}
    \caption{an example MMN $\mathcal{M}_\mathsf{ex}$. On the left we show its network $G = (V, E)$ and the alphabets $(\Sigma_e)_{e \in E}$ for edges. The component MMs are shown on the right, where state labels designate output. 
In the   transition labels, $\lnot i$ stands for all  characters other than $i$, and the symbol $\_$ matches any character}\label{fig:mmn-example}
  \end{figure}
\end{myexample}

We move on to define the semantics of MMNs. It is via a translation of an MMN $\mathcal{M}$ to an MM $\mmnMoore{\mathcal{M}}$; in the translation, the component MMs in $\mathcal{M}$ operate in a fully synchronized manner. The definition is intuitively straightforward, although its precise description below involves somewhat heavy notations.

The set of \emph{(system-level) configurations} of $\mathcal{M}$, denoted by $\mathbf{Q}$, is defined by $\mathbf{Q} = \prod_{c \in \componentNodes} Q_c$.
The \emph{initial configuration} 
is  $\mathbf{q}_{0} = 
(q_{0,c})_{c \in \componentNodes}
$.

Given a configuration $\mathbf{q} = (q_c)_{c \in \componentNodes} \in \mathbf{Q}$, the \emph{total output} of $\mathcal{M}$ at $\mathbf{q}$, denoted by $\totalOutputFunction(\mathbf{q}) \in \totalOutputAlphabet$, is defined by $\totalOutputFunction(\mathbf{q}) = \bigl(\componentOutputFunction{c}(q_c)\bigr)_{c \in \componentNodes} $. Similarly, the \emph{system-level output} of $\mathcal{M}$ at $\mathbf{q}$ is
 defined by $\outputFunction(\mathbf{q}) = \restrictTo{\totalOutputFunction(\mathbf{q})}{\outputEdges}\in \outputAlphabet$. (Recall the restriction operation $\mid$ from \cref{sec:definitions}.) 

Given a configuration $\mathbf{q} = (q_c)_{c \in \componentNodes} \in \mathbf{Q}$ and a system-level input character $\mathbf{i} \in \inputAlphabet$, we define $\Delta$, the \emph{system-level transition function} of $\mathcal{M}$, by
 $\Delta(\mathbf{q}, \mathbf{i}) = 
\Bigl(\,
 \Delta_c\bigl(q_c, \restrictTo{(\mathbf{i}, \totalOutputFunction(\mathbf{q}))}{\incomingEdges{c}}\bigr)\,\Bigr)_{c \in \componentNodes}$. 
Intuitively: the tuple $\totalOutputFunction(\mathbf{q})$ of characters is output from the current states
 $\mathbf{q}=(q_c)_{c \in \componentNodes}$; it is combined with the system-level input $\mathbf{i}$ and fed to each component's transition function $\Delta_{c}$.

We formalize the following definition, using the above constructions.


\begin{mydefinition}[Moore machine $\mmnMoore{\mathcal{M}}$]
Let $\mathcal{M}$ be an MMN. The \emph{Moore machine $\mmnMoore{\mathcal{M}}$ induced by} $\mathcal{M}$ is $\mmnMoore{\mathcal{M}} = (\mathbf{Q}, \mathbf{q}_{0}, \inputAlphabet, \outputAlphabet, \Delta, \outputFunction)$.
 \end{mydefinition}

The semantics of $\mathcal{M}$ is defined by
$\semMoore{\mathcal{M}}_{\mathbf{q}} = \semMoore{\mmnMoore{\mathcal{M}}}_{\mathbf{q}}$ for any $\mathbf{q} \in \mathbf{Q}$.


When we turn to completeness of MMs (\cref{def:moore-machine}),  the completeness of $\mmnMoore{\mathcal{M}}$ does not necessarily imply that of each component MM $M_c$. This is obvious from \cref{ex:counterInit}---it does not matter even if some transitions are not defined in $M'$, since $M'$ is never invoked. This is an implication of component redundancies (\cref{sec:intro}).




\section{An L*-Style Algorithm}\label{sec:lstar}
We review the classic L* algorithm from \cite{DBLP:journals/iandc/Angluin87}. 
We formulate 
it in such a way that it easily adapts to our componentwise and contextual algorithm in \cref{sec:ccwlstar}. Consequently, the algorithm we present (\cref{alg:lstar})  differs slightly  from the original L*; nevertheless, for simplicity, we call it L* throughout the paper.

We start with formulating the problem.
\begin{myproblem}[Moore machine learning]\label{prob:automata-learning}
  \noindent\begin{description}
    \item[Input:] the problem takes two alphabets $I, O$ and two
oracles $\mathsf{OQ}, \mathsf{EQ}$ as inputs:
      \begin{enumerate*}[label=\arabic*)]
        \item $I$ and $O$ are input and output alphabets, respectively;
        \item the \emph{output query} oracle $\mathsf{OQ}$, given an input string $w \in I^\ast$, returns a string $\mathsf{OQ}(w) \in O^\ast$; and
        \item  the \emph{equivalence query} oracle $\mathsf{EQ}$, given a (hypothesis) Moore machine $H$, returns ``yes'' or a \emph{counterexample} sequence $w \in I^\ast$.
      \end{enumerate*}
    \item[Output:] a Moore machine $M$.
  \end{description}
\end{myproblem}
Here, $\mathsf{OQ}$ and $\mathsf{EQ}$ form an abstraction of a black-box SUL, and the goal is to learn $M$ that behaves the same as the SUL.



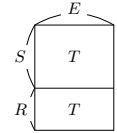
\begin{wrapfigure}[8]{r}[0pt]{.19\textwidth}
\vspace{-2em}
\centering
  \scalebox{0.7}{\begin{tikzpicture}
    \coordinate (lb) at (0,0);
    \coordinate (lt) at (0,2);
    \coordinate (rb) at (1.5,0);
    \coordinate (rt) at (1.5,2);
    \coordinate (lm) at (0,0.8);
    \coordinate (rm) at (1.5,0.8);
    \draw (lb) rectangle (rt);
    \draw (lm) -- (rm);
    \draw (lb) edge [bend left] node[fill=white, xshift=-0.15cm] {\small $R$} (lm);
    \draw (lm) edge [bend left] node[fill=white, xshift=-0.1cm] {\small $S$} (lt);
    \draw (lt) edge [bend left] node[fill=white, yshift=0.1cm] {\small $E$} (rt);
    \node (TS) at (0.75,1.4) {\small $T$};
    \node (TR) at (0.75,0.4) {\small $T$};
  \end{tikzpicture}}
\caption{an observation table}
\label{fig:obsTable}
\end{wrapfigure}

Our L*-style algorithm (\cref{alg:lstar}), henceforth simply called L*,  uses an \emph{observation table} $(S, R, E, T)$. Here
\begin{itemize}
  \item $S \subseteq I^\ast$ is a set of \emph{prefixes},
  \item $R \subseteq (S \cdot I) \setminus S$ is a set of \emph{1-step extensions} of $S$,
  \item $E \subseteq I^\ast$ is a set of \emph{suffixes}, and
  \item $T\colon (S \cup R) \cdot E \to O$ is the \emph{entry map}, where $(S \cup R) \cdot E$ collects  all concatenations of strings from $S\cup R$ and those from $E$. 
\end{itemize}
See \cref{fig:obsTable}. We initialize an observation table as
 $(\{ \varepsilon \}, \emptyset, \{ \varepsilon \},  (\varepsilon\mapsto \mathsf{OQ}(\varepsilon)) )$, and both $S$ and $E$ will always contain the empty string $\varepsilon$.
We write $\mathsf{row}(s)$ for the row of $s$ in the observation table, i.e.\  $\mathsf{row}(s): e \in E \mapsto T(s \cdot e)$.
An observation table is \emph{closed} if every $r \in R$ has some $s \in S$ such that $\mathsf{row}(r) = \mathsf{row}(s)$.


From a closed observation table $(S, R, E, T)$, we can construct a deterministic Moore machine $H = (Q, q_0, I, O, \delta, \lambda)$ with $Q = \{ \mathsf{row}(s) \mid s \in S \}$, $q_0 = \mathsf{row}(\varepsilon)$, $\delta(\mathsf{row}(s), i) = \mathsf{row}(s \cdot i)$ for $s \cdot i \in R$, and $\lambda(\mathsf{row}(s)) = T(s)$.
This MM is called the \emph{hypothesis Moore machine} from the observation table $(S, R, E, T)$.

The L* algorithm \cref{alg:lstar} works by initializing an observation table, growing it using $\mathsf{OQ}$ till it is closed, making a hypothesis MM $H$, checking if $H$ is good using $\mathsf{EQ}$, and if $H$ is not good, using the counterexample to further grow the table. When $\mathsf{OQ}$ and $\mathsf{EQ}$ are based on some MM $M$, \cref{alg:lstar} terminates and returns a MM equivalent to $M$.


The differences between \cref{alg:lstar}  and the original L*~\cite{DBLP:journals/iandc/Angluin87} are as follows.
\begin{enumerate}

  \item L* grows an observation table by 1) ``extending input words'' by picking $s\in S$ and $i\in I$ and adding $s\cdot i$ to $R$ (Lines~\ref{line:lstar-extend-begin}--\ref{line:lstar-extend-end}), and 2) ``closing the table'' by moving rows from $R$ to $S$ (Lines~\ref{line:lstar-close-begin}--\ref{line:lstar-close-end}). We parameterize the first part with a function $\textsc{1Ext}^\text{L*}$. This parameter is set in the usual L* manner in \cref{alg:lstar} (namely $\textsc{1Ext}^\text{L*}(H)=\{(s, i)\mid \mathsf{row}(s) \in Q_{H} \land i\in I\}$ where $Q_{H}$ is the state space of $H$). Changing this parameter will be central in the next section.

  \item In the original L*~\cite{DBLP:journals/iandc/Angluin87}, all the prefixes of the counterexample are added to $S$, but this can break a property called consistency. We, instead, add to $E$ a suffix $d$ that satisfies the condition in Line~\ref{line:lstar-analyzecex-find}. This maintains consistency. This is a well-known technique; see e.g.~\cite{DBLP:conf/icgi/IsbernerS14}.
\end{enumerate}

\begin{algorithm}[t]
  \caption{an L*-style Moore machine learning algorithm (simply called L*)}\label{alg:lstar}
  \scalebox{.8}{\begin{minipage}{1.25\textwidth}
		 \begin{algorithmic}[1]
    		 \Procedure{$\text{L*}$}{$I, O, \mathsf{OQ}, \mathsf{EQ}$}
      		 \State $S \gets \{ \varepsilon \}$, $R \gets \emptyset$, $E \gets \{ \varepsilon \}$, 
              and $T(\varepsilon) \gets \mathsf{OQ}(\varepsilon)$
      		 \Repeat
        	 \While{$(S, R, E, T)$ is not closed}\label{line:lstar-close-begin}
          	 \State Find $s \cdot i \in R$ s.t. $\mathsf{row}(s \cdot i) \ne \mathsf{row}(t)$ for all $t \in S$
          	 \State $S \gets S \cup \{ s \cdot i \}$ and $R \gets R \setminus \{ s \cdot i \}$
        	 \EndWhile\label{line:lstar-close-end}
        	 \State Let $H$ be the hypothesis Moore machine constructed from $(S, R, E, T)$
        	 \State $D \gets \{ (s, i) \in \Call{$\textsc{1Ext}^\text{L*}$}{H} \mid s \cdot i \notin S \cup R \}$\label{line:lstar-extend-begin}   \Comment{$\textsc{1Ext}^\text{L*}$ is defined in the main text}
        	 \If{$D \ne \emptyset$}
          	 \For{$(s, i) \in D$}
            	 \State $R \gets R \cup \{ s \cdot i \}$\label{line:lstar-extend} and 
            	  $T(s \cdot i \cdot e) \gets \mathsf{OQ}(s \cdot i \cdot e)$ for each $e \in E$
          	 \EndFor
          	 \State \Continue\label{line:lstar-extend-end}
        	 \EndIf
        	 \If{$\mathsf{EQ}(H) \ne \mathsf{true}$}
          	 \State Let $w$ be a counterexample reported by $\mathsf{EQ}(H)$
          	 \State \Call{$\textsc{AnalyzeCex}^\text{L*}$}{$H, w$}
        	 \EndIf
      		 \Until{$\mathsf{EQ}(H) = \mathsf{true}$}
    		 \EndProcedure
    		 \Statex
    		 \Procedure{$\textsc{AnalyzeCex}^\text{L*}$}{$H, w$}
      		 \State Find the decomposition $(s, i, d)$ of the counterexample $w$ s.t. $s \cdot i \cdot d = w$ and\label{line:lstar-analyzecex-find}
      		 \Statex\quad $\mathsf{OQ}(t \cdot i \cdot d) \ne \mathsf{OQ}(t' \cdot d)$ where $\mathsf{row}(t) = \delta_H(s)$ and $\mathsf{row}(t') = \delta_H(s \cdot i)$
      		 \State $E \gets E \cup \{ d \}$ and
      		  $T(s \cdot d) \gets \mathsf{OQ}(s \cdot d)$ for each $s \in S \cup R$
    		 \EndProcedure
 		\end{algorithmic}
  \end{minipage}}
\end{algorithm}

 An  appropriate suffix $d$ in Line~\ref{line:lstar-analyzecex-find} is effectively searched
by the binary search~\cite{DBLP:conf/icgi/IsbernerS14,DBLP:journals/iandc/RivestS93}; this leads to the 
following complexity bounds.

\begin{mytheorem}[$\mathsf{OQ}$ and $\mathsf{EQ}$ complexities of L* (\cref{alg:lstar})]\label{lem:lstar-complexity}
Assume that $\mathsf{OQ}$ and $\mathsf{EQ}$ are implemented using an MM $M$. Then
\cref{alg:lstar} can correctly infer $M$ with at most $O(\ell n^2 + n \log m)$ output queries and $O(n)$ equivalence queries, where $n$ is the number of states of $M$, $m$ is the maximal length of counterexamples, and $\ell$ is the input alphabet size.
\end{mytheorem}

\section{Our Contextual Componentwise Learning Algorithm}\label{sec:ccwlstar}
\noindent We introduce our main contribution, a \emph{contextual componentwise L* algorithm}  CCwL*. Two baselines are \emph{monolithic L*} (MnL*) and \emph{componentwise L*} (CwL*). 

\myparagraph{Problem}
We formulate the problem. As discussed in~\cref{sec:intro}, it is tailored to system integration applications where the learner has more access to the SUL.

\begin{myproblem}[componentwise automata learning]\label{prob:componentwise-learning}
  \noindent\begin{description}
    \item[Input:] the problem takes the following inputs:
\begin{enumerate*}[label=\arabic*)]
 \item a directed graph $G = (V, E)$;
 \item alphabets $(\Sigma_e)_{e \in E}$;
 \item system-level oracles $\mathsf{OQ}$ and $\mathsf{EQ}$;
 \item component-level oracles $\mathsf{OQ}_{c}$ and $\mathsf{EQ}_{c}$ for each $c \in \componentNodes$
 \end{enumerate*}
    \item[Output:] a Moore machine $M$ 
  \end{description}
\end{myproblem}

Typically, all the oracles are implemented by a black-box MMN $\mathcal{M}$, and the goal is to learn an MM $M$ that is equivalent to $\mmnMoore{\mathcal{M}}$.

\myparagraph{Two Baselines}
The monolithic L* (MnL*) simply applies L* (\cref{alg:lstar}) to $\mathsf{OQ}$ and $\mathsf{EQ}$, ignoring the network structure $G$ and the component-level oracles. This way one has to learn a large MM, as discussed in \cref{sec:intro}. 

The (naive) componentwise L* (CwL*), in contrast,  runs L* (\cref{alg:lstar}) with the component-level oracles $\mathsf{OQ}_{c}$ and $\mathsf{EQ}_{c}$, and learns each component MM $\mathcal{M}_{c}$ separately. Once it is done, it combines the learned MMs along the graph $G$, gets an MMN $\mathcal{H}$. This can exploit compositionality and decrease the states to learn; yet it may still suffer from \emph{component redundancies} (the cost of learning  parts of components that are not relevant system-level). See \cref{sec:intro}.


\myparagraph{Our Algorithm CCwL*}
Our contextual algorithm CCwL* is shown  in \cref{alg:ccwlstar}; it 
aims to alleviate component redundancies.  We list its core features.
\begin{enumerate}
 \item CCwL* learns  components separately.
This is much like CwL*. Therefore it keeps an observation table $(S_c, R_c, E_c, T_c)$ for each component $c \in \componentNodes$.
 \item CCwL* only uses (component-level) $\mathsf{OQ}_{c}$ and (system-level) $\mathsf{EQ}$. This is unlike CwL* that uses only component-level $\mathsf{OQ}_{c}$  and $\mathsf{EQ}_{c}$.
 \item Learning each component $c$ is much like L* (\cref{alg:lstar}), but CCwL* uses different procedure/function there (namely, $\OneExtER$ and $\textsc{AnalyzeCex}^\text{C}$ in Lines~\ref{line:ccwlstar-1ext-call} and \ref{line:ccwlstar-analyzecex-end}). Notably, these $\OneExtER$ and $\textsc{AnalyzeCex}^\text{C}$ are \emph{contextual}---they depend not only on the component $c$ but also on the other components.
\end{enumerate}


 The oracle $\overline{\mathsf{OQ}}\colon {(\inputAlphabet)}^\ast \to \totalOutputAlphabet^\ast$ in Line~\ref{line:ccwlstar-analyzecex-tildeoq} is for \emph{total output queries}: it answers what strings are observed \emph{at all edges} (including system-level output and inter-component edges), given an input string. The learner can compute it using component-level output query oracles $(\mathsf{OQ}_c)_{c \in \componentNodes}$ in a natural way.

\myparagraph{On $\textsc{AnalyzeCex}^\text{C}$}
Overall, \cref{alg:ccwlstar} mirrors the structure of \cref{alg:lstar}, with differences only  in $\OneExtER$ and $\textsc{AnalyzeCex}^\text{C}$ (Lines~\ref{line:ccwlstar-1ext-call} and \ref{line:ccwlstar-analyzecex-end})). To motivate the latter, recall that  CCwL* uses only system-level EQs---using component-level EQs means we try to learn everything about a component and thus goes against our goal of eliminating component redundancies. Therefore the counterexamples obtained from EQs are system-level input strings $\mathbf{w}$. The procedure $\textsc{AnalyzeCex}^\text{C}$ in \cref{alg:ccwlstar} lets such $\mathbf{w}$  generate a component-level input string $\mathbf{w}'$ for  $c$ in Line~\ref{line:ccwlstar-analyzecex-tildeoq}, and passes it to the analysis routine in \cref{alg:lstar}  (namely $\textsc{AnalyzeCex}^\text{L*}$).

\begin{auxproof}
 This procedure takes input parameters described in \cref{prob:componentwise-learning} and correctly learns the target system as an MMN.
 The learner initializes the observation tables $(S_c, R_c, E_c, T_c)$ for each component $c \in \componentNodes$ (Line~\ref{line:ccwlstar-init-begin}--\ref{line:ccwlstar-init-end}).
 A main loop iterates until the hypothesis MMN $\mathcal{H}$ is equivalent to the target system.
 The learner iteratively extends $S_c$ of the observation tables until they are closed (Line~\ref{line:ccwlstar-closed-begin}--\ref{line:ccwlstar-closed-end}), and constructs the hypothesis MMN $\mathcal{H}$ from the observation tables (Line~\ref{line:ccwlstar-hypothesis}).
 The learner uses the \textsc{1Ext} function to find the set of pairs $(c, \mathbf{s}, \widehat{\mathbf{i}})$, which are not in $S_c \cup R_c$, to make $\mmnMoore{\mathcal{H}}$ complete (Line~\ref{line:ccwlstar-1ext-call}).
 If the set $D$ is not empty (Line~\ref{line:ccwlstar-1ext-if}), since $\mmnMoore{\mathcal{H}}$ is not complete and the learner needs to extends some $R_c$, it updates the observation tables and continues the loop (Line~\ref{line:ccwlstar-1ext-begin}--\ref{line:ccwlstar-1ext-end}).
 Finally, the learner uses the system-level equivalence query $\mathsf{EQ}(\mathcal{H})$ to check the equivalence of $\mathcal{H}$ and the target system (Line~\ref{line:ccwlstar-eq}), and performs counterexample analysis if $\mathsf{EQ}(\mathcal{H})$ reports a counterexample (Line~\ref{line:ccwlstar-analyzecex-begin}--\ref{line:ccwlstar-analyzecex-end}).
 In the $\textsc{AnalyzeCex}^\text{C}$ procedure, the learner finds a component that produces an incorrect output, constructs a counterexample for that component and performs counterexample analysis for that component.
\end{auxproof}

\begin{algorithm}[t!]
  \caption{our contextual componentwise L* algorithm  CCwL* }\label{alg:ccwlstar}
  \scalebox{.8}{\begin{minipage}{1.25\textwidth}
  \begin{algorithmic}[1]
    \Procedure{$\text{CCwL*}$}{$(V, E), (\Sigma_e)_{e \in E}, \mathsf{OQ}, \mathsf{EQ}, (\mathsf{OQ_c})_{c \in \componentNodes}, (\mathsf{EQ_c})_{c \in \componentNodes}$}
      \For{$c \in \componentNodes$}\label{line:ccwlstar-init-begin}
        \State $S_c \gets \{ \varepsilon \}$, $R_c \gets \emptyset$, $E_c \gets \{ \varepsilon \}$, and $T_c(\varepsilon) \gets \mathsf{OQ}_c(\varepsilon)$
      \EndFor\label{line:ccwlstar-init-end}
      \Repeat
        \While{$(S_c, R_c, E_c, T_c)$ is not closed for some $c \in \componentNodes$}\label{line:ccwlstar-closed-begin}
          \State Find $\mathbf{s} \cdot \mathbf{i} \in R_c$
            s.t. $\mathsf{row}_c(\mathbf{s} \cdot \mathbf{i}) \ne \mathsf{row}_c(\mathbf{t})$ for all $\mathbf{t} \in S_c$
          \State $S_c \gets S_c \cup \{ \mathbf{s} \cdot \mathbf{i} \}$ and $R_c \gets R_c \setminus \{ \mathbf{s} \cdot \mathbf{i} \}$
        \EndWhile\label{line:ccwlstar-closed-end}
        \State Let $\mathcal{H}$ be the hypothesis MMN constructed from $(S_c, R_c, E_c, T_c)_{c \in \componentNodes}$\label{line:ccwlstar-hypothesis}
        \State $D \gets \{ (c, \mathbf{s}, \widehat{\mathbf{i}}) \in \Call{$\OneExtER$}{\mathcal{H}} \mid \mathbf{s} \cdot \widehat{\mathbf{i}} \notin S_c \cup R_c \}$\label{line:ccwlstar-1ext-call}
        \If{$D \ne \emptyset$}\label{line:ccwlstar-1ext-if}
          \For{$(c, \mathbf{s}, \widehat{\mathbf{i}}) \in D$}\label{line:ccwlstar-1ext-begin}
            \State $R_c \gets R_c \cup \{ \mathbf{s} \cdot \widehat{\mathbf{i}} \}$ and
                   $T_c(\mathbf{s} \cdot \widehat{\mathbf{i}} \cdot \mathbf{e}) \gets \mathsf{OQ}_c(\mathbf{s} \cdot \widehat{\mathbf{i}} \cdot \mathbf{e})$ for each $\mathbf{e} \in E_c$
          \EndFor
          \Continue\label{line:ccwlstar-1ext-end}
        \EndIf
        \If{$\mathsf{EQ}(\mathcal{H}) \ne \mathsf{true}$}\label{line:ccwlstar-eq}
          \State Let $\mathbf{w}$ be a counterexample reported by $\mathsf{EQ}(\mathcal{H})$\label{line:ccwlstar-analyzecex-begin}
          \State \Call{$\textsc{AnalyzeCex}^\text{C}$}{$\mathcal{H}, \mathbf{w}$}\label{line:ccwlstar-analyzecex-end}
        \EndIf
      \Until{$\mathsf{EQ}(\mathcal{H}) = \mathsf{true}$}
    \EndProcedure
    \Statex
    \Function{$\OneExtER$}{$\mathcal{H}$}
      \State $\widetilde{\mathcal{H}} \gets$ the quotient MMN $\mathcal{H} / \mathcal{E}$ (\FinalOrArxivVersion{\cite[Thm.~A.3]{FujinamiH25ATVA_arxiv_extended_ver}}{\cref{def:quotientMM}}) with respect to $(\mathcal{E}(c))_{c \in \componentNodes}$ \label{line:OneExtERQuotient} 
      \State
             $D \gets \emptyset$
      \For{$\widetilde{\mathbf{q}} = ([\mathsf{row}(\mathbf{s}_c)]_{\mathcal{E}(c)})_{c \in \componentNodes} \in \mathcal{R}(\widetilde{\mathcal{H}})$} \label{line:OneExtERFindTildeQ}
        \For{$\mathbf{i} \in \inputAlphabet$, $c \in \componentNodes$, $\overline{\mathbf{o}} \in \totalOutputFunction(\widetilde{\mathbf{q}})$, and $\mathsf{row}(\mathbf{s}') \in [\mathsf{row}(\mathbf{s}_c)]_{\mathcal{E}(c)}$} \label{line:OneExtERGenOutput}
          \State Let $\widehat{\mathbf{i}}$ be a possible input character to $c$ in $\mathcal{H}$ on $\mathbf{q}$ and $\mathbf{i}$, \ie $\widehat{\mathbf{i}} = \restrictTo{(\mathbf{i}, \overline{\mathbf{o}})}{\incomingEdges{c}}$ \label{line:OneExtERGenInput}
          \State $D \gets D \cup \{ (c, \mathbf{s}', \widehat{\mathbf{i}}) \}$ \label{line:OneExtERAddToD}
        \EndFor
      \EndFor
      \State\Return $D$
    \EndFunction

    \Statex
    \Procedure{$\textsc{AnalyzeCex}^\text{C}$}{$\mathcal{H}, \mathbf{w}$}
    \LComment{this $\textsc{AnalyzeCex}^\text{C}$ is for \emph{sound} $(\mathcal{E},\mathcal{R})$; otherwise ext. is needed  (\FinalOrArxivVersion{\cite[Appendix~D.4]{FujinamiH25ATVA_arxiv_extended_ver}}{\cref{appendix:AnalyzeCexForUnsound}})}
      \State Find a component $c \in \componentNodes$ that produces an incorrect output,
            \label{line:ccwlstar-analyzecex-tildeoq}
      \Statex\quad that is, $\restrictTo{\overline{\mathsf{OQ}}(\mathbf{w})}{\outgoingEdges{c}} \ne \restrictTo{\semMoore{\mathcal{H}}(\mathbf{w})}{\outgoingEdges{c}}$  
      \Comment{the oracle $\overline{\mathsf{OQ}}$  
             is described in the main text}
      \State Construct an input $\widehat{\mathbf{w}}$ to the component $c$ from the system-level input $w$
      \Statex\quad where $\seqIndex{\widehat{\mathbf{w}}}{k} = \restrictTo{(\seqIndex{\mathbf{w}}{k}, \seqIndex{\overline{\mathsf{OQ}}(\mathbf{w})}{k})}{\incomingEdges{c}}$ for each $k\in [0,  |\mathbf{w}|)$ 
      \State Apply \Call{$\textsc{AnalyzeCex}^\text{L*}$}{$H_c, \widehat{\mathbf{w}}$}
       \Comment{\textsc{$\textsc{AnalyzeCex}^\text{L*}$} is from \cref{alg:lstar}}
    \EndProcedure
  \end{algorithmic}
  \end{minipage}}
\end{algorithm}

\myparagraph{On $\OneExtER$}
On the other difference from L* ($\OneExtER$ in Line~\ref{line:ccwlstar-1ext-call}), we  note that the function $\OneExtER$ has two parameters $\mathcal{E}$ (called \emph{component abstraction}) and $\mathcal{R}$ (called \emph{reachability analysis bound (RA bound)}). Combined, they are called \emph{context analysis parameters (CA-parameters)}. We start with some intuitions.

Firstly, the goal of $\OneExtER$ is to find out \emph{what input character $\mathbf{i}$ a component $c$ can receive, when $c$ runs in the MMN $\mathcal{M}$ and $c$'s current state is (represented by the prefix) $\mathbf{s}'$}. It adds all such tuples $(c,\mathsf{s}',\widehat{\mathsf{i}})$ to $D$ (Line~\ref{line:OneExtERAddToD}). 

In principle, it does so via the reachability analysis $\mathcal{R}$ of the current hypothesis MMN $\mathcal{H}$. Specifically, $\OneExtER$ identifies all the combinations $\widetilde{\mathbf{q}}$ of component states that $\mathcal{H}$ can encounter (Line~\ref{line:OneExtERFindTildeQ}), collects all output characters $\overline{\mathbf{o}}$ given by such component states  $\widetilde{\mathbf{q}}$ (Line~\ref{line:OneExtERGenOutput}), and combines this $\overline{\mathbf{o}}$ with  system-level input $\mathbf{i}$ to find a possible input character $\widehat{\mathbf{i}}$ to $c$ (Line~\ref{line:OneExtERGenInput}). 

This baseline behavior of $\OneExtER$ is what happens with the most fine-grained CA-parameters ($\mathcal{E}=\mathsf{Eq},\mathcal{R}=\mathsf{D}_{\infty}$). We present this special case in \FinalOrArxivVersion{\cite[Appendix~D.3]{FujinamiH25ATVA_arxiv_extended_ver}}{\cref{appendix:oneExtEqDInftyForIllustraion}} for  illustration.

\myparagraph{CA-Parameters $\mathcal{E},\mathcal{R}$}
 However, this full reachability analysis can be very expensive; the CA-parameters $\mathcal{E}, \mathcal{R}$ are there to relieve it. The basic idea here is that we quotient hypothesis MMNs in order to ease reachability analysis. Those quotients naturally come with nondeterminism; thus we need the notions of nondeterministic MM and MMN
 (see \FinalOrArxivVersion{\cite[Appendix~A]{FujinamiH25ATVA_arxiv_extended_ver}}{\cref{appendix:nondetMM}}).

 The \emph{component abstraction}  $\mathcal{E}$ specifies how we quotient the components in the hypothesis MMN $\mathcal{H}$ (Line~\ref{line:OneExtERQuotient}). Note that it is used only within  $\OneExtER$ (i.e.\ for context analysis); in particular, it is not directly used in observation tables. 
  \begin{itemize}
     \item $\mathcal{E} = \mathsf{Eq}$ (\emph{equality}) means no quotienting.
    \item $\mathcal{E} = \mathsf{Eq}_k$ (\emph{$k$-equivalence}), with $k\in \mathbb{N}$, is
  \begin{math} \mathsf{Eq}_k=
   \{(s,t)\mid \lambda(s \cdot w) = \lambda(t \cdot w) \text{ for all strings $w$ with $|w| \le k$}\}
  \end{math}. In particular, $\mathsf{Eq}_0 = \{ (s, t) \mid \lambda(s) = \lambda(t) \}$.
    \item $\mathcal{E} = \mathsf{Uni}$ (\emph{universal}) is given by $Q_{c}\times Q_{c}$ and collapses each component MM to a single state.

  \end{itemize}
We define quotients of MMs (see~\FinalOrArxivVersion{\cite[Thm~A.3]{FujinamiH25ATVA_arxiv_extended_ver}}{\cref{def:quotientMM}})
so that 
 quotienting always leads to an \emph{overapproximation} of output behaviors. Therefore, a possible input character $\widehat{\mathbf{i}}$ (Line~\ref{line:OneExtERAddToD}) is never missed. Such a choice of CA-parameters is said to be \emph{sound}.

 The \emph{RA bound} $\mathcal{R}$ specifies how complete our reachability analysis should be (for finding $\widetilde{\mathbf{q}}$,  Line~\ref{line:OneExtERFindTildeQ}).  We do so by limiting the depth of breadth-first search.
\begin{itemize}
 \item $\mathcal{R} = \mathsf{D}_{\infty}$ means we set no bound and run full breadth-first search.
 \item $\mathcal{R} = \mathsf{D}_{d}$ means we set the limit of depth $d\in \mathbb{N}$. 
\end{itemize}
 Here, unlike with $\mathcal{E}$, the use of $\mathcal{R}\neq\mathsf{D}_{\infty}$ may lead to missing
 some $\widetilde{\mathbf{q}}$ and thus 
some possible input  $\widehat{\mathbf{i}}$. Such 
 a choice of CA-parameters
is said to be \emph{unsound}.

\myparagraph{On $\textsc{AnalyzeCex}^\text{C}$, Again}
In case unsound CA-parameters are chosen (i.e.\ $\mathcal{R}\neq\mathsf{D}_{\infty}$), a counterexample can arise not only in the usual L* way (wrong output), but also by finding out that the hypothesis MM for a component $c$ is not prepared for some input character $\widehat{\mathbf{i}}$, missing a transition for $\widehat{\mathbf{i}}$. Therefore $\textsc{AnalyzeCex}^\text{C}$ must be extended to handle such counterexamples. Doing so is not hard, and the extension is shown in \FinalOrArxivVersion{\cite[Appendix~D.4]{FujinamiH25ATVA_arxiv_extended_ver}}{\cref{appendix:AnalyzeCexForUnsound}}. The extension subsumes the one in \cref{alg:ccwlstar}; one can use the extension regardless of soundness of CA-parameters.


\begin{auxproof}
 The \textsc{1Ext} function is shown in \cref{alg:ccwlstar}.
 This function takes a hypothesis MMN $\mathcal{H}$ and returns a set of triples $(c, \mathbf{s}, \widehat{\mathbf{i}})$, where $c \in \componentNodes$, $\mathbf{s} \in S_c$, and $\widehat{\mathbf{i}} \in \componentInputAlphabet{c}$.
 A character $\widehat{\mathbf{i}}$ is a possible input character to the component $c$ in $\mathcal{H}$ on the configuration $\mathbf{q} = (\mathsf{row}(\mathbf{t}_c))_{c \in \componentNodes}$ such that $\mathbf{t}_c = \mathbf{s}$, that is, to make $\mmnMoore{\mathcal{H}}$ complete, a transition from $\mathbf{s}$ with $\widehat{\mathbf{i}}$ in $H_c$ is necessary.
\end{auxproof}


\myparagraph{Query Complexities}
We state the following result.
\begin{mytheorem}[$\mathsf{OQ}$ and $\mathsf{EQ}$ complexities of CCwL*]\label{lem:ccwlstar-complexity}
  Assume that $\mathcal{E},\mathcal{R}$ is sound (i.e.\ $\mathcal{R}=\mathsf{D}_{\infty}$). 
  The CCwL* algorithm (\cref{alg:ccwlstar}), assuming that all oracles are implemented using an MMN $\mathcal{M}$, can correctly infer $\mathcal{M}$ with at most $O(\ell n^2 + n |\componentNodes| \log m)$ component-level output queries and $O(n)$ system-level equivalence queries. Here $n$ is the sum of the numbers of states of component Moore machines in $\mathcal{M}$, $m$ is the maximal length of counterexamples, and $\ell$ is the system-level input alphabet size.

  If  $\mathcal{E},\mathcal{R}$ is unsound, then the number of component-level OQs is bounded by $O(\ell n^2 + n |\componentNodes| \log m + \ell n |\componentNodes|)$, and that of system-level EQs is $O(n + \ell n)$.
\end{mytheorem}
Here is a proof sketch. The  sound case  adapts \cref{lem:lstar-complexity}; the extra $|\componentNodes|$ factor comes from the use of total output queries $\overline{\mathsf{OQ}}(w)$ in $\textsc{AnalyzeCex}^\text{C}$. For the unsound case, 
EQs may also increase transitions (besides states, as in L*); this increase the bound for EQs. The bound for OQs grows because calls of $\textsc{AnalyzeCex}^\text{C}$ increase and OQs are used there.

\begin{auxproof}

 The $\textsc{AnalyzeCex}^\text{C}$ function calls $\overline{\mathsf{OQ}}(w)$, however, this value can be computed using the component-level output queries $(\mathsf{OQ}_c)_{c \in \componentNodes}$ as described in the comment at Line~\ref{line:ccwlstar-analyzecex-tildeoq}.
 Thus, the learner needs $|\componentNodes|$ extra output queries to compute $\overline{\mathsf{OQ}}(w)$ and we conclude the $\mathsf{OQ}$ and $\mathsf{EQ}$ complexities of the CCwL* algorithm as follows:
\end{auxproof}

\section{Implementation and Experiments}\label{sec:experiments}

The code of the implementations, as well as all experiment scripts, is available~\cite{fujinami_2025_15846781}.

\myparagraph{Implementation}
We implemented our proposal CCwL*, together with two baselines  MnL* and CwL*, in Scala.
It takes an MMN as input, which is treated as a black-box teacher and used only for answering queries.
Equivalence queries (EQs) are implemented through  testing by randomly generated input words.


\myparagraph{Benchmarks}
We used two families: \emph{random} benchmarks where random  components are arranged in a fixed network, and \emph{realistic} benchmarks. 

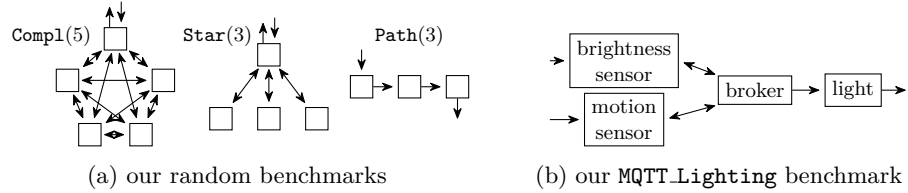
\begin{figure}[tbp]
 \begin{subfigure}{0.55\linewidth}
    \centering
  \begin{tikzpicture}[shorten >=1pt,scale=0.85, every node/.style={transform shape}, node distance=2.5cm,on grid,auto,>={Stealth[round,sep]}]
    \footnotesize
    \begin{scope}
      \node (k1label) at (-1,0.8) {$\mathtt{Compl}(5)$};

      \node[rectangle,draw,minimum width=0.35cm, minimum height=0.35cm] (k1) at (0, 0.75) {};
      \node[rectangle,draw,minimum width=0.35cm, minimum height=0.35cm] (k2) at (-0.75, 0.1) {};
      \node[rectangle,draw,minimum width=0.35cm, minimum height=0.35cm] (k3) at (0.75, 0.1) {};
      \node[rectangle,draw,minimum width=0.35cm, minimum height=0.35cm] (k4) at (-0.4, -0.75) {};
      \node[rectangle,draw,minimum width=0.35cm, minimum height=0.35cm] (k5) at (0.4, -0.75) {};

      \coordinate (kin1) at ($(k1.north east)!0.25!(k1.north west)$);
      \coordinate (kin2) at ($(kin1) + (0,0.4)$);
      \coordinate (kout1) at ($(k1.north east)!0.75!(k1.north west)$);
      \coordinate (kout2) at ($(kout1) + (0,0.4)$);

      \path[<->]
        (k1) edge (k2)
        (k1) edge (k3)
        (k1) edge (k4)
        (k1) edge (k5)
        (k2) edge (k3)
        (k2) edge (k4)
        (k2) edge (k5)
        (k3) edge (k4)
        (k3) edge (k5)
        (k4) edge (k5);
      \path[->]
        (kin2) edge (kin1)
        (kout1) edge (kout2);
    \end{scope}
    \begin{scope}[xshift=2.4cm]
      \node (k1label) at (-0.8,0.8) {$\mathtt{Star}(3)$};

      \node[rectangle,draw,minimum width=0.35cm, minimum height=0.35cm] (s1) at (0, 0.5) {};
      \node[rectangle,draw,minimum width=0.35cm, minimum height=0.35cm] (s2) at (-0.75, -0.5) {};
      \node[rectangle,draw,minimum width=0.35cm, minimum height=0.35cm] (s3) at (0, -0.5) {};
      \node[rectangle,draw,minimum width=0.35cm, minimum height=0.35cm] (s4) at (0.75, -0.5) {};

      \coordinate (sin1) at ($(s1.north east)!0.25!(s1.north west)$);
      \coordinate (sin2) at ($(sin1) + (0,0.4)$);
      \coordinate (sout1) at ($(s1.north east)!0.75!(s1.north west)$);
      \coordinate (sout2) at ($(sout1) + (0,0.4)$);

      \path[<->]
        (s1) edge (s2)
        (s1) edge (s3)
        (s1) edge (s4);

      \path[->]
        (sin2) edge (sin1)
        (sout1) edge (sout2);
    \end{scope}
    \begin{scope}[xshift=4.6cm]
      \node (k1label) at (0,0.8) {$\mathtt{Path}(3)$};

      \node[rectangle,draw,minimum width=0.35cm, minimum height=0.35cm] (p1) at (-0.75, 0) {};
      \node[rectangle,draw,minimum width=0.35cm, minimum height=0.35cm] (p2) at (0, 0) {};
      \node[rectangle,draw,minimum width=0.35cm, minimum height=0.35cm] (p3) at (0.75, 0) {};

      \coordinate (pin1) at ($(p1.north east)!0.5!(p1.north west)$);
      \coordinate (pin2) at ($(pin1) + (0,0.4)$);
      \coordinate (pout1) at ($(p3.south east)!0.5!(p3.south west)$);
      \coordinate (pout2) at ($(pout1) - (0,0.4)$);

      \path[->]
        (p1) edge (p2)
        (p2) edge (p3)
        (pin2) edge (pin1)
        (pout1) edge (pout2);
    \end{scope}
  \end{tikzpicture}

  
  \caption{our random benchmarks}
  \label{fig:topol}
  \end{subfigure}
  \hfill
 \begin{subfigure}{0.4\linewidth}
   \begin{tikzpicture}[shorten >=1pt, scale=0.87, on grid,auto,>={Stealth[round,sep]},every node/.style={transform shape}]
  \footnotesize
  \def\inX{-3.25}
  \def\sensorX{-2.0}
  \def\sensorY{0.45}
  \node (brightness_in) at (\inX, \sensorY) [align=center] {};
  \node[rectangle,draw] (brightness) at (\sensorX, \sensorY) [align=center] {brightness\\sensor};
  \node (motion_in) at (\inX, -\sensorY) [align=center] {};
  \node[rectangle,draw] (motion) at (\sensorX, -\sensorY) [align=center] {motion\\sensor};
  \node[rectangle,draw] (broker) at (0, 0) [align=center] {broker};
  \node[rectangle,draw]  (light) at (1.5, 0) [align=center] {light};
  \node (light_out) at (2.5, 0) [align=center] {};

  \path[->] 
  (brightness_in) edge (brightness)
  (motion_in) edge (motion)
  (broker) edge (light)
  (light) edge (light_out)
  ;

  \path[<->]
    (broker.north west) edge[bend left=0] (brightness.east)
    (broker.south west) edge[bend left=0] (motion.east)
  ;
 \end{tikzpicture}
  \caption{our \lighting{} benchmark}
 \label{figure:lightingTop}
 \end{subfigure}
 \caption{network topologies for our random and \lighting{} benchmarks}
\end{figure}

The random benchmarks $\mathtt{Rand(nwk,comp)}$ use the following parameters.
\begin{itemize}
 \item The parameter $\mathtt{nwk}$ specifies the network topology. We use three families of network topologies: $\mathtt{Compl}(k)$ (a complete graph of $k$ components), $\mathtt{Star}(k)$ (a ``frontend'' component interconnected with $k$ ``backend'' components), and $\mathtt{Path}(k)$ ($k$ components serially connected). See \cref{fig:topol}.
 \item The parameter $\mathtt{comp}\in\{\mathtt{LeanComp},\mathtt{RichComp}\}$ specifies how each component is randomly generated. When $\mathtt{comp}=\mathtt{LeanComp}$, each component is a Moore machine whose number of states is chosen from the normal distribution $N(10,1)$. For each (inter-component) edge, its alphabet size is picked from the uniform distribution over $\{2,3,4,5\}$. 

When $\mathtt{comp}=\mathtt{RichComp}$, we augment each component in such a way that roughly a half of it is redundant. Specifically,
\begin{enumerate*}[label=\arabic*)]
  \item each component is the interleaving product $M^{\circ}_{c}\times M^{\bullet}_{c}$ of two Moore machines generated in the above way (for $\mathtt{LeanComp}$);
  \item the two machines $M^{\circ}_{c}, M^{\bullet}_{c}$ have disjoint input and output alphabets;
  \item therefore the alphabet for each edge in the MMN is bigger than for $\mathtt{LeanComp}$;
  \item nevertheless, the system-level input alphabets as well as component output alphabets are chosen so that only the first machine $M^{\circ}_{c}$ is invoked.
\end{enumerate*}
This way we force the redundancy of $M^{\bullet}_{c}$.

\end{itemize}

Our realistic benchmarks are \lighting{} and $\mathtt{BinaryCounter}(k)$. The latter models a $k$-bit counter; its details are in \FinalOrArxivVersion{\cite[Appendix~B.2]{FujinamiH25ATVA_arxiv_extended_ver}}{\cref{appendix:binCount}}. 
In what follows, we describe \lighting{} in some detail (further details are in \FinalOrArxivVersion{\cite[Appendix~B.1]{FujinamiH25ATVA_arxiv_extended_ver}}{\cref{appendix:lighting}}).

The MMN \lighting{}  models a lighting system in which two sensors and one light communicate (\cref{figure:lightingTop}). Notably, it uses  the \emph{MQTT protocol}~\cite{oasis2019mqtt}---a protocol commonly used  for IoT applications---and thus has a component called an \emph{(MQTT) broker}. In this system,
\begin{enumerate*}[label=(\arabic*)]
  \item the brightness sensor uses \emph{QoS 1} of MQTT---meaning that, for each sensing data, four messages $\mathtt{Connect}$, $\mathtt{ConnAck}$, $\mathtt{Publish}$, $\mathtt{PubAck}$ are exchanged; and
  \item the motion sensor uses \emph{QoS 2} of MQTT. It uses six messages: $\mathtt{Connect}$, $\mathtt{ConnAck}$, $\mathtt{Publish}$, $\mathtt{PubRec}$, $\mathtt{PubRel}$, and $\mathtt{PubComp}$.
\end{enumerate*}
Different QoS levels provide different guarantees; see e.g.~\cite{oasis2019mqtt}. Our Moore machine model for the broker,\footnote{Our broker model is adapted from Automata Wiki \url{https://automata.cs.ru.nl/BenchmarkMQTT-TapplerEtAl2017/Description}. It is originally from~\cite{DBLP:conf/icst/TapplerAB17}.} without knowing who uses which QoS,  prepares for both QoS levels for each client. This redundancy (e.g.\ QoS 2 for brightness) is what we would like to eliminate via context analysis.




\myparagraph{Experiment Settings} We conducted experiments on AWS EC2 r7i.2xlarge instances, with 3.2 GHz Intel Xeon Scalable (Sapphire Rapids), 8 virtual cores, and 64GB RAM, with OpenJDK 23 (OpenJDK 64-Bit Server VM Temurin-23.0.2+7). Both the learner and an SUL were executed in the same machine. We set a timeout of 3600 seconds for the whole learning process; it returns once a system-level equivalence query succeeds. 

After a successful return, we ran extra \emph{validation} where, unlike equivalence queries during learning (these are by random-word testing), rigorous system-level equivalence verification is conducted between the learned system and the SUL. Note that this validation time is not included in the aforementioned timeout. 

For random benchmarks $\mathtt{Rand(nwk,comp)}$, we generated $10$ instances for each parameter value $\mathtt{(nwk,comp)}$, and we report the average.


In evaluation, noting that the speed of  SUL execution can vary greatly  in different applications (cf.\ \cref{sec:intro}), we are not so interested in the total execution time as in the number of queries. Following~\cite{DBLP:conf/fossacs/LabbafGHM23}, we report
\begin{itemize}
 \item the number of \emph{steps} (i.e.\ the number of input characters, but we use the word ``step'' since our input character can be a tuple $(a_{e})_{e}$ in our setting), and
 \item the number of \emph{resets} (resets can be much more costly than steps, see~\cite{DBLP:conf/fossacs/LabbafGHM23}).
\end{itemize}
We report these numbers separately for OQs and EQs. This is because the numbers for EQs depend heavily on how we choose to implement EQs, namely which method to use (random testing, conformance testing, black-box checking etc.), how many and how long words, etc.\footnote{We can imagine  application scenarios where we need a more refined view, separating the numbers of resets and steps for system-level queries and component-level ones. This is the case, for example, when a component-level interface is well-developed and fast (since a component is a commodity) but a system-level interface is slow (since it is a system under development). This  data is not shown due to limited space.
}

\begin{table}[tbp]
      \begin{adjustbox}{addcode={\begin{minipage}{\width}\caption{%
experiment results I. 
The rows are for different algorithms: two baselines (MnL*, CwL*) and our proposal (CCwL*) with different CA-parameters (\cref{sec:ccwlstar}). In the columns, 
\emph{st.} is the number of learned states, 
\emph{tr.} is that of learned transitions,
\emph{OQ reset} is the number of resets
caused by output queries (it coincides with the number of OQs), 
\emph{OQ step} is the number of steps caused by output queries,
\emph{EQ} is the number of equivalence queries,
\emph{EQ reset} and \emph{EQ step} are the numbers of resets and steps caused by EQs, respectively (an EQ conducts testing and thus uses many input words), and
 \emph{L.\ time} (``Learner time'') is the time (seconds) spent for the tasks on the learner's side (context analysis, counterexample analysis, building observation tables, etc.), and \emph{valid?} reports the numbers of instances of ``result validated,'' ``result found incorrect,'' and ``timeout.'' Note that we used 10 instances for each random benchmark.
	  }\label{table:exprI}}{\end{minipage}},rotate=90,center}
\scalebox{.8}{ 
\begin{tabular}{@{}p{2.6cm}%
*{8}{R{\dimexpr.8cm-2\tabcolsep\relax}}R{\dimexpr1.2cm-2\tabcolsep\relax}
*{8}{R{\dimexpr.8cm-2\tabcolsep\relax}}R{\dimexpr1.2cm-2\tabcolsep\relax}
*{8}{R{\dimexpr.8cm-2\tabcolsep\relax}}R{\dimexpr1.2cm-2\tabcolsep\relax}
@{}}
\toprule
&
\multicolumn{9}{c}{\texttt{Rand(Compl(5),LeanComp)}}
&
\multicolumn{9}{c}{\texttt{Rand(Star(5),LeanComp)}}
&
\multicolumn{9}{c}{\texttt{Rand(Path(5),LeanComp)}}
\\\cmidrule(lr){2-10}\cmidrule(lr){11-19}\cmidrule(lr){20-28}
algo.
& st. 
& tr. 
& OQ reset
& OQ step
& EQ
& EQ reset
& EQ step
& L. time
& valid?
& st.
& tr.
& OQ reset
& OQ step
& EQs
& EQ reset
& EQ step
& L. time
& valid?
& st.
& tr.
& OQ reset
& OQ step
& EQs
& EQ reset
& EQ step
& L. time
& valid?
\\\midrule
\rowcolor{gray!20} MnL*
& 27K & 53K & 591K & 18M & 15.0 & 129 & 32K & 1.0K & 0/1/9 & 26K & 53K & 917K & 33M & 23.5 & 417 & 105K & 1.0K & 0/2/8 & 5.5K & 21K & 496K & 15M & 19.7 & 223 & 54K & 221 & 1/8/1 \\
CwL*
& 46.9 & 12K & 15K & 32K & 6.20 & 501 & 137K & 0.86 & 10/0/0 & 55.6 & 26K & 26K & 51K & 17.6 & 613 & 165K & 1.14 & 10/0/0 & 47.7 & 166 & 570 & 2.8K & 17.9 & 513 & 137K & 0.53 & 10/0/0 \\
\rowcolor{gray!20} CCwL*($\mathsf{Eq},\mathsf{D}_{\infty}$)
& 46.9 & 9.0K & 11K & 35K & 2.00 & 101 & 27K & 30.8 & 10/0/0 & 54.8 & 18K & 18K & 46K & 7.30 & 107 & 28K & 188 & 10/0/0 & 46.8 & 140 & 457 & 2.5K & 6.70 & 110 & 29K & 6.43 & 10/0/0 \\
CCwL*($\mathsf{Eq},\mathsf{D}_{0}$)
& 46.9 & 8.3K & 16K & 911K & 5.4K & 133K & 33M & 1.4K & 0/10/0 & 54.3 & 7.0K & 14K & 1.2M & 6.8K & 231K & 59M & 1.6K & 0/9/1 & 46.3 & 133 & 490 & 6.6K & 64.3 & 203 & 41K & 1.35 & 7/3/0 \\
\rowcolor{gray!20} CCwL*($\mathsf{Eq}_0,\mathsf{D}_{\infty}$)
& 46.9 & 9.0K & 11K & 35K & 2.00 & 101 & 27K & 30.4 & 10/0/0 & 54.8 & 18K & 18K & 47K & 7.10 & 107 & 28K & 24.0 & 10/0/0 & 47.3 & 151 & 501 & 2.8K & 6.60 & 110 & 29K & 2.01 & 10/0/0 \\
CCwL*($\mathsf{Eq}_0,\mathsf{D}_{0}$)
& 46.9 & 8.3K & 16K & 911K & 5.4K & 133K & 33M & 1.3K & 0/10/0 & 54.3 & 7.0K & 14K & 1.2M & 6.8K & 231K & 59M & 1.6K & 0/9/1 & 46.4 & 134 & 490 & 6.2K & 58.0 & 191 & 39K & 1.37 & 7/3/0 \\
\rowcolor{gray!20} CCwL*($\mathsf{Uni},\mathsf{D}_{0}$)
& 46.9 & 9.0K & 11K & 29K & 2.20 & 101 & 27K & 93.0 & 10/0/0 & 54.8 & 19K & 19K & 47K & 7.10 & 107 & 28K & 32.5 & 10/0/0 & 47.3 & 152 & 506 & 2.7K & 6.60 & 110 & 28K & 2.06 & 10/0/0 \\
\midrule
&
\multicolumn{9}{c}{\texttt{Rand(Compl(5),RichComp)}}
&
\multicolumn{9}{c}{\texttt{Rand(Star(5),RichComp)}}
&
\multicolumn{9}{c}{\texttt{Rand(Path(5),RichComp)}}
\\\cmidrule(lr){2-10}\cmidrule(lr){11-19}\cmidrule(lr){20-28}
algo.
& st.
& tr.
& OQ reset
& OQ step
& EQs
& EQ reset
& EQ step
& L. time
& valid?
& st.
& tr.
& OQ reset
& OQ step
& EQs
& EQ reset
& EQ step
& L. time
& valid?
& st.
& tr.
& OQ reset
& OQ step
& EQs
& EQ reset
& EQ step
& L. time
& valid?
\\\midrule
\rowcolor{gray!20} MnL*
& --- & --- & --- & --- & --- & --- & --- & --- & 0/0/10 & 41K & 82K & 1.0M & 31M & 22.0 & 272 & 67K & 2.0K & 0/1/9 & 6.8K & 26K & 732K & 23M & 22.1 & 301 & 75K & 361 & 1/9/0 \\
CwL*
& 747 & 239K & 1.3M & 8.4M & 23.6 & 519 & 138K & 20.7 & 10/0/0 & 910 & 22K & 70K & 583K & 40.1 & 637 & 167K & 1.96 & 10/0/0 & 681 & 4.8K & 39K & 404K & 36.4 & 536 & 139K & 1.33 & 10/0/0 \\
\rowcolor{gray!20} CCwL*($\mathsf{Eq},\mathsf{D}_{\infty}$)
& 47.3 & 9.4K & 10K & 31K & 1.90 & 101 & 27K & 23.4 & 10/0/0 & 56.7 & 11K & 11K & 29K & 8.00 & 107 & 28K & 190 & 10/0/0 & 44.4 & 138 & 462 & 2.6K & 7.10 & 106 & 28K & 6.76 & 8/2/0 \\
CCwL*($\mathsf{Eq},\mathsf{D}_{0}$)
& 47.2 & 7.8K & 14K & 838K & 5.0K & 120K & 30M & 1.2K & 0/9/1 & 56.7 & 6.0K & 12K & 991K & 5.8K & 187K & 48M & 1.3K & 0/10/0 & 44.4 & 136 & 512 & 7.7K & 68.4 & 216 & 44K & 1.54 & 6/4/0 \\
\rowcolor{gray!20} CCwL*($\mathsf{Eq}_0,\mathsf{D}_{\infty}$)
& 47.3 & 9.4K & 10K & 31K & 1.90 & 101 & 27K & 22.4 & 10/0/0 & 56.7 & 11K & 11K & 29K & 8.00 & 107 & 28K & 16.1 & 10/0/0 & 44.6 & 145 & 482 & 2.7K & 7.00 & 106 & 28K & 1.89 & 8/2/0 \\
CCwL*($\mathsf{Eq}_0,\mathsf{D}_{0}$)
& 47.2 & 7.8K & 14K & 838K & 5.0K & 120K & 30M & 1.2K & 0/9/1 & 56.7 & 6.0K & 12K & 991K & 5.8K & 187K & 48M & 1.4K & 0/10/0 & 44.4 & 137 & 505 & 7.0K & 59.2 & 202 & 43K & 1.47 & 6/4/0 \\
\rowcolor{gray!20} CCwL*($\mathsf{Uni},\mathsf{D}_{0}$)
& 47.3 & 9.4K & 11K & 26K & 1.90 & 101 & 27K & 74.3 & 10/0/0 & 56.7 & 11K & 11K & 28K & 8.20 & 108 & 28K & 20.5 & 10/0/0 & 45.0 & 147 & 488 & 2.7K & 7.10 & 106 & 28K & 1.99 & 9/1/0 \\
\midrule
&
\multicolumn{9}{c}{\lighting}
&
\multicolumn{9}{c}{\texttt{BinaryCounter(5)}}
&
\multicolumn{9}{c}{\texttt{BinaryCounter(10)}}
\\\cmidrule(lr){2-10}\cmidrule(lr){11-19}\cmidrule(lr){20-28}
algo.
& st.
& tr.
& OQ reset
& OQ step
& EQs
& EQ reset
& EQ step
& L. time
& valid?
& st.
& tr.
& OQ reset
& OQ step
& EQs
& EQ reset
& EQ step
& L. time
& valid?
& st.
& tr.
& OQ reset
& OQ step
& EQs
& EQ reset
& EQ step
& L. time
& valid?
\\\midrule
\rowcolor{gray!20} MnL*
& 169 & 1.5K & 47K & 1.4M & 10.9 & 238 & 59K & 17.6 & 0/10/0 & 70.0 & 140 & 212 & 5.9K & 2.00 & 101 & 26K & 0.76 & 10/0/0 & --- & --- & --- & --- & --- & --- & --- & --- & 0/0/10 \\
CwL*
& 39.0 & 2.5K & 12K & 61K & 11.8 & 412 & 105K & 0.76 & 10/0/0 & 15.0 & 30.0 & 39.1 & 80.6 & 6.00 & 501 & 129K & 0.41 & 10/0/0 & 30.0 & 60.0 & 74.1 & 141 & 11.0 & 1.0K & 258K & 0.72 & 10/0/0 \\
\rowcolor{gray!20} CCwL*($\mathsf{Eq},\mathsf{D}_{\infty}$)
& 27.0 & 130 & 348 & 2.4K & 4.40 & 112 & 28K & 1.36 & 10/0/0 & 14.0 & 25.0 & 30.0 & 45.0 & 1.00 & 100 & 26K & 0.88 & 10/0/0 & 29.0 & 50.0 & 60.0 & 90.0 & 1.00 & 100 & 26K & 2.27 & 10/0/0 \\
CCwL*($\mathsf{Eq},\mathsf{D}_{0}$)
& 27.0 & 129 & 412 & 7.4K & 72.5 & 292 & 59K & 1.94 & 10/0/0 & 14.0 & 25.0 & 44.4 & 357 & 15.4 & 115 & 26K & 0.88 & 10/0/0 & 25.0 & 41.0 & 73.4 & 2.0K & 23.4 & 128 & 28K & 1.73 & 0/10/0 \\
\rowcolor{gray!20} CCwL*($\mathsf{Eq}_0,\mathsf{D}_{\infty}$)
& 27.0 & 134 & 362 & 2.8K & 4.60 & 116 & 29K & 1.21 & 10/0/0 & 14.0 & 25.0 & 30.0 & 45.0 & 1.00 & 100 & 26K & 0.87 & 10/0/0 & 29.0 & 50.0 & 60.0 & 90.0 & 1.00 & 100 & 26K & 2.27 & 10/0/0 \\
CCwL*($\mathsf{Eq}_0,\mathsf{D}_{0}$)
& 27.0 & 129 & 413 & 7.6K & 71.5 & 311 & 64K & 1.96 & 9/1/0 & 14.0 & 25.0 & 44.4 & 357 & 15.4 & 115 & 26K & 0.87 & 10/0/0 & 25.0 & 41.0 & 73.4 & 2.0K & 23.4 & 128 & 28K & 1.78 & 0/10/0 \\
\rowcolor{gray!20} CCwL*($\mathsf{Uni},\mathsf{D}_{0}$)
& 28.0 & 488 & 1.3K & 7.2K & 4.60 & 116 & 29K & 2.54 & 10/0/0 & 14.0 & 28.0 & 33.0 & 54.0 & 1.00 & 100 & 26K & 0.87 & 10/0/0 & 29.0 & 58.0 & 68.0 & 114 & 1.00 & 100 & 26K & 9.73 & 10/0/0 \\
\bottomrule
\end{tabular}
}
\end{adjustbox}
\end{table}

\begin{table}[t]
\caption{%
experiment results II. 
The legend is the same as \cref{table:exprI}
}\label{table:exprII} 
\scalebox{.7}{ 
\begin{tabular}{@{}p{2.6cm}%
*{8}{R{\dimexpr.8cm-2\tabcolsep\relax}}R{\dimexpr1.2cm-2\tabcolsep\relax}
*{8}{R{\dimexpr.8cm-2\tabcolsep\relax}}R{\dimexpr1.2cm-2\tabcolsep\relax}
*{8}{R{\dimexpr.8cm-2\tabcolsep\relax}}R{\dimexpr1.2cm-2\tabcolsep\relax}
@{}}
\toprule
&
\multicolumn{9}{c}{\texttt{Rand(Star(3),LeanComp)}}
&
\multicolumn{9}{c}{\texttt{Rand(Star(7),LeanComp)}}
\\\cmidrule(lr){2-10}\cmidrule(lr){11-19}
algo.
& st. 
& tr. 
& OQ reset
& OQ step
& EQ
& EQ reset
& EQ step
& L. time
& valid?
& st.
& tr.
& OQ reset
& OQ step
& EQs
& EQ reset
& EQ step
& L. time
& valid?
\\\midrule
\rowcolor{gray!20} MnL*
& 3.3K & 13K & 347K & 6.8M & 26.8 & 474 & 119K & 89.1 & 1/9/0 & --- & --- & --- & --- & --- & --- & --- & --- & 0/0/10 \\
CwL*
& 38.4 & 1.7K & 2.6K & 6.9K & 12.9 & 409 & 110K & 0.50 & 10/0/0 & 74.7 & 280K & 280K & 534K & 26.6 & 819 & 220K & 5.01 & 10/0/0 \\
\rowcolor{gray!20} CCwL*($\mathsf{Eq},\mathsf{D}_{\infty}$)
& 38.4 & 1.3K & 2.0K & 6.5K & 6.80 & 106 & 28K & 3.57 & 10/0/0 & 66.0 & 9.4K & 10K & 27K & 9.00 & 108 & 28K & 1.1K & 1/0/9 \\
CCwL*($\mathsf{Eq},\mathsf{D}_{0}$)
& 38.4 & 1.1K & 2.7K & 160K & 1.0K & 16K & 3.9M & 79.8 & 2/8/0 & 66.0 & 6.4K & 13K & 1.0M & 6.2K & 200K & 51M & 1.9K & 0/1/9 \\
\rowcolor{gray!20} CCwL*($\mathsf{Eq}_0,\mathsf{D}_{\infty}$)
& 38.4 & 1.3K & 2.1K & 6.6K & 6.90 & 106 & 28K & 1.65 & 10/0/0 & 74.5 & 179K & 179K & 447K & 10.0 & 109 & 28K & 646 & 10/0/0 \\
CCwL*($\mathsf{Eq}_0,\mathsf{D}_{0}$)
& 38.4 & 1.1K & 2.7K & 160K & 1.0K & 16K & 3.9M & 77.9 & 2/8/0 & 66.0 & 6.4K & 13K & 1.0M & 6.2K & 200K & 51M & 2.0K & 0/1/9 \\
\rowcolor{gray!20} CCwL*($\mathsf{Uni},\mathsf{D}_{0}$)
& 38.4 & 1.4K & 2.2K & 6.8K & 6.40 & 106 & 28K & 1.98 & 10/0/0 & 74.5 & 183K & 183K & 415K & 9.50 & 109 & 28K & 693 & 10/0/0 \\
\midrule
&
\multicolumn{9}{c}{\texttt{Rand(Star(3),RichComp)}}
&
\multicolumn{9}{c}{\texttt{Rand(Star(7),RichComp)}}
\\\cmidrule(lr){2-10}\cmidrule(lr){11-19}
algo.
& st.
& tr.
& OQ reset
& OQ step
& EQs
& EQ reset
& EQ step
& L. time
& valid?
& st.
& tr.
& OQ reset
& OQ step
& EQs
& EQ reset
& EQ step
& L. time
& valid?
\\\midrule
\rowcolor{gray!20} MnL*
& 3.2K & 12K & 238K & 4.1M & 22.1 & 298 & 74K & 56.0 & 1/9/0 & --- & --- & --- & --- & --- & --- & --- & --- & 0/0/10 \\
CwL*
& 540 & 5.6K & 37K & 360K & 27.4 & 430 & 112K & 1.21 & 10/0/0 & 1.2K & 471K & 535K & 1.6M & 57.2 & 851 & 222K & 9.01 & 10/0/0 \\
\rowcolor{gray!20} CCwL*($\mathsf{Eq},\mathsf{D}_{\infty}$)
& 37.4 & 1.3K & 2.2K & 7.0K & 4.90 & 104 & 28K & 2.76 & 10/0/0 & 73.5 & 100K & 101K & 279K & 10.5 & 110 & 28K & 2.3K & 4/0/6 \\
CCwL*($\mathsf{Eq},\mathsf{D}_{0}$)
& 37.4 & 1.2K & 3.0K & 163K & 1.1K & 16K & 4.0M & 85.4 & 1/9/0 & 75.0 & 8.3K & 17K & 1.4M & 8.1K & 297K & 76M & 3.5K & 0/1/9 \\
\rowcolor{gray!20} CCwL*($\mathsf{Eq}_0,\mathsf{D}_{\infty}$)
& 37.4 & 1.3K & 2.3K & 7.5K & 5.30 & 105 & 28K & 1.54 & 10/0/0 & 74.1 & 204K & 204K & 520K & 9.50 & 108 & 28K & 616 & 10/0/0 \\
CCwL*($\mathsf{Eq}_0,\mathsf{D}_{0}$)
& 37.4 & 1.2K & 3.0K & 163K & 1.1K & 16K & 4.0M & 83.4 & 1/9/0 & 75.0 & 8.3K & 17K & 1.4M & 8.1K & 297K & 76M & 3.0K & 0/1/9 \\
\rowcolor{gray!20} CCwL*($\mathsf{Uni},\mathsf{D}_{0}$)
& 37.4 & 1.4K & 2.4K & 7.6K & 5.00 & 104 & 28K & 1.73 & 10/0/0 & 74.2 & 214K & 215K & 487K & 9.80 & 109 & 28K & 676 & 10/0/0 \\
\bottomrule
\end{tabular}
}
\end{table}

\myparagraph{Results and Discussions} We report the results in \cref{table:exprI,table:exprII}. We discuss them along some research questions (RQs).


\noindent\textbf{RQ1:} Is the flexibility of CA-parameters useful? Which parameter to use?

A natural theoretical expectation of  benefit, and also the learner's computational cost (\emph{L.\ time}), is $\mathsf{Eq}>\mathsf{Eq}_{0}>\mathsf{Uni}$ (on $\mathcal{E}$) and $\mathsf{D}_{\infty}>\mathsf{D}_{0}$ (on $\mathcal{R}$). The experimental results confirm that this expectation is largely correct.

On benefit, indeed,  finer-grained CA (e.g.\ $(\mathsf{Eq},\mathsf{D}_{\infty})$) yielded smaller automata with fewer  resets and steps. This  is more notable in $\mathcal{R}$  than in $\mathcal{E}$. 

As an anomaly,  $(\mathsf{Uni}, \mathsf{D}_{0})$ performed pretty well on $\mathtt{Rand(\_\,,RichComp)}$. But it did not  on \lighting. This is natural: the redundancy in $\mathtt{Rand(\_\,,RichComp)}$ is non-temporal (some input characters are  never used) and even coarse-grained  $(\mathsf{Uni}, \mathsf{D}_{0})$ could detect it; but the redundancy in \lighting{} is temporal (what input characters are not used changes over time) and finding it was harder.

On the learner's cost (\emph{L.\ time}), the above expectation is not always correct: coarser CA often led to explosion of queries, which incurred the learner's bookkeeping cost. That said, the coarsest $\mathcal{E}=\mathsf{Uni}$ did not suffer from this problem. 

Overall, these observations suggest the following. There are different classes of SULs: in one class (e.g.\ \lighting{}), component redundancies are temporal,  and only fine-grained CA  e.g.\ with $(\mathsf{Eq},\mathsf{D}_{\infty})$ can detect them; in another class (e.g.\ $\mathtt{Rand(\_\,,RichComp)}$), redundancies are totally not temporal, and coarse-grained CA with e.g.\  $(\mathsf{Uni}, \mathsf{D}_{0})$ can detect them without much overhead.
This will guide a choice of CA-parameters when the nature of an SUL is known (which class it belongs to?). When an SUL's nature is unknown,  one can try some intermediate CA-parameters; in~\FinalOrArxivVersion{\cite[Appendix~D.5]{FujinamiH25ATVA_arxiv_extended_ver}}{\cref{appendix:moreExpResults}}, we introduce three such ($\mathcal{R}=\mathsf{D}_{\mathsf{sum}},\mathsf{D}_{\mathsf{max}},\mathsf{D}_{\mathsf{min}}$) and evaluate them.


\noindent\textbf{RQ2:} How does CCwL*'s performance compare with that of CwL* or MnL*?

 Henceforth, we follow the suggestion in RQ1 and focus on the CA-parameters  CCwL*($\mathsf{Eq}$,$\mathsf{D}_{\infty}$) and CCwL*($\mathsf{Uni}$, $\mathsf{D}_{0}$).

The advantages of  CwL* and CCwL*---both are componentwise---over  monolithic MnL* are observed in general. This is as expected (cf.\ \cref{sec:intro}). 

In the comparison of CCwL* and (naive) CwL*, we observe that our goal (CA for eliminating  component redundancies) is fulfilled: in the benchmarks with such redundancies ($\mathtt{Rand(\_\,,RichComp)}$, \lighting{}, $\mathtt{BinaryCounter}(k)$), CCwL* clearly outperformed CwL* in terms of automata size, resets and steps. 

On the other benchmarks  (namely $\mathtt{Rand(\_\,,LeanComp)}$), we still observe that 1) CCwL* and CwL* perform similarly, and 2) CCwL* can reduce the cost of EQs. The latter is because EQs in CCwL* are system-level, unlike  component-level EQs in CwL*; a counterexample from the former can be reused for multiple components and suggest many new states.

\noindent\textbf{RQ3:} What is the cost of context analysis? Is it tolerable?

The additional cost for context analysis is part of \emph{L.\ time}. This cost is on the learner's side and  can often be discounted (an SUL is usually slower and  is more likely to be a bottleneck); still we want to confirm that the cost is tolerable.

Indeed,  \emph{L.\ time} is  often much larger for CCwL* than for CwL*: in a large benchmark $\mathtt{Rand(Star(7),LeanComp)}$, a few seconds for CwL* but hundreds of seconds for CCwL*. Whether this cost is tolerable depends on the cost model. For example, in embedded or HILS applications, taking 1 sec.\ for a reset and 10 ms.\ for a step is a norm. The gap of \emph{L.\ time} then becomes ignorable.

\noindent\textbf{RQ4:} How does CCwL* scale to complex SULs? What SULs are suited?

CCwL* is designed to exploit redundancy of components. We have seen that, indeed, it performs well with benchmarks with redundancy.

The scalability question can be interpreted in two ways. One is \emph{whether CCwL* can extract a small essence from a seemingly complex system}; then the answer is yes. For example, on \lighting{} and $\mathtt{Rand(Star(7),LeanComp)}$, it learned much smaller automata than MnL* and CwL* did.

The other possible question is \emph{whether CCwL* can extract an essence even if it is large} and our experiments do not allow us to answer yes. The largest MMN learned by CCwL* so far is of dozens of states, not more. The challenge here is the alphabet size---it grows exponentially with respect to the number of incoming edges---and the cost of CA that is  impacted by it. As future work, we plan to work on deal with such large alphabets, e.g.\ by abstracting alphabets.

\ih{I added this because of what we promised in the rebuttal}
Summarizing the above discussions along RQ1--4, we conclude that 1) we are yet to investigate in-depth the practical scalability of our redundancy elimination methods, but 2) with the experimental results that show the efficiency of CCwL* for several benchmarks, the current work definitely opens promising avenues for future research. Regarding the first point, the main difficulty is that there are no existing benchmarks suited for our purpose, namely large real-world compositional systems whose network structures are known and formalized. We are currently mining IoT and robotics applications for such benchmarks.

\section{Conclusions and Future Work}
For compositional automata learning, we identified a new application domain of \emph{system integration}, formalized its problem setting using \emph{Moore machine networks}, and presented a novel \emph{contextual componentwise learning} algorithm CCwL*. It assumes that both system-level and component-level queries are available; to cope with the challenge of complexities due to redundancies in components (some parts do not contribute to the whole system), CCwL* performs \emph{context analysis}. Our experimental evaluation shows its practical relevance. 

One important future direction is to deal with large alphabets, as mentioned in \cref{sec:experiments}. For example, in those applications where inter-component interactions are \emph{sparse}---the characters transmitted are $\bot$ (``do nothing'') most of the time---a theoretical framework that exploits this sparseness will be useful. We are also considering abstraction using \emph{symbolic automata}~\cite{DBLP:conf/tacas/DrewsD17}.
\bibliographystyle{splncs04}
\bibliography{main}

\begin{thebibliography}{10}
\providecommand{\url}[1]{\texttt{#1}}
\providecommand{\urlprefix}{URL }
\providecommand{\doi}[1]{https://doi.org/#1}

\bibitem{DBLP:conf/stoc/AlurM04}
Alur, R., Madhusudan, P.: Visibly pushdown languages. In: Babai, L. (ed.)
  Proceedings of the 36th Annual {ACM} Symposium on Theory of Computing,
  Chicago, IL, USA, June 13-16, 2004. pp. 202--211. {ACM} (2004).
  \doi{10.1145/1007352.1007390}

\bibitem{DBLP:journals/iandc/Angluin87}
Angluin, D.: Learning regular sets from queries and counterexamples. Inf.
  Comput.  \textbf{75}(2),  87--106 (1987). \doi{10.1016/0890-5401(87)90052-6}

\bibitem{DBLP:conf/cav/ArgyrosD18}
Argyros, G., D'Antoni, L.: The learnability of symbolic automata. In: Chockler,
  H., Weissenbacher, G. (eds.) Computer Aided Verification - 30th International
  Conference, {CAV} 2018, Held as Part of the Federated Logic Conference, FloC
  2018, Oxford, UK, July 14-17, 2018, Proceedings, Part {I}. Lecture Notes in
  Computer Science, vol. 10981, pp. 427--445. Springer (2018).
  \doi{10.1007/978-3-319-96145-3\_23}

\bibitem{DBLP:conf/birthday/BainczykSSH17}
Bainczyk, A., Schieweck, A., Steffen, B., Howar, F.: Model-based testing
  without models: The todomvc case study. In: Katoen, J., Langerak, R.,
  Rensink, A. (eds.) ModelEd, TestEd, TrustEd - Essays Dedicated to Ed Brinksma
  on the Occasion of His 60th Birthday. Lecture Notes in Computer Science, vol.
  10500, pp. 125--144. Springer (2017). \doi{10.1007/978-3-319-68270-9\_7}

\bibitem{DBLP:conf/tacas/DrewsD17}
Drews, S., D'Antoni, L.: Learning symbolic automata. In: Legay, A., Margaria,
  T. (eds.) Tools and Algorithms for the Construction and Analysis of Systems -
  23rd International Conference, {TACAS} 2017, Held as Part of the European
  Joint Conferences on Theory and Practice of Software, {ETAPS} 2017, Uppsala,
  Sweden, April 22-29, 2017, Proceedings, Part {I}. Lecture Notes in Computer
  Science, vol. 10205, pp. 173--189 (2017). \doi{10.1007/978-3-662-54577-5\_10}

\bibitem{DBLP:conf/icse/DuhaibyG20}
al~Duhaiby, O., Groote, J.F.: Active learning of decomposable systems. In: Bae,
  K., Bianculli, D., Gnesi, S., Plat, N. (eds.) FormaliSE@ICSE 2020: 8th
  International Conference on Formal Methods in Software Engineering, Seoul,
  Republic of Korea, July 13, 2020. pp. 1--10. {ACM} (2020).
  \doi{10.1145/3372020.3391560}

\bibitem{DBLP:journals/lmcs/FismanFZ23}
Fisman, D., Frenkel, H., Zilles, S.: Inferring symbolic automata. Log. Methods
  Comput. Sci.  \textbf{19}(2) (2023). \doi{10.46298/LMCS-19(2:5)2023}

\bibitem{DBLP:journals/sttt/FrohmeS21}
Frohme, M., Steffen, B.: Compositional learning of mutually recursive
  procedural systems. Int. J. Softw. Tools Technol. Transf.  \textbf{23}(4),
  521--543 (2021). \doi{10.1007/S10009-021-00634-Y}

\bibitem{fujinami_2025_15846781}
Fujinami, H., Waga, M., Hasuo, I.: Artifact archive for "componentwise automata
  learning for system integration" (Jul 2025). \doi{10.5281/zenodo.15846781}

\bibitem{DBLP:conf/fossacs/HeerdtKR020}
van Heerdt, G., Kupke, C., Rot, J., Silva, A.: Learning weighted automata over
  principal ideal domains. In: Goubault{-}Larrecq, J., K{\"{o}}nig, B. (eds.)
  Foundations of Software Science and Computation Structures - 23rd
  International Conference, {FOSSACS} 2020, Held as Part of the European Joint
  Conferences on Theory and Practice of Software, {ETAPS} 2020, Dublin,
  Ireland, April 25-30, 2020, Proceedings. Lecture Notes in Computer Science,
  vol. 12077, pp. 602--621. Springer (2020).
  \doi{10.1007/978-3-030-45231-5\_31}

\bibitem{DBLP:conf/rv/IsbernerHS14}
Isberner, M., Howar, F., Steffen, B.: The {TTT} algorithm: {A} redundancy-free
  approach to active automata learning. In: Bonakdarpour, B., Smolka, S.A.
  (eds.) Runtime Verification - 5th International Conference, {RV} 2014,
  Toronto, ON, Canada, September 22-25, 2014. Proceedings. Lecture Notes in
  Computer Science, vol.~8734, pp. 307--322. Springer (2014).
  \doi{10.1007/978-3-319-11164-3\_26}

\bibitem{DBLP:conf/cav/IsbernerHS15}
Isberner, M., Howar, F., Steffen, B.: The open-source learnlib - {A} framework
  for active automata learning. In: Kroening, D., Pasareanu, C.S. (eds.)
  Computer Aided Verification - 27th International Conference, {CAV} 2015, San
  Francisco, CA, USA, July 18-24, 2015, Proceedings, Part {I}. Lecture Notes in
  Computer Science, vol.~9206, pp. 487--495. Springer (2015).
  \doi{10.1007/978-3-319-21690-4\_32}

\bibitem{DBLP:conf/icgi/IsbernerS14}
Isberner, M., Steffen, B.: An abstract framework for counterexample analysis in
  active automata learning. In: Clark, A., Kanazawa, M., Yoshinaka, R. (eds.)
  Proceedings of the 12th International Conference on Grammatical Inference,
  {ICGI} 2014, Kyoto, Japan, September 17-19, 2014. {JMLR} Workshop and
  Conference Proceedings, vol.~34, pp. 79--93. JMLR.org (2014),
  \url{http://proceedings.mlr.press/v34/isberner14a.html}

\bibitem{DBLP:journals/corr/abs-2405-08647}
Koenders, R., Moerman, J.: Output-decomposed learning of mealy machines. CoRR
  \textbf{abs/2405.08647} (2024). \doi{10.48550/ARXIV.2405.08647}, presented at
  LearnAut 2024

\bibitem{DBLP:conf/fossacs/LabbafGHM23}
Labbaf, F., Groote, J.F., Hojjat, H., Mousavi, M.R.: Compositional learning for
  interleaving parallel automata. In: Kupferman, O., Sobocinski, P. (eds.)
  Foundations of Software Science and Computation Structures - 26th
  International Conference, FoSSaCS 2023, Held as Part of the European Joint
  Conferences on Theory and Practice of Software, {ETAPS} 2023, Paris, France,
  April 22-27, 2023, Proceedings. Lecture Notes in Computer Science, vol.
  13992, pp. 413--435. Springer (2023). \doi{10.1007/978-3-031-30829-1\_20}

\bibitem{DBLP:journals/sttt/LustigV13}
Lustig, Y., Vardi, M.Y.: Synthesis from component libraries. Int. J. Softw.
  Tools Technol. Transf.  \textbf{15}(5-6),  603--618 (2013).
  \doi{10.1007/S10009-012-0236-Z}

\bibitem{MalavoltaNSRLSL23}
Malavolta, I., Nirghin, K., Scoccia, G.L., Romano, S., Lombardi, S.,
  Scanniello, G., Lago, P.: Javascript dead code identification, elimination,
  and empirical assessment. IEEE Transactions on Software Engineering
  \textbf{49}(7),  3692--3714 (2023). \doi{10.1109/TSE.2023.3267848}

\bibitem{DBLP:conf/cai/2015}
Maletti, A. (ed.): Algebraic Informatics - 6th International Conference, {CAI}
  2015, Stuttgart, Germany, September 1-4, 2015. Proceedings, Lecture Notes in
  Computer Science, vol.~9270. Springer (2015). \doi{10.1007/978-3-319-23021-4}

\bibitem{MP19}
Meijer, J., van~de Pol, J.: Sound black-box checking in the learnlib. Innov.
  Syst. Softw. Eng.  \textbf{15}(3-4),  267--287 (2019).
  \doi{10.1007/s11334-019-00342-6}

\bibitem{DBLP:conf/fase/NeeleS23}
Neele, T., Sammartino, M.: Compositional automata learning of synchronous
  systems. In: Lambers, L., Uchitel, S. (eds.) Fundamental Approaches to
  Software Engineering - 26th International Conference, {FASE} 2023, Held as
  Part of the European Joint Conferences on Theory and Practice of Software,
  {ETAPS} 2023, Paris, France, April 22-27, 2023, Proceedings. Lecture Notes in
  Computer Science, vol. 13991, pp. 47--66. Springer (2023).
  \doi{10.1007/978-3-031-30826-0\_3}

\bibitem{DBLP:phd/de/Niese2003}
Niese, O.: An integrated approach to testing complex systems. Ph.D. thesis,
  Technical University of Dortmund, Germany (2003),
  \url{http://eldorado.uni-dortmund.de:8080/0x81d98002\_0x0007b62b}

\bibitem{oasis2019mqtt}
OASIS: {MQTT Version 5} (March 07 2019),
  \url{https://docs.oasis-open.org/mqtt/mqtt/v5.0/mqtt-v5.0.pdf}

\bibitem{DBLP:conf/forte/PeledVY99}
Peled, D.A., Vardi, M.Y., Yannakakis, M.: Black box checking. In: Wu, J.,
  Chanson, S.T., Gao, Q. (eds.) Formal Methods for Protocol Engineering and
  Distributed Systems, {FORTE} {XII} / {PSTV} XIX'99, {IFIP} {TC6} {WG6.1}
  Joint International Conference on Formal Description Techniques for
  Distributed Systems and Communication Protocols {(FORTE} {XII)} and Protocol
  Specification, Testing and Verification {(PSTV} XIX), October 5-8, 1999,
  Beijing, China. {IFIP} Conference Proceedings, vol.~156, pp. 225--240. Kluwer
  (1999)

\bibitem{DBLP:conf/focs/PnueliR90}
Pnueli, A., Rosner, R.: Distributed reactive systems are hard to synthesize.
  In: 31st Annual Symposium on Foundations of Computer Science, St. Louis,
  Missouri, USA, October 22-24, 1990, Volume {II}. pp. 746--757. {IEEE}
  Computer Society (1990). \doi{10.1109/FSCS.1990.89597}

\bibitem{DBLP:journals/iandc/RivestS93}
Rivest, R.L., Schapire, R.E.: Inference of finite automata using homing
  sequences. Inf. Comput.  \textbf{103}(2),  299--347 (1993).
  \doi{10.1006/INCO.1993.1021}

\bibitem{DBLP:journals/tecs/ShijuboWS23}
Shijubo, J., Waga, M., Suenaga, K.: Probabilistic black-box checking via active
  {MDP} learning. {ACM} Trans. Embed. Comput. Syst.  \textbf{22}(5s),
  148:1--148:26 (2023). \doi{10.1145/3609127}

\bibitem{DBLP:conf/icst/TapplerAB17}
Tappler, M., Aichernig, B.K., Bloem, R.: Model-based testing iot communication
  via active automata learning. In: 2017 {IEEE} International Conference on
  Software Testing, Verification and Validation, {ICST} 2017, Tokyo, Japan,
  March 13-17, 2017. pp. 276--287. {IEEE} Computer Society (2017).
  \doi{10.1109/ICST.2017.32}

\bibitem{DBLP:conf/tacas/VaandragerGRW22}
Vaandrager, F.W., Garhewal, B., Rot, J., Wi{\ss}mann, T.: A new approach for
  active automata learning based on apartness. In: Fisman, D., Rosu, G. (eds.)
  Tools and Algorithms for the Construction and Analysis of Systems - 28th
  International Conference, {TACAS} 2022, Held as Part of the European Joint
  Conferences on Theory and Practice of Software, {ETAPS} 2022, Munich,
  Germany, April 2-7, 2022, Proceedings, Part {I}. Lecture Notes in Computer
  Science, vol. 13243, pp. 223--243. Springer (2022).
  \doi{10.1007/978-3-030-99524-9\_12}

\bibitem{DBLP:journals/tase/ZhangFL20}
Zhang, H., Feng, L., Li, Z.: Control of black-box embedded systems by
  integrating automaton learning and supervisory control theory of
  discrete-event systems. {IEEE} Trans Autom. Sci. Eng.  \textbf{17}(1),
  361--374 (2020). \doi{10.1109/TASE.2019.2929563}

\end{thebibliography}

\appendix
\crefalias{section}{appendix}
\crefalias{subsection}{appendix}

\begin{ArxivBlock}

\section{Nondeterministic Moore Machines and MMNs}\label{appendix:nondetMM}
\subsection{Nondeterministic MMs}\label{appendix:mmND}
\begin{mydefinition}[Moore machine]\label{def:moore-machineND}
  A (nondeterministic) \emph{Moore machine (MM)} is a tuple $M = (Q, Q_0, I, O, \Delta, \lambda)$, where
  \begin{itemize}
    \item $Q$ is a finite set of states, $Q_0 \subseteq Q$ is a set of initial states,
    \item $I$ is an input alphabet, $O$ is an output alphabet,
    \item $\Delta\colon Q \times I \to 2^Q$ is a transition function, and
    \item $\lambda\colon Q \to (2^O \setminus \{ \emptyset \})$ is an output function that assigns a nonempty set of output symbols to each state.
  \end{itemize}

A Moore machine is \emph{deterministic} if $|\Delta(q, i)| \le 1$ for all $q \in Q$ and $i \in I$, $|Q_0| = 1$, and $|\lambda(q)| = 1$ for all $q \in Q$.
For a deterministic Moore machine $M$, the transition function in $M$ is denoted as a partial function by $\delta\colon Q \times I \rightharpoonup Q$ and the initial state is $q_0 \in Q$.
A deterministic Moore machine is \emph{complete} if $\delta(q, i)\!\!\downarrow$ for all $q \in Q$ and $i \in I$; otherwise, it is called \emph{partial}.
\end{mydefinition}

The following definitions  are standard, but embracing nondeterminism has incurred some notational complications. The reader can first skim through them and come back later for checking details.

As usual, the transition function $\Delta$ can be extended to a set $P\subseteq Q$ of states and an input string $w\in I^{\ast}$. Precisely,
$\Delta(P, \varepsilon) = P$ and $\Delta(P, wi) = \bigcup_{q \in \Delta(P, w)} \Delta(q, i)$.
Similarly, the output function is extended by $\lambda(P, w) = \bigcup_{q \in \Delta(P, w)} \lambda(q)$.
When starting from the set  $Q_0$ of initial states, often we simply write $\Delta(w) = \Delta(Q_0, w)$ and $\lambda(w) = \lambda(Q_0, w)$.

Given a Moore machine $M = (Q, Q_0, I, O, \Delta, \lambda)$ and a set  $P \subseteq Q$ of states, the \emph{semantics} of $M$,
denoted by  $\semMoore{M}_P\colon I^\ast \to {(2^O)}^\ast$ and defined below, represents  the behavior of the machine when starting from $P$:
\begin{equation}\label{eq:semMooreND}
  \semMoore{M}_P(w) \;=\; \lambda(P, \seqSlice{w}{0}{0})\;\lambda(P, \seqSlice{w}{0}{1}) \cdots \lambda(P, \seqSlice{w}{0}{|w|})\quad\text{for each $w \in I^\ast$.}
\end{equation}
This is the  nondeterministic adaptation of the usual definition.
Note that $|\semMoore{M}_P(w)| = |w| + 1$,
as $\seqIndex{\semMoore{M}_P(w)}{0}$ is the output for $P$ without consuming any input characters.
If
 $\seqIndex{\semMoore{M}_P(w)}{k+1} = \emptyset$ for some $0 \le k < |w|$, then $\seqIndex{\semMoore{M}_P(w)}{j} = \emptyset$ for all $k < j \le |w|$; this intuitively means, according to \cref{def:moore-machine}, that 1) a Moore machine gets stuck only because of $\Delta$ ($\lambda$ is nonempty), and 2) once stuck, it remains stuck.
When $k$ is the largest number with $\seqIndex{\semMoore{M}_P(w)}{k} \ne \emptyset$,
we say that $\semMoore{M}_P$ is \emph{defined} on $w$ \emph{up to} $k$.
When starting from the set $Q_0$  of initial states, we write $\semMoore{M}(w)$ for $\semMoore{M}_{Q_0}(w)$ (much like for $\Delta$ and $\lambda$).

Using this semantics, we define the equivalence of Moore machines.

\begin{mydefinition}[equivalence of Moore machines]\label{def:moore-machine-equivalenceND}
 Two Moore machines $M_1$ and $M_2$ are said to be \emph{equivalent} if and only if $\semMoore{M_1} = \semMoore{M_2}$.
\end{mydefinition}

For a deterministic Moore machine $M$ and a state $q \in Q$, we adopt the convention that its semantics is a \emph{total} function $\semMoore{M}_q\colon I^\ast \to O^\ast$. This can be done by 1) specializing~\cref{eq:semMooreND} to  deterministic $M$, and 2) saying that, if $\lambda(P, \seqSlice{w}{0}{k})=\emptyset$, it is interpreted as $\varepsilon$ rather than the empty set. (That is, while $\lambda(P, \seqSlice{w}{0}{k})=\emptyset$ does not make the right-hand side longer, it does not make it undefined.)



The following construction, used in~\cref{sec:ccwlstar}, is our central reason for accommodating nondeterminism. In the usual theory of \emph{deterministic} automata, a quotient is taken only wrt.\ a suffix-closed equivalence, and this ensures that the \emph{deterministic} quotient is well-defined. In contrast, we would like flexible quotient automata (wrt.\ any equivalence) in~\cref{sec:ccwlstar}; we can make them well-defined thanks to nondeterminism.
\begin{mydefinition}[quotient Moore machine $M/{\sim}$]\label{def:quotientMM}
 Let $\sim$ be an   equivalence relation on $Q$. The \emph{quotient Moore machine}  $M/{\sim}$ of $M$ with respect to $\sim$  is a Moore machine $(Q/{\sim}, Q_0/{\sim}, I, O, \Delta', \lambda')$ with
 \begin{math}
  \Delta'([q]_{\sim}, i) = \bigl(\textstyle\bigcup_{q' \in [q]_{\sim}} \Delta(q', i)\bigr)/{\sim}
 \end{math}
 and
 \begin{math}
  \lambda'([q]_{\sim}) = \textstyle\bigcup_{q' \in [q]_{\sim}} \lambda(q').
 \end{math}
\end{mydefinition}

For a Moore machine $M$ and states $q, q' \in Q$, a state $q$ is said to be \emph{reachable} from $q'$ in $M$ if there exists a string $w \in I^\ast$ such that $q \in \Delta(q', w)$.
When $q'$ is one of the initial state $Q_0$, we simply say that $q$ is reachable in $M$.


\subsection{Nondeterministic MMNs}\label{appendix:mmnND}

\begin{mydefinition}[Moore machine network]\label{def:mmnND}
  A (nondeterministic) \emph{Moore machine network (MMN)} is a tuple $\mathcal{M} = (G, {(\Sigma_e)}_{e \in E}, {(M_c)}_{c \in \componentNodes})$, where
  \begin{itemize}
    \item $G = (V, E)$ is a directed graph representing the network structure,
    \item $\Sigma_e$ is an alphabet associated with each edge $e \in E$, and
    \item $M_c$ is a Moore machine associated with each component $c \in \componentNodes$.
  \end{itemize}
 On each component Moore machine $M_c = (Q_c, Q_{0,c}, \componentInputAlphabet{c}, \componentOutputAlphabet{c}, \Delta_c, \componentOutputFunction{c})$, we require that its input and output alphabets are in  accordance with the edge alphabets $\Sigma_{e}$. Specifically, we require $\componentInputAlphabet{c} = \prod_{e \in \incomingEdges{c}} \Sigma_e$ (the product of the alphabets of all incoming edges) and, similarly,  $\componentOutputAlphabet{c} = \prod_{e \in \outgoingEdges{c}} \Sigma_e$.
\end{mydefinition}

The definition of the semantics of MMNs is adapted as follows, to the current nondeterministic setting. This adaptation is easy. 

The set of \emph{(system-level) configurations} of $\mathcal{M}$, denoted by $\mathbf{Q}$, is defined by $\mathbf{Q} = \prod_{c \in \componentNodes} Q_c$.
The set of \emph{initial configurations} 
is  $\mathbf{Q}_{0} = \prod_{c \in \componentNodes} Q_{0,c}$.

Given a configuration $\mathbf{q} = (q_c)_{c \in \componentNodes} \in \mathbf{Q}$, the \emph{total output} of $\mathcal{M}$ at $\mathbf{q}$, denoted by $\totalOutputFunction(\mathbf{q}) \subseteq \totalOutputAlphabet$, is defined by $\totalOutputFunction(\mathbf{q}) = \prod_{c \in \componentNodes} \componentOutputFunction{c}(q_c)$. Similarly, the \emph{system-level output} of $\mathcal{M}$ at $\mathbf{q}$ is
 defined by $\outputFunction(\mathbf{q}) = \restrictTo{\totalOutputFunction(\mathbf{q})}{\outputEdges}\subseteq \outputAlphabet$. (Recall the restriction operation $\mid$ from \cref{sec:definitions}.) 

Given a configuration $\mathbf{q} = (q_c)_{c \in \componentNodes} \in \mathbf{Q}$ and a system-level input character $\mathbf{i} \in \inputAlphabet$, we define $\Delta$, the \emph{system-level transition function} of $\mathcal{M}$, by $\Delta(\mathbf{q}, \mathbf{i}) = \prod_{c \in \componentNodes} \bigcup_{\overline{\mathbf{o}} \in \totalOutputFunction(\mathbf{q})} \Delta_c(q_c, \restrictTo{(\mathbf{i}, \overline{\mathbf{o}})}{\incomingEdges{c}})$. Intuitively: $\overline{\mathbf{o}}$ is a tuple of characters that can be output from the current states $(q_c)_{c \in \componentNodes}$; it is combined with the system-level input $\mathbf{i}$ and fed to each component's transition function $\Delta_{c}$; we take the union $\bigcup$ over all possible output $\overline{o}$; and this happens at every component ($\prod_{c}$).

We formalize the following definition, using the above constructions.


\begin{mydefinition}[Moore machine $\mmnMoore{\mathcal{M}}$]
Let $\mathcal{M}$ be an MMN. The \emph{Moore machine $\mmnMoore{\mathcal{M}}$ induced by} $\mathcal{M}$ is $\mmnMoore{\mathcal{M}} = (\mathbf{Q}, \mathbf{Q}_{0}, \inputAlphabet, \outputAlphabet, \Delta, \outputFunction)$.
 \end{mydefinition}

The semantics of $\mathcal{M}$ is defined by
$\semMoore{\mathcal{M}}_{\mathbf{P}} = \semMoore{\mmnMoore{\mathcal{M}}}_{\mathbf{P}}$ for any $\mathbf{P} \subseteq \mathbf{Q}$.

An MMN $\mathcal{M}$ is \emph{deterministic} if every component Moore machine $M_c$ 
 is deterministic. In this case, obviously,
the  Moore machine $\mmnMoore{\mathcal{M}}$ is also deterministic.

Given an MMN $\mathcal{M} = (G, (\Sigma_e)_{e \in E}, (M_c)_{c \in \componentNodes})$ and an indexed family of equivalence relations $(\sim_c)_{c \in \componentNodes}$ such that ${\sim_c}\subseteq Q_c \times Q_c$, the \emph{quotient MMN} $\mathcal{M}/{\sim}$ of $\mathcal{M}$ with respect to $(\sim_c)_{c \in \componentNodes}$ is the MMN $(G, {(\Sigma_e)}_{e \in E}, {(M_c/{\sim_c})}_{c \in \componentNodes})$. Here $M_c/{\sim_c}$ is defined in \cref{def:quotientMM}. 

\section{Details of the benchmarks}\label{section:detail_benchmarks}

\subsection{The Benchmark \lighting{}}\label{appendix:lighting}

\begin{figure}[tbp]
 \centering
 \begin{tikzpicture}[shorten >=1pt,node distance=2.5cm,on grid,auto,>={Stealth[round,sep]}] 
  \small
  \node (brightness_in) at (-7.5, 0.75) [align=center] {};
  \node[rectangle,draw] (brightness) at (-5, 0.75) [align=center] {Brightness Sensor $\brightnessMoore$};
  \node (motion_in) at (-7.5, -0.75) [align=center] {};
  \node[rectangle,draw] (motion) at (-5, -0.75) [align=center] {Motion Sensor $\motionMoore$};
  \node[rectangle,draw] (broker) at (0, 0) [align=center] {Broker $\brokerMoore$};
  \node[rectangle,draw, node distance=3cm, right=of broker]  (light) [align=center] {Light $\lightMoore$};
  \node (light_out) at (4.5, 0) [align=center] {};

  \path[->] 
  (brightness_in) edge (brightness)
  (brightness.east) edge[bend left=10] (broker.north west)
  (broker.north west) edge[bend left=10] (brightness.east)
  (motion_in) edge (motion)
  (motion.east) edge[bend left=10] (broker.south west)
  (broker.south west) edge[bend left=10] (motion.east)
  (broker) edge (light)
  (light) edge (light_out)
  ;
 \end{tikzpicture}
 \caption{outline of the MMN $\MMNlighting$ in \lighting{}.}
 \label{figure:lighting:MMN}
\end{figure}
\lighting{} is our original benchmark modeling a reactive lighting system.
\cref{figure:lighting:MMN} outlines the MMN $\MMNlighting$ in \lighting{}.
The MMN $\MMNlighting$ consists of the following four components in addition to the input and output nodes:
the brightness sensor component $\brightnessComponent$,
the motion sensor component $\motionComponent$,
the broker component $\brokerComponent$, and
the light component $\lightComponent$.
The sensor components ($\brightnessComponent$ and $\motionComponent$) observe the current status of the environment and publish it to the broker component.
Specifically, the brightness sensor component observes if the environment is bright or not, and the motion sensor component observes if any motion is detected.
The broker component receives messages from the sensor components and forwards them to the light component.
The light component receives messages from the broker component and controls the light based on the current environment's status.
We use a communication protocol inspired by MQTT~\cite{oasis2019mqtt} for inter-component communication.
Our encoding is inspired by the Mealy machines that model the MQTT protocol on Automata Wiki\footnote{Automata Wiki: \url{https://automata.cs.ru.nl/BenchmarkMQTT-TapplerEtAl2017/Description}}, which is originally from~\cite{DBLP:conf/icst/TapplerAB17}.

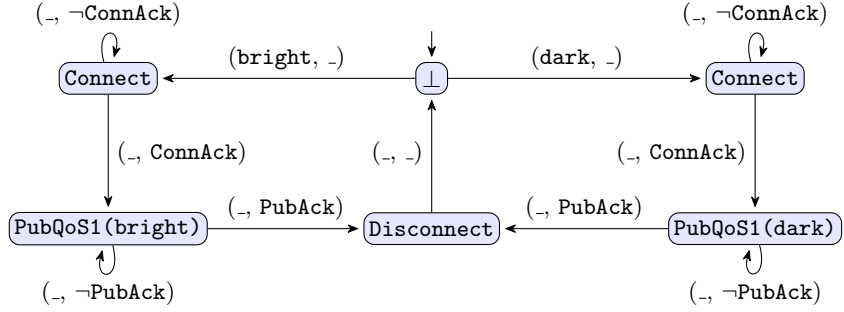
\begin{figure}[tbp]
 \centering
 \begin{tikzpicture}[shorten >=1pt,node distance=2.0cm,on grid,auto,>={Stealth[round,sep]}]
  \small
  \node[location, initial above] (initial) [align=center] {$\bot$};
  \node[location, node distance=4.3cm, left=of initial] (bright_connect) [align=center] {\texttt{Connect}};
  \node[location, below=of bright_connect] (bright_publish) [align=center] {\texttt{PubQoS1(bright)}};
  \node[location, node distance=4.3cm, right=of initial] (dark_connect) [align=center] {\texttt{Connect}};
  \node[location, below=of dark_connect] (dark_publish) [align=center] {\texttt{PubQoS1(dark)}};
  \node[location, below=of initial] (disconnect) [align=center] {\texttt{Disconnect}};

  \path[->] 
  (initial) edge node[above] {(\texttt{bright}, $\_$)} (bright_connect)
  (bright_connect) edge node[right] {($\_$, \texttt{ConnAck})} (bright_publish)
  (bright_connect) edge[loop above] node[above] {($\_$, $\neg \mathtt{ConnAck}$)} (bright_connect)
  (bright_publish) edge node[above] {($\_$, \texttt{PubAck})} (disconnect)
  (bright_publish) edge[loop below] node[below] {($\_$, $\neg\mathtt{PubAck}$)} (bright_publish)
  (initial) edge node[above] {(\texttt{dark}, $\_$)} (dark_connect)
  (dark_connect) edge node[left] {($\_$, \texttt{ConnAck})} (dark_publish)
  (dark_connect) edge[loop above] node[above] {($\_$, $\neg \mathtt{ConnAck}$)} (dark_connect)
  (dark_publish) edge node[above] {($\_$, \texttt{PubAck})} (disconnect)
  (dark_publish) edge[loop below] node[below] {($\_$, $\neg\mathtt{PubAck}$)} (dark_publish)
  (disconnect) edge node[left] {($\_$, $\_$)} (initial)
  ;
 \end{tikzpicture}
 \caption{the Moore machine $\brightnessMoore$ for the brightness sensor component in \lighting{}. Each transition is labeled with a pair $(\tilde{i}_{\mathrm{in}, \brightnessComponent}, \tilde{i}_{\brokerComponent, \brightnessComponent})$ of symbols representing the subset of $\Sigma_{\mathsf{in}, \brightnessComponent} \times \Sigma_{\brokerComponent, \brightnessComponent}$: the symbol $\_$ represents any character, and for any character $i$, $\neg i$ represents the character other than $i$. For instance, $(\_\,, \neg \mathtt{ConnAck})$ represents $\{(i_{\mathrm{in}, \brightnessComponent}, i_{\brokerComponent, \brightnessComponent}) \in \Sigma_{\mathsf{in}, \brightnessComponent} \times \Sigma_{\brokerComponent, \brightnessComponent} \mid i_{\brokerComponent, \brightnessComponent} \neq \mathtt{ConnAck}\}$}
 \label{figure:lighting:brightness_moore}
\end{figure}

\paragraph{Brightness sensor component}
\cref{figure:lighting:brightness_moore} illustrates the Moore machine $\brightnessMoore$ for the brightness sensor component.
The brightness sensor observes whether the environment is bright or dark (\ie{} $\Sigma_{\mathsf{in}, \brightnessComponent} = \{\mathtt{bright}, \mathtt{dark}\}$) and publishes it to the broker.
The messages are sent via a protocol based on MQTT QoS 1:
\begin{renumeration}
 \item The publisher (\ie{} the brightness sensor component) tries to establish a connection by sending \texttt{Connect} to the broker;
 \item The broker returns \texttt{ConnAck} when the connection is established;
 \item Then, the publisher publishes the current status by sending either \texttt{PubQoS1(bright)} or \texttt{PubQoS1(dark)} to the broker;
 \item The broker returns \texttt{PubAck} to make sure that the publication was successful;
 \item Finally, the publisher sends \texttt{Disconnect} to close the connection.
\end{renumeration}

In the above illustration,
the brightness sensor component sends the following messages: \texttt{Connect}, \texttt{PubQoS1(bright)}, \texttt{PubQoS1(dark)}, and \texttt{Disconnect}.
We assume that the broker does not know the QoS used by the publisher in advance and we include the messages for other QoS in the alphabet.
Moreover, we allow the publisher to not send any message.
In total, the alphabet $\Sigma_{\brightnessComponent, \brokerComponent}$ from $\brightnessComponent$ to $\brokerComponent$ is as follows:
$\Sigma_{\brightnessComponent, \brokerComponent} = \{\mathtt{Connect}, \mathtt{PubQoS0(bright)}, \mathtt{PubQoS0(dark)}, \mathtt{PubQoS1(bright)},\\ \mathtt{PubQoS1(dark)}, \mathtt{PubQoS2(bright)}, \mathtt{PubQoS2(dark)}, \mathtt{PubRel}, \mathtt{Disconnect}, \bot\}$.

In the above illustration,
the broker component sends \texttt{ConnAck} or \texttt{PubAck} to the brightness sensor component.
Including the messages not used in QoS 1 and the special character $\bot$ that shows that no message is sent,
the alphabet $\Sigma_{\brokerComponent, \brightnessComponent}$ from $\brokerComponent$ to $\brightnessComponent$ is as follows:
$\Sigma_{\brokerComponent, \brightnessComponent} = \{\mathtt{ConnAck}, \mathtt{PubAck}, \mathtt{PubRec}, \mathtt{PubComp}, \bot\}$.

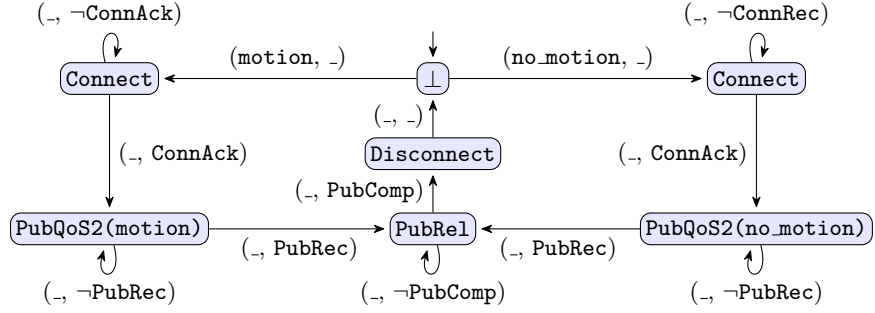
\begin{figure}[tbp]
 \centering
 \begin{tikzpicture}[shorten >=1pt,node distance=2.0cm,on grid,auto,>={Stealth[round,sep]}]
  \small
  \node[location, initial above] (initial) [align=center] {$\bot$};
  \node[location, node distance=4.3cm, left=of initial] (motion_connect) [align=center] {\texttt{Connect}};
  \node[location, below=of motion_connect] (motion_publish) [align=center] {\texttt{PubQoS2(motion)}};
  \node[location, node distance=4.3cm, right=of initial] (no_motion_connect) [align=center] {\texttt{Connect}};
  \node[location, below=of no_motion_connect] (no_motion_publish) [align=center] {\texttt{PubQoS2(no\_motion)}};
  \node[location, node distance=1.0cm, below=of initial] (disconnect) [align=center] {\texttt{Disconnect}};
  \node[location, node distance=1.0cm, below=of disconnect] (pubrel) [align=center] {\texttt{PubRel}};

  \path[->] 
  (initial) edge node[above] {(\texttt{motion}, $\_$)} (motion_connect)
  (motion_connect) edge node[right] {($\_$, \texttt{ConnAck})} (motion_publish)
  (motion_connect) edge[loop above] node[above] {($\_$, $\neg \mathtt{ConnAck}$)} (motion_connect)
  (motion_publish) edge node[below] {($\_$, \texttt{PubRec})} (pubrel)
  (motion_publish) edge[loop below] node[below] {($\_$, $\neg\mathtt{PubRec}$)} (motion_publish)
  (initial) edge node[above] {(\texttt{no\_motion}, $\_$)} (no_motion_connect)
  (no_motion_connect) edge node[left] {($\_$, \texttt{ConnAck})} (no_motion_publish)
  (no_motion_connect) edge[loop above] node[above] {($\_$, $\neg \mathtt{ConnRec}$)} (no_motion_connect)
  (no_motion_publish) edge node[below] {($\_$, \texttt{PubRec})} (pubrel)
  (pubrel) edge node[left] {($\_$, \texttt{PubComp})} (disconnect)
  (no_motion_publish) edge[loop below] node[below] {($\_$, $\neg\mathtt{PubRec}$)} (no_motion_publish)
  (pubrel) edge[loop below] node[below] {($\_$, $\neg\mathtt{PubComp}$)} (pubrel)
  (disconnect) edge node[left] {($\_$, $\_$)} (initial)
  ;
 \end{tikzpicture}
 \caption{the Moore machine $\motionMoore$ for the motion sensor component in \lighting{}. We use the same notation as in \cref{figure:lighting:motion_moore}}
 \label{figure:lighting:motion_moore}
\end{figure}

\paragraph{Motion sensor component}
\cref{figure:lighting:motion_moore} illustrates the Moore machine $\motionMoore$ for the motion sensor component.
The motion sensor observes whether or not a motion is detected (\ie{} $\Sigma_{\mathsf{in}, \motionComponent} = \{\mathtt{motion}, \mathtt{no\_motion}\}$) and publishes it to the broker.
The messages are sent via a protocol based on MQTT QoS 2:
\begin{renumeration}
 \item The publisher (\ie{} the motion sensor component) tries to establish a connection by sending \texttt{Connect} to the broker;
 \item The broker returns \texttt{ConnAck} when the connection is established;
 \item Then, the publisher publishes the current status by sending either \texttt{PubQoS2(bright)} or \texttt{PubQoS2(dark)} to the broker;
 \item The broker returns \texttt{PubRec} to make sure that the publication was successful;
 \item Then, the publisher sends \texttt{PubRel} to show that \texttt{PubRec} was successfully received;
 \item The broker returns \texttt{PubComp} to complete the publication;
 \item Finally, the publisher sends \texttt{Disconnect} to close the connection.
\end{renumeration}

In the above illustration,
the motion sensor component sends the following messages: \texttt{Connect}, \texttt{PubQoS2(motion)}, \texttt{PubQoS2(no\_motion)}, \texttt{PubRel}, and \texttt{Disconnect}, and
the broker component sends \texttt{ConnAck}, \texttt{PubRec}, and \texttt{PubComp}.
Similarly to $\Sigma_{\brightnessComponent,\brokerComponent}$ and $\Sigma_{\brokerComponent,\brightnessComponent}$,
$\Sigma_{\motionComponent,\brokerComponent}$ and $\Sigma_{\motionComponent,\brightnessComponent}$ are as follows:

\begin{align*}
 \Sigma_{\motionComponent,\brokerComponent} &= \{\mathtt{Connect}, \mathtt{PubQoS0(motion)}, \mathtt{PubQoS0(no\_motion)}, \mathtt{PubQoS1(motion)},\\&
 \qquad\mathtt{PubQoS1(no\_motion)}, \mathtt{PubQoS2(motion)}, \mathtt{PubQoS2(no\_motion)}, \mathtt{PubRel},\\&
 \qquad \mathtt{Disconnect}, \bot\} \\
 \Sigma_{\brokerComponent,\motionComponent} &= \{\mathtt{ConnAck}, \mathtt{PubAck}, \mathtt{PubRec}, \mathtt{PubComp}, \bot\}
\end{align*}

\begin{figure}[tbp]
 \centering
 \begin{tikzpicture}[shorten >=1pt,node distance=3.0cm,on grid,auto,>={Stealth[round,sep]}]
  \small
  \node[location] (dark_motion) [align=center] {\texttt{ON}};
  \node[location, below=of dark_motion] (bright_motion) [align=center] {\texttt{OFF}};
  \node[location, initial above, right=of dark_motion] (dark_no_motion) [align=center] {\texttt{OFF}};
  \node[location, below=of dark_no_motion] (bright_no_motion) [align=center] {\texttt{OFF}};

  \path[->] 
  (dark_motion) edge[bend left=15] node[above] {\texttt{no\_motion}} (dark_no_motion)
  (dark_no_motion) edge[bend left=15] node[below] {\texttt{motion}} (dark_motion)
  (bright_motion) edge[bend left=15] node[above] {\texttt{no\_motion}} (bright_no_motion)
  (bright_no_motion) edge[bend left=15] node[below] {\texttt{motion}} (bright_motion)
  (dark_motion) edge[bend left=15] node[right] {\texttt{bright}} (bright_motion)
  (dark_no_motion) edge[bend left=15] node[right] {\texttt{bright}} (bright_no_motion)
  (bright_motion) edge[bend left=15] node[left] {\texttt{dark}} (dark_motion)
  (bright_no_motion) edge[bend left=15] node[left] {\texttt{dark}} (dark_no_motion)
  (dark_motion) edge[loop left] node[left] {$\bot$} (dark_motion)
  (dark_no_motion) edge[loop right] node[right] {$\bot$} (dark_no_motion)
  (bright_motion) edge[loop left] node[left] {$\bot$} (bright_motion)
  (bright_no_motion) edge[loop right] node[right] {$\bot$} (bright_no_motion)
  ;
 \end{tikzpicture}
 \caption{the Moore machine $\lightMoore$ for the light component in \lighting{}. We use the same notation as in \cref{figure:lighting:motion_moore}}
 \label{figure:lighting:light_moore}
\end{figure}
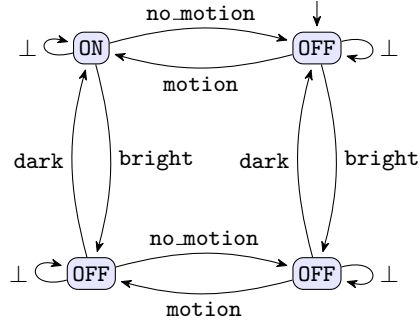

\paragraph{Light component}
\cref{figure:lighting:light_moore} illustrates the Moore machine $\lightMoore$ for the light component.
The light component receives the current environment's status (\ie{} the brightness and the detection of the motion) and controls the light.
Thus,
the input alphabet is $\Sigma_{\brokerComponent, \lightComponent} = \{\mathtt{bright}, \mathtt{dark}, \mathtt{motion}, \mathtt{no\_motion}\}$ and
the output alphabet is $\Sigma_{\lightComponent,\mathsf{out}} = \{\mathtt{ON}, \mathtt{OFF}\}$.
We assume that initially, the environment is dark, and no motion is detected.

\paragraph{Broker component}

The broker component $\brokerComponent$ manages the aforementioned communication, mainly the data transmission from the sensor components to the light component.
We assume that once a connection is established, the broker component only reads the messages from the connected component until the connection is explicitly closed by the publisher.
For simplicity, we assume that we statically have two publishers (the sensor components $\brightnessComponent$ and $\motionComponent$) and one subscriber (the light component $\lightComponent$).
For the fairness of the broker, if both publishers try to establish the connection at the same time,
the broker establishes the connection with the publisher that was not the most recently communicated with.

\subsection{The Benchmark $\mathtt{BinaryCounter}(k)$}\label{appendix:binCount}

The MMN for our realistic benchmark $\mathtt{BinaryCounter}(k)$ is shown in \cref{fig:binCount}.
Note that the MMN shown in \cref{fig:binCount} is for $\mathtt{BinaryCounter}(2)$, but the generalization to $\mathtt{BinaryCounter}(k)$ is straightforward.
The MMN counts the number of $1$s in an input string and outputs a binary representation of it.
Due to the delay in the output of Moore machines, $k$-times $0$s between each $1$ are required for correctly counting.
It consists of $k$ Moore machines, with each component in charge of each of the $k$-bit output, and inter-component edges address carry out.
The Moore machine for each component consists of 3 states: two states $(0,0)$ and $(0,1)$ to hold the current output value and $(1,0)$ to pass the carry-in bit to the next component.
Please note that the last component has no output to the next component.

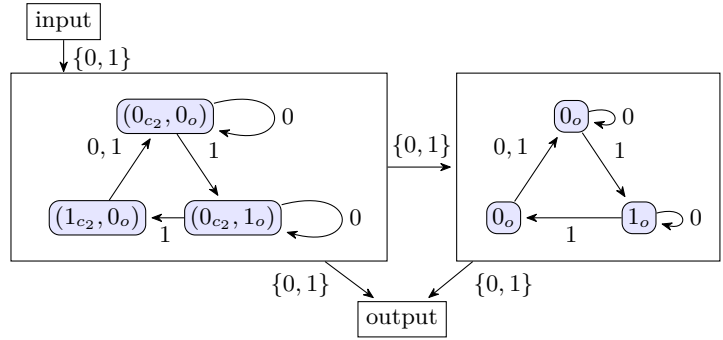
\begin{figure}[tbp]
  \centering
  \begin{tikzpicture}[scale=0.9,shorten >=1pt,node distance=2.5cm,on grid,auto,>={Stealth[round,sep]}]
    \node [rectangle,draw] (in) at (-1.5,2.4) {input};

    \coordinate (inc1_1) at ($(in.south west)!0.5!(in.south east)$);
    \coordinate (inc1_2) at ($(inc1_1)+(0,-0.55)$);

    \path[->]
      (inc1_1) edge node {$\{0,1\}$} (inc1_2);

    \begin{scope}
      \node (c1box) at (0.5,0.25) [draw,minimum width=5cm,minimum height=2.5cm] {};

      \node (c1_00) at (0,1) [location] {$(0_{c_2},0_o)$};
      \node (c1_01) at (1,-0.5) [location] {$(0_{c_2},1_o)$};
      \node (c1_10) at (-1,-0.5) [location] {$(1_{c_2},0_o)$};

      \path[->]
        (c1_00) edge node {$1$} (c1_01)
        (c1_01) edge node {$1$} (c1_10)
        (c1_10) edge node {$0,1$} (c1_00);
      \path[->]
        (c1_00) edge [loop right] node {$0$} (c1_00)
        (c1_01) edge [loop right] node {$0$} (c1_01);
    \end{scope}

    \begin{scope}[xshift=6cm]
      \node (c2box) at (0.25,0.25) [draw,minimum width=3.5cm,minimum height=2.5cm] {};

      \node (c2_00) at (0,1) [location] {$0_o$};
      \node (c2_01) at (1,-0.5) [location] {$1_o$};
      \node (c2_10) at (-1,-0.5) [location] {$0_o$};

      \path[->]
        (c2_00) edge node {$1$} (c2_01)
        (c2_01) edge node {$1$} (c2_10)
        (c2_10) edge node {$0,1$} (c2_00);
      \path[->]
        (c2_00) edge [loop right] node {$0$} (c2_00)
        (c2_01) edge [loop right] node {$0$} (c2_01);
    \end{scope}

    \coordinate (c1c2_1) at ($(c1box.north east)!0.5!(c1box.south east)$);
    \coordinate (c1c2_2) at ($(c2box.north west)!0.5!(c2box.south west)$);

    \path[->]
      (c1c2_1) edge node {$\{0,1\}$} (c1c2_2);

    \node [rectangle,draw] (out) at (3.5,-2) {output};
    \path[->]
      (c1box) edge (out)
      (c2box) edge (out);
    \node (out1_label) at (2,-1.5) {$\{ 0, 1 \}$};
    \node (out2_label) at (5,-1.5) {$\{ 0, 1 \}$};
  \end{tikzpicture}
  \caption{the MMN of $\mathtt{BinaryCounter}(2)$}
  \label{fig:binCount}
\end{figure}

\section{Omitted Proofs and Remarks}

\subsection{Moore Machines over Mealy Machines}\label{appendix:whyMoore}

\begin{figure}[tbp]
\centering
\includegraphics[width=.7\textwidth]{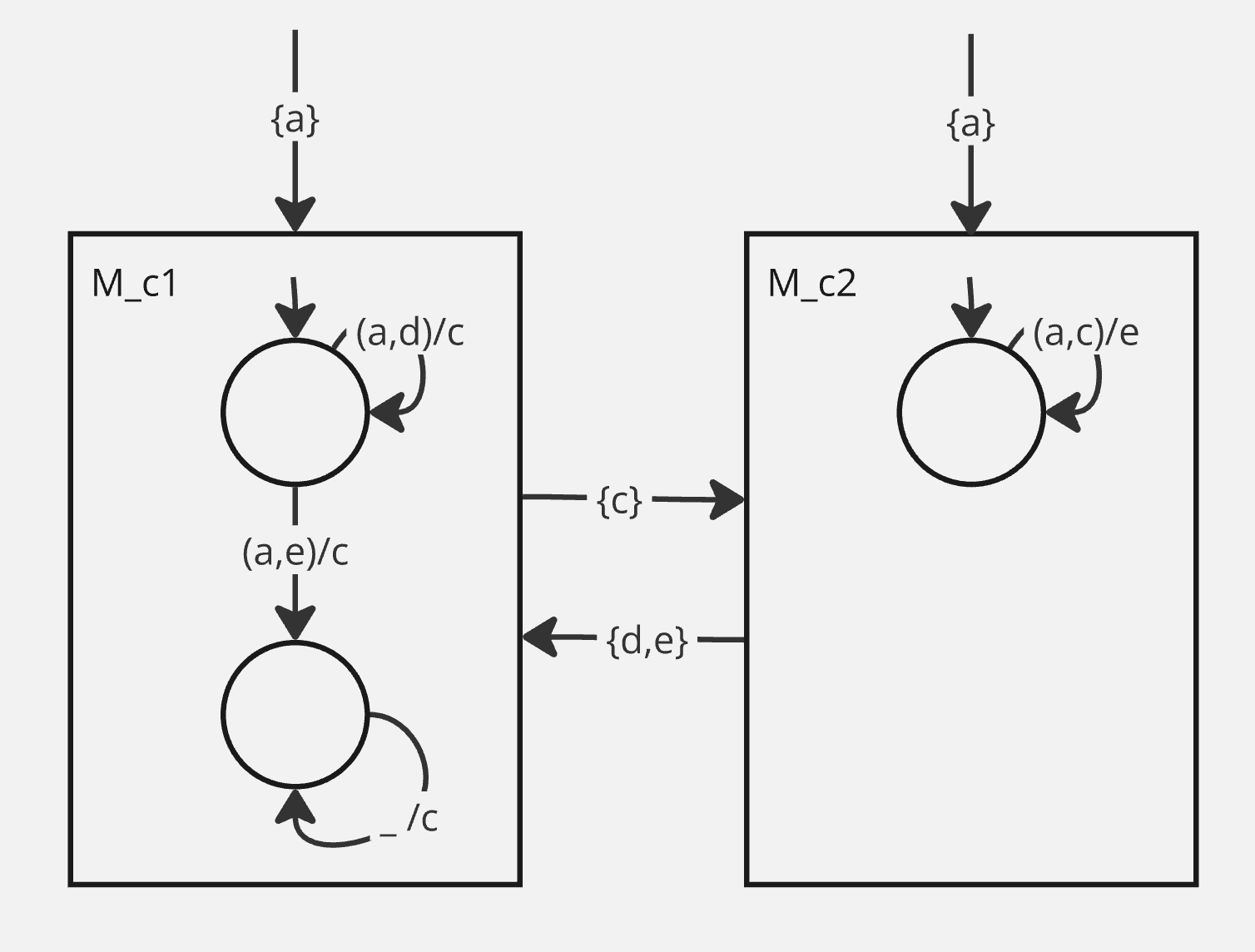}
\caption{``Mealy machine network'' is problematic}
\label{fig:mealy}
\end{figure}

\begin{myremark}\label{rem:whyMoore}
Our choice of Moore machines as models of components is crucial for the above operational semantics of MMNs. In particular, we find that Mealy machines are not suited for our purpose. 

An example of a ``Mealy machine network'' is in \cref{fig:mealy}. Assume that at some tick,  $M_{c_1}$ is at its top state and receives $d$ from $M_{c_2}$.
 Assuming that it also receives the system-level input $a$, the loop transition $(a,d)/c$ is fired and an output $c$ is sent to $M_{c_2}$. This fires the transition $(a,c)/e$ in $M_{c_2}$, leading to a character $e$ sent to $M_{c_{1}}$, which then fires the transition $(a,e)/c$ to the bottom state of $M_{c_{1}}$.

The problem is that all these should happen within a single tick---so $M_{c_{1}}$ should make two transitions (the loop and the downward) at the same time. This is strange.

The key difference between Moore and Mealy machines is that, in the former, the effect of  input is reflected on  output with a one-tick delay. Indeed, with Moore machines, the above ``chain of immediate consequences'' never arises. 

\end{myremark}

\subsection{Details of \cref{exa:mmn-network-alphabet}}\label{appendix:mmn-network-alphabet-detailed}

\begin{myexample}
\label{exa:mmn-network-alphabet-detailed}
  Consider the MMN $\mathcal{M}_\mathsf{ex}$ in \cref{fig:mmn-example}, the network structure $G = (V, E)$ is
  \begin{equation*}
    V = \{ i_1, i_2, c_1, c_2, o_1, o_2 \} \text{ and }
    E = \{ i_1 c_1, i_2 c_2, c_1 c_2, c_2 c_1, c_1 o_1, c_2 o_2 \}\text{.}
  \end{equation*}
  Furthermore, the set of nodes $V$ is partitioned by $\inputNodes = \{ i_1, i_2 \}$, $\outputNodes = \{ o_1, o_2 \}$, and $\componentNodes = \{ c_1, c_2 \}$.
  The alphabets of the system and components are:
  \begin{align*}
    \inputAlphabet &= \Sigma_{i_1 c_1} \times \Sigma_{i_2 c_2} = \{ (a, c), (a, d), (b, c), (b, d) \} \\
    \outputAlphabet &= \Sigma_{c_1 o_1} \times \Sigma_{c_2 o_2} = \{ (x, z), (x, w), (y, z), (y, w) \} \\
    \componentInputAlphabet{c_1} &= \Sigma_{i_1 c_1} \times \Sigma_{c_2 c_1} = \{ (a, 3), (a, 4), (b, 3), (b, 4) \} \\
    \componentOutputAlphabet{c_1} &= \Sigma_{c_1 o_1} \times \Sigma_{c_1 c_2} = \{ (x, 1), (x, 2), (y, 1), (y, 2) \} \\
    \componentInputAlphabet{c_2} &= \Sigma_{i_2 c_2} \times \Sigma_{c_1 c_2} = \{ (c, 1), (c, 2), (d, 1), (d, 2) \} \\
    \componentOutputAlphabet{c_2} &= \Sigma_{c_2 o_2} \times \Sigma_{c_2 c_1} = \{ (z, 3), (z, 4), (w, 3), (w, 4) \} \\
    \totalOutputAlphabet &= \componentOutputAlphabet{c_1} \times \componentOutputAlphabet{c_2} = \{ (x, 1, z, 3), (x, 1, z, 4), (x, 1, w, 3), \ldots, (y, 2, w, 4) \}
  \end{align*}

  \begin{figure}[tbp]
    \centering
    \begin{tikzpicture}[>={Stealth[round,sep]}]
      \begin{scope}[xshift=-0.2cm]
        \node (desc) at (-1.2,1.8) {\scriptsize\underline{$G = (V, E)$ and $(\Sigma)_{e \in E}$}};

        \node (i1) at (-1,1) {\scriptsize $i_1$};
        \node (i2) at (1,1)  {\scriptsize $i_2$};
    
        \node (c1) at (-1,0) [draw] {\scriptsize $c_1$};
        \node (c2) at (1,0) [draw] {\scriptsize $c_2$};
    
        \node (o1) at (-1,-1) {\scriptsize $o_1$};
        \node (o2) at (1,-1)  {\scriptsize $o_2$};
    
        \node (i1c1) at (-1.8,0.5) {\scriptsize $\Sigma_{i_1\!c_1}\!\! = \!\!\{ a, b \}$};
        \node (i2c2) at (1.8,0.5) {\scriptsize $\Sigma_{i_2\!c_2}\!\! = \!\!\{ c, d \}$};
    
        \coordinate (c1_1) at ($(c1.north east)!0.3!(c1.south east)$);
        \coordinate (c2_1) at ($(c2.north west)!0.3!(c2.south west)$);
        \coordinate (c1_2) at ($(c1.north east)!0.7!(c1.south east)$);
        \coordinate (c2_2) at ($(c2.north west)!0.7!(c2.south west)$);
    
        \node (c1c2) at (0,0.28) {\scriptsize $\Sigma_{c_1\!c_2}\!\! = \!\!\{ 1, 2 \}$};
        \node (c2c1) at (0,-0.28) {\scriptsize $\Sigma_{c2\!c1}\!\! = \!\!\{ 3, 4 \}$};
    
        \node (o1c1) at (-1.83,-0.5) {\scriptsize $\Sigma_{o_1\!c_1}\!\! = \!\!\{ x, y \}$};
        \node (o2c2) at (1.83,-0.5) {\scriptsize $\Sigma_{o_2\!c_2}\!\! = \!\!\{ z, w \}$};
      \end{scope}
  
      \graph {
        (i1) -> (c1);
        (i2) -> (c2);
        (c1_1) -> (c2_1);
        (c2_2) -> (c1_2);
        (c1) -> (o1);
        (c2) -> (o2);
      };

      \begin{scope}[xshift=-1.1cm]
        \node (mc1) at (4.2,1.8) {\scriptsize\underline{$M_{c_1}$}}; 

        \node (mc1q0) at (4.2,0.5) {};
        \node (mc1q1) at (5,0.5) [location] {\scriptsize $(x,1)$};
        \node (mc1q2) at (5,-0.5) [location] {\scriptsize $(y,2)$};
  
        \node (mc1q1q1) at (5,1.3) {\scriptsize $\lnot (a, 3)$};
        \node (mc1q1q2) at (4.6,0) {\scriptsize $(a,3)$};
        \node (mc1q2q2) at (5,-1.3) {\scriptsize $(\_,\_)$};
  
        \path
          (mc1q0) edge [->] (mc1q1)
          (mc1q1) edge [loop above] (mc1q1)
          (mc1q1) edge [->] (mc1q2)
          (mc1q2) edge [loop below] (mc1q2);

        \node (mc2) at (6.5,1.8) {\scriptsize\underline{$M_{c_2}$}};

        \node (mc2q0) at (7.2,1) {}; 
        \node (mc2q1) at (8,1) [location] {\scriptsize $(z,3)$};
        \node (mc2q2) at (8,0) [location] {\scriptsize $(w,3)$};
        \node (mc2q3) at (7.5,-1) [location] {\scriptsize $(z,4)$};
        \node (mc2q4) at (8.5,-1) [location] {\scriptsize $(w,4)$};
  
        \node (mc2q1q1) at (8,1.8) {\scriptsize $\lnot (c, 2)$};
        \node (mc2q1q2) at (7.6,0.5) {\scriptsize $(c,2)$};
        \node (mc2q2q3) at (7.3,-0.5) {\scriptsize $(\_,1)$};
        \node (mc2q2q4) at (8.7,-0.5) {\scriptsize $(\_,2)$};
        \node (mc2q4q4) at (9.6,-1.3) {\scriptsize $(\_,\_)$};
  
        \path
          (mc2q0) edge [->] (mc2q1)
          (mc2q1) edge [loop above] (mc2q1)
          (mc2q1) edge [->] (mc2q2)
          (mc2q2) edge [->] (mc2q3)
          (mc2q2) edge [->] (mc2q4)
          (mc2q4) edge [loop right] (mc2q4);
      \end{scope}
    \end{tikzpicture}
    \caption{an example MMN $\mathcal{M}_\mathsf{ex}$. On the left we show its network $G = (V, E)$ and the alphabets $(\Sigma_e)_{e \in E}$ for edges. The component MMs are shown on the right, where state labels designate output. 
In the   transition labels, $\lnot i$ stands for all  characters other than $i$, and the symbol $\_$ matches any character}\label{fig:mmn-exampleAppendix}
  \end{figure}
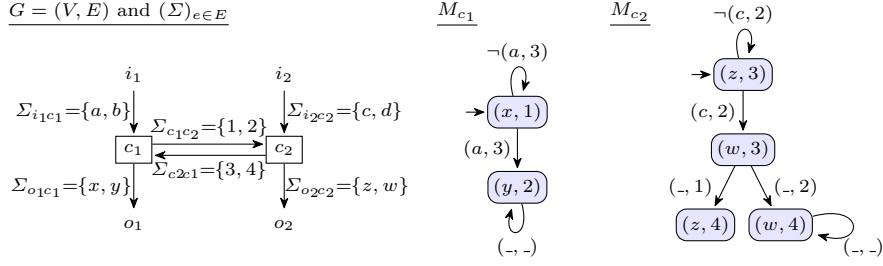
\end{myexample}

\subsection{The Function  $\textsc{1Ext}_{\mathsf{Eq},\mathsf{D}_{\infty}}$, for Illustration}\label{appendix:oneExtEqDInftyForIllustraion}

The special case $\textsc{1Ext}_{\mathsf{Eq},\mathsf{D}_{\infty}}$ of $\OneExtER$ is shown in \cref{alg:oneExtEqDInftyForIllustraion}.
\begin{algorithm}[tb]
  \caption{the function $\textsc{1Ext}_{\mathsf{Eq},\mathsf{D}_{\infty}}$, a special case of  $\textsc{1Ext}_{\mathcal{E},\mathcal{R}}$, presented for illustration
} \label{alg:oneExtEqDInftyForIllustraion}
  \scalebox{.9}{\begin{minipage}{1.1\textwidth}
  \begin{algorithmic}[1]
    \Function{$\textsc{1Ext}_{\mathsf{Eq},\mathsf{D}_{\infty}}$}{$\mathcal{H}$}
      \State $D \gets \emptyset$
      \For{all reachable $\mathbf{q} = (\mathsf{row}(\mathbf{s}_c))_{c \in \componentNodes} \in \mathbf{Q}$ in $\mathcal{H}$}
        \For{$\mathbf{i} \in \inputAlphabet$ and $c \in \componentNodes$}
          \State Let $\widehat{\mathbf{i}}$ be a possible input character to $c$ in $\mathcal{H}$ on $\mathbf{q}$ and $\mathbf{i}$, \ie $\widehat{\mathbf{i}} = \restrictTo{(\mathbf{i}, \totalOutputFunction(\mathbf{q}))}{\incomingEdges{c}}$
          \State $D \gets D \cup \{ (c, \mathbf{s}_c, \widehat{\mathbf{i}}) \}$
        \EndFor
      \EndFor
      \State\Return $D$
    \EndFunction
  \end{algorithmic}
  \end{minipage}}
\end{algorithm}

\subsection{The Extended Procedure  $\textsc{AnalyzeCex}^\text{C}$ That Accommodates Unsound $(\mathcal{E},\mathcal{R})$}\label{appendix:AnalyzeCexForUnsound}
The procedure is in \cref{alg:analyzecex-approx}.

\begin{algorithm}[tbp]
  \caption{an extension of $\textsc{AnalyzeCex}^\text{C}$ in \cref{alg:ccwlstar} for accommodating unsound $(\mathcal{E},\mathcal{R})$}\label{alg:analyzecex-approx}
  \scalebox{.9}{\begin{minipage}{1.1\textwidth}
  \begin{algorithmic}[1]
    \Procedure{$\textsc{AnalyzeCex}^\text{C}$}{$\mathcal{H}, \mathbf{w}$}
      \If{$\delta(\seqSlice{\mathbf{w}}{0}{k})\!\!\downarrow$ and $\delta(\seqSlice{\mathbf{w}}{0}{k+1})\!\!\uparrow$ for some $0 \le k < |\mathbf{w}|$}\label{line:analyzecex-approx-handle-begin}
        \State Let $\delta(\seqSlice{\mathbf{w}}{0}{k})$ be $\mathbf{q} = (\mathbf{s}_c)_{c \in \componentNodes}$ and $\mathbf{i}$ be $\seqIndex{w}{k}$.
        \State Find a component $c \in \componentNodes$ and an input character $\widehat{\mathbf{i}}$
        \Statex\quad in which a transition is missing, that is, $\delta_c(\mathbf{s}_c, \widehat{\mathbf{i}})\!\!\uparrow$ and $\widehat{\mathbf{i}} = \restrictTo{(\mathbf{i}, \totalOutputFunction(\mathbf{q}))}{\incomingEdges{c}}$
        \State $R_c \gets R_c \cup \{ (\mathbf{s}_c, \widehat{\mathbf{i}}) \}$
        \State $T(\mathbf{s}_c \cdot \widehat{\mathbf{i}} \cdot \mathbf{e}) \gets \mathsf{OQ}_c(\mathbf{s}_c \cdot \widehat{\mathbf{i}} \cdot \mathbf{e})$ for each $\mathbf{e} \in E_c$
        \State \Return
      \EndIf\label{line:analyzecex-approx-handle-end}
      \State Find a component $c \in \componentNodes$ that produces an incorrect output, \label{line:analyzecex-approx-find-component}
      \Statex\quad that is, $\restrictTo{\overline{\mathsf{OQ}}(\mathbf{w})}{\outgoingEdges{c}} \ne \restrictTo{\semMoore{\mathcal{H}}(\mathbf{w})}{\outgoingEdges{c}}$
      \State Construct an input $\widehat{\mathbf{w}}$ to the component $c$ from the system-level input $w$
      \Statex\quad where $\seqIndex{\widehat{\mathbf{w}}}{k} = \restrictTo{(\seqIndex{\mathbf{w}}{k}, \seqIndex{\overline{\mathsf{OQ}}(\mathbf{w})}{k})}{\incomingEdges{c}}$ for $0 \le k < |\mathbf{w}|$
      \State Apply \Call{$\text{L*}$-AnalyzeCex}{$H_c, \widehat{\mathbf{w}}$}
    \EndProcedure
  \end{algorithmic}
  \end{minipage}}
\end{algorithm}

\subsection{Full Experimental Results with Intermediate CA-parameters}
\label{appendix:moreExpResults}
Full experimental results, with additional \emph{intermediate} CA-parameters, are in \cref{table:exprIext,table:exprIIext}. 

They extend \cref{table:exprI,table:exprII} by additional values
 $\mathcal{R}=\mathsf{D}_{\mathsf{sum}},\mathsf{D}_{\mathsf{max}},\mathsf{D}_{\mathsf{min}}$ for  $\mathcal{R}$.
Here
$\mathcal{R}=\mathsf{D}_{\mathsf{sum}}$ means that we limit the depth of breadth-first search by the sum of the learned components' numbers of states (cf.\ \cref{sec:ccwlstar}). Similarly, $\mathcal{R}=\mathsf{D}_{\mathsf{max}}$ means that the limit is the maximum of the learned components' numbers of states; and $\mathcal{R}=\mathsf{D}_{\mathsf{min}}$ means we take the minimum. 

An overall tendency in \cref{table:exprIext,table:exprIIext} is that these intermediate values for $\mathcal{R}$ indeed achieved intermediate performance between the two extremes ($\mathsf{D}_{0}$ and $\mathsf{D}_{\infty}$). As discussed in \cref{sec:experiments} (RQ1), the best performer is likely to be one of these two extremes, but it depends on the nature of an SUL which is. Therefore, when the nature of an SUL is unknown, trying an intermediate parameter value is a viable option.

\begin{table}[tbp]
  \begin{adjustbox}{addcode={\begin{minipage}{\width}\caption{%
experiment results I (extended). 
The rows are for different algorithms: two baselines (MnL*, CwL*) and our proposal (CCwL*) with different CA-parameters (\cref{sec:ccwlstar}). In the columns, 
\emph{st.} is the number of learned states, 
\emph{tr.} is that of learned transitions,
\emph{OQ reset} is the number of resets
caused by output queries (it coincides with the number of OQs), 
\emph{OQ step} is the number of steps caused by output queries,
\emph{EQ} is the number of equivalence queries,
\emph{EQ reset} and \emph{EQ step} are the numbers of resets and steps caused by EQs, respectively (an EQ conducts testing and thus uses many input words), and
\emph{L.\ time} (``Learner time'') is the time (seconds) spent for the tasks on the learner's side (context analysis, counterexample analysis, building observation tables, etc.), and \emph{valid?} reports the numbers of instances of ``result validated,'' ``result found incorrect,'' and ``timeout.'' Note that we used 10 instances for each random benchmark.
}\label{table:exprIext}}{\end{minipage}},rotate=90,center}
\scalebox{.8}{ 
\begin{tabular}{@{}p{2.6cm}%
*{8}{R{\dimexpr.8cm-2\tabcolsep\relax}}R{\dimexpr1.2cm-2\tabcolsep\relax}
*{8}{R{\dimexpr.8cm-2\tabcolsep\relax}}R{\dimexpr1.2cm-2\tabcolsep\relax}
*{8}{R{\dimexpr.8cm-2\tabcolsep\relax}}R{\dimexpr1.2cm-2\tabcolsep\relax}
@{}}
\toprule
&
\multicolumn{9}{c}{\texttt{Rand(Compl(5),LeanComp)}}
&
\multicolumn{9}{c}{\texttt{Rand(Star(5),LeanComp)}}
&
\multicolumn{9}{c}{\texttt{Rand(Path(5),LeanComp)}}
\\\cmidrule(lr){2-10}\cmidrule(lr){11-19}\cmidrule(lr){20-28}
algo.
& st. 
& tr. 
& OQ reset
& OQ step
& EQ
& EQ reset
& EQ step
& L. time
& valid?
& st.
& tr.
& OQ reset
& OQ step
& EQs
& EQ reset
& EQ step
& L. time
& valid?
& st.
& tr.
& OQ reset
& OQ step
& EQs
& EQ reset
& EQ step
& L. time
& valid?
\\\midrule
\rowcolor{gray!20} MnL*
& 27K & 53K & 591K & 18M & 15.0 & 129 & 32K & 1.0K & 0/1/9 & 26K & 53K & 917K & 33M & 23.5 & 417 & 105K & 1.0K & 0/2/8 & 5.5K & 21K & 496K & 15M & 19.7 & 223 & 54K & 221 & 1/8/1 \\
CwL*
& 46.9 & 12K & 15K & 32K & 6.20 & 501 & 137K & 0.86 & 10/0/0 & 55.6 & 26K & 26K & 51K & 17.6 & 613 & 165K & 1.14 & 10/0/0 & 47.7 & 166 & 570 & 2.8K & 17.9 & 513 & 137K & 0.53 & 10/0/0 \\
\rowcolor{gray!20} CCwL*($\mathsf{Eq},\mathsf{D}_{\infty}$)
& 46.9 & 9.0K & 11K & 35K & 2.00 & 101 & 27K & 30.8 & 10/0/0 & 54.8 & 18K & 18K & 46K & 7.30 & 107 & 28K & 188 & 10/0/0 & 46.8 & 140 & 457 & 2.5K & 6.70 & 110 & 29K & 6.43 & 10/0/0 \\
CCwL*($\mathsf{Eq},\mathsf{D}_{\mathsf{sum}}$)
& 46.9 & 9.0K & 11K & 35K & 2.00 & 101 & 27K & 30.6 & 10/0/0 & 54.8 & 18K & 18K & 46K & 7.30 & 107 & 28K & 182 & 10/0/0 & 46.8 & 140 & 457 & 2.5K & 6.70 & 110 & 29K & 6.16 & 10/0/0 \\
\rowcolor{gray!20} CCwL*($\mathsf{Eq},\mathsf{D}_{\mathsf{max}}$)
& 46.9 & 8.9K & 11K & 120K & 514 & 13K & 3.2M & 171 & 3/7/0 & 54.8 & 17K & 18K & 130K & 506 & 19K & 4.9M & 268 & 1/9/0 & 46.8 & 138 & 456 & 2.9K & 11.2 & 120 & 30K & 3.21 & 9/1/0 \\
CCwL*($\mathsf{Eq},\mathsf{D}_{\mathsf{min}}$)
& 46.9 & 8.8K & 12K & 185K & 914 & 22K & 5.6M & 335 & 1/9/0 & 54.8 & 14K & 16K & 337K & 1.8K & 63K & 16M & 730 & 0/10/0 & 46.4 & 134 & 466 & 3.9K & 30.5 & 153 & 34K & 2.12 & 7/3/0 \\
\rowcolor{gray!20} CCwL*($\mathsf{Eq},\mathsf{D}_{0}$)
& 46.9 & 8.3K & 16K & 911K & 5.4K & 133K & 33M & 1.4K & 0/10/0 & 54.3 & 7.0K & 14K & 1.2M & 6.8K & 231K & 59M & 1.6K & 0/9/1 & 46.3 & 133 & 490 & 6.6K & 64.3 & 203 & 41K & 1.35 & 7/3/0 \\
CCwL*($\mathsf{Eq}_0,\mathsf{D}_{\infty}$)
& 46.9 & 9.0K & 11K & 35K & 2.00 & 101 & 27K & 30.4 & 10/0/0 & 54.8 & 18K & 18K & 47K & 7.10 & 107 & 28K & 24.0 & 10/0/0 & 47.3 & 151 & 501 & 2.8K & 6.60 & 110 & 29K & 2.01 & 10/0/0 \\
\rowcolor{gray!20} CCwL*($\mathsf{Eq}_0,\mathsf{D}_{\mathsf{sum}}$)
& 46.9 & 9.0K & 11K & 35K & 2.00 & 101 & 27K & 29.2 & 10/0/0 & 54.8 & 18K & 18K & 47K & 7.10 & 107 & 28K & 23.7 & 10/0/0 & 47.3 & 151 & 501 & 2.8K & 6.60 & 110 & 29K & 1.99 & 10/0/0 \\
CCwL*($\mathsf{Eq}_0,\mathsf{D}_{\mathsf{max}}$)
& 46.9 & 9.0K & 11K & 48K & 85.7 & 2.3K & 565K & 64.2 & 3/7/0 & 54.8 & 18K & 18K & 47K & 7.10 & 107 & 28K & 22.8 & 10/0/0 & 47.3 & 151 & 501 & 2.8K & 6.60 & 110 & 29K & 1.89 & 10/0/0 \\
\rowcolor{gray!20} CCwL*($\mathsf{Eq}_0,\mathsf{D}_{\mathsf{min}}$)
& 46.9 & 8.9K & 11K & 92K & 352 & 8.7K & 2.2M & 196 & 1/9/0 & 54.8 & 14K & 18K & 533K & 3.0K & 109K & 28M & 1.1K & 0/10/0 & 47.3 & 148 & 493 & 2.9K & 10.5 & 116 & 29K & 1.46 & 10/0/0 \\
CCwL*($\mathsf{Eq}_0,\mathsf{D}_{0}$)
& 46.9 & 8.3K & 16K & 911K & 5.4K & 133K & 33M & 1.3K & 0/10/0 & 54.3 & 7.0K & 14K & 1.2M & 6.8K & 231K & 59M & 1.6K & 0/9/1 & 46.4 & 134 & 490 & 6.2K & 58.0 & 191 & 39K & 1.37 & 7/3/0 \\
\rowcolor{gray!20} CCwL*($\mathsf{Uni},\mathsf{D}_{0}$)
& 46.9 & 9.0K & 11K & 29K & 2.20 & 101 & 27K & 93.0 & 10/0/0 & 54.8 & 19K & 19K & 47K & 7.10 & 107 & 28K & 32.5 & 10/0/0 & 47.3 & 152 & 506 & 2.7K & 6.60 & 110 & 28K & 2.06 & 10/0/0 \\
\midrule
&
\multicolumn{9}{c}{\texttt{Rand(Compl(5),RichComp)}}
&
\multicolumn{9}{c}{\texttt{Rand(Star(5),RichComp)}}
&
\multicolumn{9}{c}{\texttt{Rand(Path(5),RichComp)}}
\\\cmidrule(lr){2-10}\cmidrule(lr){11-19}\cmidrule(lr){20-28}
algo.
& st.
& tr.
& OQ reset
& OQ step
& EQs
& EQ reset
& EQ step
& L. time
& valid?
& st.
& tr.
& OQ reset
& OQ step
& EQs
& EQ reset
& EQ step
& L. time
& valid?
& st.
& tr.
& OQ reset
& OQ step
& EQs
& EQ reset
& EQ step
& L. time
& valid?
\\\midrule
\rowcolor{gray!20} MnL*
& --- & --- & --- & --- & --- & --- & --- & --- & 0/0/10 & 41K & 82K & 1.0M & 31M & 22.0 & 272 & 67K & 2.0K & 0/1/9 & 6.8K & 26K & 732K & 23M & 22.1 & 301 & 75K & 361 & 1/9/0 \\
CwL*
& 747 & 239K & 1.3M & 8.4M & 23.6 & 519 & 138K & 20.7 & 10/0/0 & 910 & 22K & 70K & 583K & 40.1 & 637 & 167K & 1.96 & 10/0/0 & 681 & 4.8K & 39K & 404K & 36.4 & 536 & 139K & 1.33 & 10/0/0 \\
\rowcolor{gray!20} CCwL*($\mathsf{Eq},\mathsf{D}_{\infty}$)
& 47.3 & 9.4K & 10K & 31K & 1.90 & 101 & 27K & 23.4 & 10/0/0 & 56.7 & 11K & 11K & 29K & 8.00 & 107 & 28K & 190 & 10/0/0 & 44.4 & 138 & 462 & 2.6K & 7.10 & 106 & 28K & 6.76 & 8/2/0 \\
CCwL*($\mathsf{Eq},\mathsf{D}_{\mathsf{sum}}$)
& 47.3 & 9.4K & 10K & 31K & 1.90 & 101 & 27K & 23.3 & 10/0/0 & 56.7 & 11K & 11K & 29K & 8.00 & 107 & 28K & 182 & 10/0/0 & 44.4 & 138 & 462 & 2.6K & 7.10 & 106 & 28K & 6.60 & 8/2/0 \\
\rowcolor{gray!20} CCwL*($\mathsf{Eq},\mathsf{D}_{\mathsf{max}}$)
& 47.3 & 9.2K & 11K & 178K & 872 & 24K & 6.1M & 343 & 5/5/0 & 56.7 & 9.8K & 10K & 63K & 219 & 8.6K & 2.2M & 162 & 2/8/0 & 44.4 & 138 & 461 & 2.9K & 10.3 & 114 & 29K & 3.33 & 6/4/0 \\
CCwL*($\mathsf{Eq},\mathsf{D}_{\mathsf{min}}$)
& 47.3 & 9.1K & 12K & 308K & 1.7K & 40K & 10M & 523 & 1/9/0 & 56.7 & 8.2K & 9.7K & 214K & 1.1K & 39K & 10M & 529 & 0/10/0 & 44.4 & 137 & 475 & 3.7K & 24.7 & 157 & 37K & 2.06 & 6/4/0 \\
\rowcolor{gray!20} CCwL*($\mathsf{Eq},\mathsf{D}_{0}$)
& 47.2 & 7.8K & 14K & 838K & 5.0K & 120K & 30M & 1.2K & 0/9/1 & 56.7 & 6.0K & 12K & 991K & 5.8K & 187K & 48M & 1.3K & 0/10/0 & 44.4 & 136 & 512 & 7.7K & 68.4 & 216 & 44K & 1.54 & 6/4/0 \\
CCwL*($\mathsf{Eq}_0,\mathsf{D}_{\infty}$)
& 47.3 & 9.4K & 10K & 31K & 1.90 & 101 & 27K & 22.4 & 10/0/0 & 56.7 & 11K & 11K & 29K & 8.00 & 107 & 28K & 16.1 & 10/0/0 & 44.6 & 145 & 482 & 2.7K & 7.00 & 106 & 28K & 1.89 & 8/2/0 \\
\rowcolor{gray!20} CCwL*($\mathsf{Eq}_0,\mathsf{D}_{\mathsf{sum}}$)
& 47.3 & 9.4K & 10K & 31K & 1.90 & 101 & 27K & 22.1 & 10/0/0 & 56.7 & 11K & 11K & 29K & 8.00 & 107 & 28K & 16.3 & 10/0/0 & 44.6 & 145 & 482 & 2.7K & 7.00 & 106 & 28K & 1.93 & 8/2/0 \\
CCwL*($\mathsf{Eq}_0,\mathsf{D}_{\mathsf{max}}$)
& 47.3 & 9.2K & 11K & 131K & 587 & 17K & 4.4M & 297 & 5/5/0 & 56.7 & 11K & 11K & 29K & 8.00 & 107 & 28K & 16.4 & 10/0/0 & 44.6 & 145 & 482 & 2.7K & 7.00 & 106 & 28K & 1.85 & 8/2/0 \\
\rowcolor{gray!20} CCwL*($\mathsf{Eq}_0,\mathsf{D}_{\mathsf{min}}$)
& 47.3 & 9.1K & 12K & 290K & 1.6K & 38K & 9.6M & 535 & 1/9/0 & 56.7 & 8.5K & 10K & 289K & 1.6K & 55K & 14M & 574 & 0/10/0 & 44.6 & 144 & 486 & 2.9K & 11.0 & 112 & 28K & 1.48 & 8/2/0 \\
CCwL*($\mathsf{Eq}_0,\mathsf{D}_{0}$)
& 47.2 & 7.8K & 14K & 838K & 5.0K & 120K & 30M & 1.2K & 0/9/1 & 56.7 & 6.0K & 12K & 991K & 5.8K & 187K & 48M & 1.4K & 0/10/0 & 44.4 & 137 & 505 & 7.0K & 59.2 & 202 & 43K & 1.47 & 6/4/0 \\
\rowcolor{gray!20} CCwL*($\mathsf{Uni},\mathsf{D}_{0}$)
& 47.3 & 9.4K & 11K & 26K & 1.90 & 101 & 27K & 74.3 & 10/0/0 & 56.7 & 11K & 11K & 28K & 8.20 & 108 & 28K & 20.5 & 10/0/0 & 45.0 & 147 & 488 & 2.7K & 7.10 & 106 & 28K & 1.99 & 9/1/0 \\
\midrule
&
\multicolumn{9}{c}{\lighting}
&
\multicolumn{9}{c}{\texttt{BinaryCounter(5)}}
&
\multicolumn{9}{c}{\texttt{BinaryCounter(10)}}
\\\cmidrule(lr){2-10}\cmidrule(lr){11-19}\cmidrule(lr){20-28}
algo.
& st.
& tr.
& OQ reset
& OQ step
& EQs
& EQ reset
& EQ step
& L. time
& valid?
& st.
& tr.
& OQ reset
& OQ step
& EQs
& EQ reset
& EQ step
& L. time
& valid?
& st.
& tr.
& OQ reset
& OQ step
& EQs
& EQ reset
& EQ step
& L. time
& valid?
\\\midrule
\rowcolor{gray!20} MnL*
& 169 & 1.5K & 47K & 1.4M & 10.9 & 238 & 59K & 17.6 & 0/10/0 & 70.0 & 140 & 212 & 5.9K & 2.00 & 101 & 26K & 0.76 & 10/0/0 & --- & --- & --- & --- & --- & --- & --- & --- & 0/0/10 \\
CwL*
& 39.0 & 2.5K & 12K & 61K & 11.8 & 412 & 105K & 0.76 & 10/0/0 & 15.0 & 30.0 & 39.1 & 80.6 & 6.00 & 501 & 129K & 0.41 & 10/0/0 & 30.0 & 60.0 & 74.1 & 141 & 11.0 & 1.0K & 258K & 0.72 & 10/0/0 \\
\rowcolor{gray!20} CCwL*($\mathsf{Eq},\mathsf{D}_{\infty}$)
& 27.0 & 130 & 348 & 2.4K & 4.40 & 112 & 28K & 1.36 & 10/0/0 & 14.0 & 25.0 & 30.0 & 45.0 & 1.00 & 100 & 26K & 0.88 & 10/0/0 & 29.0 & 50.0 & 60.0 & 90.0 & 1.00 & 100 & 26K & 2.27 & 10/0/0 \\
CCwL*($\mathsf{Eq},\mathsf{D}_{\mathsf{sum}}$)
& 27.0 & 130 & 349 & 2.7K & 4.10 & 110 & 28K & 1.34 & 10/0/0 & 14.0 & 25.0 & 37.0 & 298 & 8.00 & 107 & 26K & 0.88 & 10/0/0 & 25.0 & 41.0 & 63.0 & 1.9K & 13.0 & 117 & 28K & 1.80 & 0/10/0 \\
\rowcolor{gray!20} CCwL*($\mathsf{Eq},\mathsf{D}_{\mathsf{max}}$)
& 27.0 & 130 & 354 & 3.1K & 9.60 & 116 & 28K & 1.35 & 10/0/0 & 14.0 & 25.0 & 40.0 & 329 & 11.0 & 110 & 26K & 0.87 & 10/0/0 & 25.0 & 41.0 & 69.0 & 1.9K & 19.0 & 123 & 28K & 1.76 & 0/10/0 \\
CCwL*($\mathsf{Eq},\mathsf{D}_{\mathsf{min}}$)
& 27.0 & 129 & 386 & 5.2K & 45.6 & 227 & 48K & 1.77 & 8/2/0 & 14.0 & 25.0 & 42.5 & 347 & 13.5 & 113 & 26K & 0.86 & 10/0/0 & 25.0 & 41.0 & 71.5 & 2.0K & 21.5 & 126 & 28K & 1.76 & 0/10/0 \\
\rowcolor{gray!20} CCwL*($\mathsf{Eq},\mathsf{D}_{0}$)
& 27.0 & 129 & 412 & 7.4K & 72.5 & 292 & 59K & 1.94 & 10/0/0 & 14.0 & 25.0 & 44.4 & 357 & 15.4 & 115 & 26K & 0.88 & 10/0/0 & 25.0 & 41.0 & 73.4 & 2.0K & 23.4 & 128 & 28K & 1.73 & 0/10/0 \\
CCwL*($\mathsf{Eq}_0,\mathsf{D}_{\infty}$)
& 27.0 & 134 & 362 & 2.8K & 4.60 & 116 & 29K & 1.21 & 10/0/0 & 14.0 & 25.0 & 30.0 & 45.0 & 1.00 & 100 & 26K & 0.87 & 10/0/0 & 29.0 & 50.0 & 60.0 & 90.0 & 1.00 & 100 & 26K & 2.27 & 10/0/0 \\
\rowcolor{gray!20} CCwL*($\mathsf{Eq}_0,\mathsf{D}_{\mathsf{sum}}$)
& 27.0 & 134 & 360 & 2.6K & 4.40 & 119 & 30K & 1.16 & 10/0/0 & 14.0 & 25.0 & 37.0 & 298 & 8.00 & 107 & 26K & 0.88 & 10/0/0 & 25.0 & 41.0 & 63.0 & 1.9K & 13.0 & 117 & 28K & 1.81 & 0/10/0 \\
CCwL*($\mathsf{Eq}_0,\mathsf{D}_{\mathsf{max}}$)
& 27.0 & 134 & 364 & 2.9K & 8.60 & 123 & 30K & 1.14 & 10/0/0 & 14.0 & 25.0 & 40.0 & 329 & 11.0 & 110 & 26K & 0.88 & 10/0/0 & 25.0 & 41.0 & 69.0 & 1.9K & 19.0 & 123 & 28K & 1.79 & 0/10/0 \\
\rowcolor{gray!20} CCwL*($\mathsf{Eq}_0,\mathsf{D}_{\mathsf{min}}$)
& 27.0 & 129 & 399 & 6.8K & 59.7 & 267 & 56K & 1.83 & 9/1/0 & 14.0 & 25.0 & 42.5 & 347 & 13.5 & 113 & 26K & 0.84 & 10/0/0 & 25.0 & 41.0 & 71.5 & 2.0K & 21.5 & 126 & 28K & 1.74 & 0/10/0 \\
CCwL*($\mathsf{Eq}_0,\mathsf{D}_{0}$)
& 27.0 & 129 & 413 & 7.6K & 71.5 & 311 & 64K & 1.96 & 9/1/0 & 14.0 & 25.0 & 44.4 & 357 & 15.4 & 115 & 26K & 0.87 & 10/0/0 & 25.0 & 41.0 & 73.4 & 2.0K & 23.4 & 128 & 28K & 1.78 & 0/10/0 \\
\rowcolor{gray!20} CCwL*($\mathsf{Uni},\mathsf{D}_{0}$)
& 28.0 & 488 & 1.3K & 7.2K & 4.60 & 116 & 29K & 2.54 & 10/0/0 & 14.0 & 28.0 & 33.0 & 54.0 & 1.00 & 100 & 26K & 0.87 & 10/0/0 & 29.0 & 58.0 & 68.0 & 114 & 1.00 & 100 & 26K & 9.73 & 10/0/0 \\
\bottomrule
\end{tabular}
}
\end{adjustbox}
\end{table}

\begin{table}[t]
\caption{%
experiment results II (extended). 
The legend is the same as \cref{table:exprI}
}\label{table:exprIIext} 
\scalebox{.7}{ 
\begin{tabular}{@{}p{2.6cm}%
*{8}{R{\dimexpr.8cm-2\tabcolsep\relax}}R{\dimexpr1.2cm-2\tabcolsep\relax}
*{8}{R{\dimexpr.8cm-2\tabcolsep\relax}}R{\dimexpr1.2cm-2\tabcolsep\relax}
@{}}
\toprule
&
\multicolumn{9}{c}{\texttt{Rand(Star(3),LeanComp)}}
&
\multicolumn{9}{c}{\texttt{Rand(Star(7),LeanComp)}}
\\\cmidrule(lr){2-10}\cmidrule(lr){11-19}
algo.
& st. 
& tr. 
& OQ reset
& OQ step
& EQ
& EQ reset
& EQ step
& L. time
& valid?
& st.
& tr.
& OQ reset
& OQ step
& EQs
& EQ reset
& EQ step
& L. time
& valid?
\\\midrule
\rowcolor{gray!20} MnL*
& 3.3K & 13K & 347K & 6.8M & 26.8 & 474 & 119K & 89.1 & 1/9/0 & --- & --- & --- & --- & --- & --- & --- & --- & 0/0/10 \\
CwL*
& 38.4 & 1.7K & 2.6K & 6.9K & 12.9 & 409 & 110K & 0.50 & 10/0/0 & 74.7 & 280K & 280K & 534K & 26.6 & 819 & 220K & 5.01 & 10/0/0 \\
\rowcolor{gray!20} CCwL*($\mathsf{Eq},\mathsf{D}_{\infty}$)
& 38.4 & 1.3K & 2.0K & 6.5K & 6.80 & 106 & 28K & 3.57 & 10/0/0 & 66.0 & 9.4K & 10K & 27K & 9.00 & 108 & 28K & 1.1K & 1/0/9 \\
CCwL*($\mathsf{Eq},\mathsf{D}_{\mathsf{sum}}$)
& 38.4 & 1.3K & 2.0K & 6.5K & 6.80 & 106 & 28K & 3.38 & 10/0/0 & 66.0 & 9.4K & 10K & 27K & 9.00 & 108 & 28K & 1.2K & 1/0/9 \\
\rowcolor{gray!20} CCwL*($\mathsf{Eq},\mathsf{D}_{\mathsf{max}}$)
& 38.4 & 1.3K & 2.0K & 8.0K & 16.8 & 367 & 93K & 4.61 & 7/3/0 & 73.2 & 53K & 55K & 305K & 886 & 44K & 12M & 1.8K & 0/4/6 \\
CCwL*($\mathsf{Eq},\mathsf{D}_{\mathsf{min}}$)
& 38.4 & 1.3K & 2.1K & 15K & 68.2 & 830 & 199K & 6.06 & 2/8/0 & 66.0 & 6.7K & 12K & 742K & 4.3K & 155K & 40M & 1.7K & 0/1/9 \\
\rowcolor{gray!20} CCwL*($\mathsf{Eq},\mathsf{D}_{0}$)
& 38.4 & 1.1K & 2.7K & 160K & 1.0K & 16K & 3.9M & 79.8 & 2/8/0 & 66.0 & 6.4K & 13K & 1.0M & 6.2K & 200K & 51M & 1.9K & 0/1/9 \\
CCwL*($\mathsf{Eq}_0,\mathsf{D}_{\infty}$)
& 38.4 & 1.3K & 2.1K & 6.6K & 6.90 & 106 & 28K & 1.65 & 10/0/0 & 74.5 & 179K & 179K & 447K & 10.0 & 109 & 28K & 646 & 10/0/0 \\
\rowcolor{gray!20} CCwL*($\mathsf{Eq}_0,\mathsf{D}_{\mathsf{sum}}$)
& 38.4 & 1.3K & 2.1K & 6.6K & 6.90 & 106 & 28K & 1.67 & 10/0/0 & 74.5 & 179K & 179K & 447K & 10.0 & 109 & 28K & 654 & 10/0/0 \\
CCwL*($\mathsf{Eq}_0,\mathsf{D}_{\mathsf{max}}$)
& 38.4 & 1.3K & 2.1K & 6.6K & 6.90 & 106 & 28K & 1.67 & 10/0/0 & 74.5 & 179K & 179K & 447K & 10.0 & 109 & 28K & 638 & 10/0/0 \\
\rowcolor{gray!20} CCwL*($\mathsf{Eq}_0,\mathsf{D}_{\mathsf{min}}$)
& 38.4 & 1.3K & 2.2K & 32K & 178 & 2.2K & 511K & 12.4 & 4/6/0 & 71.7 & 212K & 213K & 596K & 445 & 14K & 3.7M & 1.4K & 0/3/7 \\
CCwL*($\mathsf{Eq}_0,\mathsf{D}_{0}$)
& 38.4 & 1.1K & 2.7K & 160K & 1.0K & 16K & 3.9M & 77.9 & 2/8/0 & 66.0 & 6.4K & 13K & 1.0M & 6.2K & 200K & 51M & 2.0K & 0/1/9 \\
\rowcolor{gray!20} CCwL*($\mathsf{Uni},\mathsf{D}_{0}$)
& 38.4 & 1.4K & 2.2K & 6.8K & 6.40 & 106 & 28K & 1.98 & 10/0/0 & 74.5 & 183K & 183K & 415K & 9.50 & 109 & 28K & 693 & 10/0/0 \\
\midrule
&
\multicolumn{9}{c}{\texttt{Rand(Star(3),RichComp)}}
&
\multicolumn{9}{c}{\texttt{Rand(Star(7),RichComp)}}
\\\cmidrule(lr){2-10}\cmidrule(lr){11-19}
algo.
& st.
& tr.
& OQ reset
& OQ step
& EQs
& EQ reset
& EQ step
& L. time
& valid?
& st.
& tr.
& OQ reset
& OQ step
& EQs
& EQ reset
& EQ step
& L. time
& valid?
\\\midrule
\rowcolor{gray!20} MnL*
& 3.2K & 12K & 238K & 4.1M & 22.1 & 298 & 74K & 56.0 & 1/9/0 & --- & --- & --- & --- & --- & --- & --- & --- & 0/0/10 \\
CwL*
& 540 & 5.6K & 37K & 360K & 27.4 & 430 & 112K & 1.21 & 10/0/0 & 1.2K & 471K & 535K & 1.6M & 57.2 & 851 & 222K & 9.01 & 10/0/0 \\
\rowcolor{gray!20} CCwL*($\mathsf{Eq},\mathsf{D}_{\infty}$)
& 37.4 & 1.3K & 2.2K & 7.0K & 4.90 & 104 & 28K & 2.76 & 10/0/0 & 73.5 & 100K & 101K & 279K & 10.5 & 110 & 28K & 2.3K & 4/0/6 \\
CCwL*($\mathsf{Eq},\mathsf{D}_{\mathsf{sum}}$)
& 37.4 & 1.3K & 2.2K & 7.0K & 4.90 & 104 & 28K & 2.67 & 10/0/0 & 73.5 & 100K & 101K & 279K & 10.5 & 110 & 28K & 2.2K & 4/0/6 \\
\rowcolor{gray!20} CCwL*($\mathsf{Eq},\mathsf{D}_{\mathsf{max}}$)
& 37.4 & 1.3K & 2.2K & 11K & 29.6 & 828 & 211K & 6.85 & 6/4/0 & 74.5 & 37K & 38K & 198K & 584 & 30K & 8.0M & 1.5K & 0/2/8 \\
CCwL*($\mathsf{Eq},\mathsf{D}_{\mathsf{min}}$)
& 37.4 & 1.2K & 2.3K & 30K & 153 & 3.4K & 853K & 21.2 & 2/8/0 & 75.0 & 10K & 14K & 664K & 3.7K & 162K & 42M & 2.5K & 0/1/9 \\
\rowcolor{gray!20} CCwL*($\mathsf{Eq},\mathsf{D}_{0}$)
& 37.4 & 1.2K & 3.0K & 163K & 1.1K & 16K & 4.0M & 85.4 & 1/9/0 & 75.0 & 8.3K & 17K & 1.4M & 8.1K & 297K & 76M & 3.5K & 0/1/9 \\
CCwL*($\mathsf{Eq}_0,\mathsf{D}_{\infty}$)
& 37.4 & 1.3K & 2.3K & 7.5K & 5.30 & 105 & 28K & 1.54 & 10/0/0 & 74.1 & 204K & 204K & 520K & 9.50 & 108 & 28K & 616 & 10/0/0 \\
\rowcolor{gray!20} CCwL*($\mathsf{Eq}_0,\mathsf{D}_{\mathsf{sum}}$)
& 37.4 & 1.3K & 2.3K & 7.5K & 5.30 & 105 & 28K & 1.53 & 10/0/0 & 74.1 & 204K & 204K & 520K & 9.50 & 108 & 28K & 607 & 10/0/0 \\
CCwL*($\mathsf{Eq}_0,\mathsf{D}_{\mathsf{max}}$)
& 37.4 & 1.3K & 2.3K & 7.5K & 5.30 & 105 & 28K & 1.47 & 10/0/0 & 74.1 & 204K & 204K & 520K & 9.50 & 108 & 28K & 603 & 10/0/0 \\
\rowcolor{gray!20} CCwL*($\mathsf{Eq}_0,\mathsf{D}_{\mathsf{min}}$)
& 37.4 & 1.3K & 2.3K & 22K & 102 & 1.7K & 433K & 10.6 & 3/7/0 & 73.5 & 216K & 219K & 921K & 2.4K & 93K & 24M & 1.9K & 0/2/8 \\
CCwL*($\mathsf{Eq}_0,\mathsf{D}_{0}$)
& 37.4 & 1.2K & 3.0K & 163K & 1.1K & 16K & 4.0M & 83.4 & 1/9/0 & 75.0 & 8.3K & 17K & 1.4M & 8.1K & 297K & 76M & 3.0K & 0/1/9 \\
\rowcolor{gray!20} CCwL*($\mathsf{Uni},\mathsf{D}_{0}$)
& 37.4 & 1.4K & 2.4K & 7.6K & 5.00 & 104 & 28K & 1.73 & 10/0/0 & 74.2 & 214K & 215K & 487K & 9.80 & 109 & 28K & 676 & 10/0/0 \\
\bottomrule
\end{tabular}
}
\end{table}

\end{ArxivBlock}

\end{document}